\documentclass[prd,10pt,nofootinbib,twocolumn,superscriptaddress,preprintnumbers,balancelastpage,longbibliography]{revtex4-1}

\usepackage{amsmath,amssymb}	
\usepackage{mathtools}
\usepackage{slashed}
\usepackage{xspace}
\usepackage{comment}
\usepackage{braket}
\usepackage{graphicx}
\usepackage{float}
\usepackage{siunitx}
\usepackage{physics}
\usepackage{fontawesome}
\usepackage{booktabs}
\usepackage{anyfontsize}
\usepackage{aas_macros}
\usepackage{listings}
\usepackage{inconsolata}
\usepackage{multirow}
\usepackage{tikz}
\tikzset{every picture/.style={line width=0.75pt}}
\usetikzlibrary{calc}
\usetikzlibrary{decorations.pathmorphing}

\definecolor{linkcolor}{rgb}{0.0, 0.28, 0.67}
\definecolor{userInputColor}{HTML}{4287f5}
\definecolor{userCallColor}{HTML}{fcba03}
\definecolor{codeCallColor}{HTML}{039c33}

\usepackage[
   colorlinks=true,
    urlcolor=linkcolor,
   anchorcolor=linkcolor,
    citecolor=linkcolor,
    filecolor=linkcolor,
    linkcolor=linkcolor,
    menucolor=linkcolor,
    linktocpage=true,
    pdfproducer=medialab,
    pdfa=true
]{hyperref}

\usepackage{listings}

\definecolor{softgreen}{rgb}{0.85,0.95,0.85}
\definecolor{softblue}{rgb}{0.99,0.99,1.0}
\definecolor{softpurple}{rgb}{0.95,0.90,1.0}
\definecolor{darkblue}{rgb}{0,0.2,0.4}
\definecolor{darkgreen}{rgb}{0.1,0.4,0.1}
\definecolor{darkpurple}{rgb}{0.4,0.1,0.4}

\lstdefinestyle{pystyle}{
    language = Python,  
    backgroundcolor=\color{softblue},
    commentstyle=\color{darkgreen},
    keywordstyle=\color{darkblue},
    stringstyle=\color{darkpurple},
    basicstyle=\ttfamily\small\selectfont,,
    breakatwhitespace=false,         
    breaklines=true,                 
    captionpos=b,                    
    keepspaces=true,                 
    numbers=none,                    
    numbersep=10pt,                  
    numberstyle=\tiny\color{darkblue},
    showspaces=false,                
    showstringspaces=false,
    showtabs=false,                  
    tabsize=2,
    otherkeywords={
        as,
        len,True,False},
    columns=flexible,
    frame=single,
    framesep=2pt,
    framerule=0.4pt,
    rulecolor=\color{darkblue},
    xleftmargin=\parindent,
    aboveskip=10pt,
    belowskip=10pt,
}

\lstdefinestyle{mathstyle}{
    language = Mathematica,  
    commentstyle=\color{codegreen},
    keywordstyle=\color[rgb]{0,0,0.75},
    keywordstyle=[1]\color[rgb]{0,0,0.75},
    keywordstyle=[2]\color[rgb]{0,0,0.75},
    keywordstyle=[3]\color[rgb]{0,0,0.75},
    keywordstyle=[4]\color[rgb]{0,0,0.75},
    commentstyle=\color[rgb]{0.133,0.545,0.133},
    stringstyle=\color{codepurple},
    basicstyle={\ttfamily \small},
    breakatwhitespace=false,         
    breaklines=true,                 
    captionpos=b,                    
    keepspaces=false,                 
    numbers=none,                    
    numbersep=5pt,                  
    showspaces=false,                
    showstringspaces=false,
    showtabs=false,                  
    tabsize=2,
    morekeywords={True, False, len},
    columns=flexible
}

\DeclareSIUnit\electronvolt{e\kern-.05em V}
\newcommand{\githubicon}{\href{https://github.com/cgiovanetti/LINX}{\faGithub}\xspace}

\defcitealias{Giovanetti_2024}{Paper~II}

\begin{document}
\title{LINX: A Fast, Differentiable, and Extensible Big Bang Nucleosynthesis Package}
\author{Cara Giovanetti}
\email{cg3566@nyu.edu}
\thanks{ORCID: \href{https://orcid.org/0000-0003-1611-3379}{0000-0003-1611-3379}}
\affiliation{Center for Cosmology and Particle Physics, Department of Physics, New York University, New York, NY 10003, USA}

\author{Mariangela Lisanti}
\email{mlisanti@princeton.edu}
\thanks{ORCID: \href{https://orcid.org/0000-0002-8495-8659}{0000-0002-8495-8659}}
\affiliation{Department of Physics, Princeton University, Princeton, NJ 08544, USA}
\affiliation{Center for Computational Astrophysics, Flatiron Institute, 162 Fifth Ave, New York, NY 10010, USA}

\author{Hongwan Liu}
\email{hongwan@bu.edu}
\thanks{ORCID: \href{https://orcid.org/0000-0003-2486-0681}{0000-0003-2486-0681}}
\affiliation{Physics Department, Boston University, Boston, MA 02215, USA}
\affiliation{Kavli Institute for Cosmological Physics, University of Chicago, Chicago, IL 60637}
\affiliation{Theoretical Physics Department, Fermi National Accelerator Laboratory, Batavia, IL 60510}

\author{Siddharth Mishra-Sharma}
\email{smsharma@mit.edu}
\thanks{ORCID: \href{https://orcid.org/0000-0001-9088-7845}{0000-0001-9088-7845}}
\thanks{Currently at Anthropic; work performed while at MIT/IAIFI\@.}
\affiliation{The NSF AI Institute for Artificial Intelligence and Fundamental Interactions}
\affiliation{Center for Theoretical Physics, Massachusetts Institute of Technology, Cambridge, MA 02139, USA}

\author{Joshua T. Ruderman}
\email{ruderman@nyu.edu}
\thanks{ORCID: \href{https://orcid.org/0000-0001-6051-9216}{0000-0001-6051-9216}}
\affiliation{Center for Cosmology and Particle Physics, Department of Physics, New York University, New York, NY 10003, USA}

\preprint{MIT-CTP/5736}

\date{\today}
\begin{abstract} 
We introduce LINX (Light Isotope Nucleosynthesis with JAX), a new differentiable public Big Bang Nucleosynthesis (BBN) code designed for fast parameter estimation. By leveraging JAX, LINX achieves both speed and differentiability, enabling the use of Bayesian inference, including gradient-based methods. We discuss the formalism used in LINX for rapid primordial elemental abundance predictions and give examples of how LINX can be used. When combined with differentiable Cosmic Microwave Background (CMB) power spectrum emulators, LINX can be used for joint CMB and BBN analyses without requiring extensive computational resources, including on personal hardware.~\githubicon
\end{abstract}
\maketitle

\section{Introduction}

The era of precision cosmology has created demand for detailed numerical estimates of physical quantities, often achieved through sophisticated numerical codes.  These tools combine with the unprecedented sensitivity of cosmological experiments and surveys to test both the prevailing cosmological model, Lambda Cold Dark Matter~($\Lambda$CDM), and alternative scenarios with great precision.

Many such tests rely on probes of the Cosmic Microwave Background (CMB), Big Bang Nucleosynthesis (BBN), or some combination of the two, for discoveries and constraints (see Refs.~\cite{Berlin:2017ftj,Sabti:2019mhn,Depta:2020zbh,Sabti:2021reh,Giovanetti:2021izc,Adshead:2022ovo,An:2022sva,Burns_2022,Giovanetti:2024orj} for recent examples).  
Predicting the CMB anisotropy power spectrum in a variety of cosmological scenarios can be done using Boltzmann codes such as CLASS~\cite{lesgourgues_2011a,lesgourgues_2011b,lesgourgues_2011c,lesgourgues_2011d} or CAMB~\cite{Lewis_1999,Howlett_2012}.
Meanwhile, a number of public options exist for BBN codes as well, including AlterBBN \cite{Arbey_2012,arbey2019}, PArthENoPE \cite{Consiglio_2018}, PRIMAT \cite{Pitrou_2018}, and PRyMordial \cite{burns_2023}, which variously provide the user with options to predict the abundances of light elements in standard cosmology and in select new-physics scenarios. 

BBN and CMB data can place comparable constraints on important cosmological parameters; for example, both datasets are able to determine the baryon density $\Omega_b h^2$ and the effective number of neutrinos $N_\text{eff}$ with percent-level and ten-percent-level uncertainty, respectively. 
This makes a joint analysis involving data from both epochs desirable for understanding both $\Lambda$CDM and new physics.  
Parameter inference with CMB data typically involves estimating a high-dimensional posterior probability distribution, or constructing likelihoods profiled over a large number of parameters. 
Nuisance parameters stemming from measurement uncertainties associated with CMB experiments like Planck~\cite{Planck:2018vyg}  are often accounted for and marginalized over.

In both CMB and BBN analyses, the use of Monte Carlo (MC) samplers and other efficient computational methods in high-dimensional spaces is crucial. 
Moreover, the CMB power spectrum prediction itself depends on the primordial abundance of helium-4, which needs to be self-consistently computed by a BBN code before a Boltzmann code can correctly make a power spectrum prediction. 
A consistent, joint CMB and BBN analysis---especially with the inclusion of new physics---can therefore only be performed with a BBN code that is at least as fast as currently available Boltzmann codes, and that is amenable to the techniques used in CMB analyses. 

In this paper, we introduce a new, accurate BBN code, Light Isotope Nucleosynthesis with JAX (LINX), which is fast enough to use with the kinds of MC samplers that have become commonplace in CMB studies.  
Written in JAX~\cite{jax2018github,deepmind2020jax}, a Python library designed to perform fast and differentiable numerical computation, LINX makes it possible to compute primordial elemental abundances in tandem with the CMB power spectrum given a large set of varying cosmological and nuisance parameters. 
It exploits the ability of JAX code to be compiled into low-level, performant instructions at runtime and to be easily vectorized.  
At the same time, the high-level, Python-based syntax of LINX makes the code easy to understand and modify.

The differentiable nature of LINX allows users to employ gradient-based methods of parameter inference, such as Hamiltonian MC or variational inference, greatly improving the efficiency of data analysis pipelines. 
These are more popular than ever in cosmology and particle physics, with several options for users to choose from for a variety of calculations---for example, MadJax~\cite{Heinrich_2023} for matrix elements in particle physics, the differentiable cosmology library JAX-COSMO~\cite{Campagne_2023}, the Boltzmann solver DISCO-DJ~\cite{hahn2023discodj}, and differentiable CMB emulators like CosmoPower~\cite{Spurio_Mancini_2022} and Capse.jl~\cite{Bonici_2024}.  
LINX's differentiability makes it possible to perform efficient gradient-based sampling, either as a standalone code, or as part of fully-differentiable pipelines combined with differentiable CMB codes.  

In Ref.~\cite{Giovanetti_2024} (hereinafter \citetalias{Giovanetti_2024}), we demonstrate the capabilities of LINX by using nested sampling to perform a consistent joint CMB and BBN fit to the $\Lambda$CDM parameters with LINX and CLASS. This paper presents the structure of the code and the formalism required for these kinds of  joint analyses.  The goal of LINX is to provide a fast, accurate, and differentiable BBN code, which is user-friendly and extensible for new physics.  Following a brief overview of BBN and other public codes in Section~\ref{sec:BBN}, we address user-friendliness and extensibility in Section~\ref{sec:structure}, which details the structure and philosophy of the code. We address the questions of speed and accuracy next, assessing performance and validating the assumptions and new techniques used in LINX by comparing its output to existing BBN codes in Section~\ref{sec:validate_limitate}\@.  In Section~\ref{sec:examples}, we provide pedagogical examples of calculations that can be performed efficiently with LINX, namely exploring the impact of nuclear rate uncertainties on BBN predictions~(Section~\ref{sec:uncert_impact}) and making the Schramm plot, which displays the abundances as a function of the baryon density, including uncertainties on nuclear rates (Section~\ref{sec:schramm}).  We finally demonstrate the benefits of differentiability in LINX in Section~\ref{sec:gradients}, where we review gradient-assisted inference methods, new methods of sampling opened up to BBN analyses by LINX\@. 
We then perform a joint CMB and BBN fit for the $\Lambda$CDM + $N_\text{eff}$ model by combining LINX with CosmoPower, in order to demonstrate the capabilities of LINX in a fully-differentiable joint-analysis pipeline.  
We summarize and discuss the implications of these results in Section~\ref{sec:discussion}.

This paper also includes three appendices to supplement the main text.  Appendix~\ref{app:special_functions} includes details about the LINX implementation of a number of special functions that are not included with JAX libraries.  Appendix~\ref{app:series} includes details about the series method used to speed up the computation of thermodynamic integrals.  Finally, Appendix~\ref{app:gradients} includes details about gradient-based sampling methods meant to supplement Section~\ref{sec:gradients}.

The units used in LINX are \SI{}{\mega\eV} for all thermodynamics quantities (temperature, pressure, energy density, {\it etc.}), but seconds for time and rates, including the Hubble parameter. 
Exceptions to this are made in an effort to respect nuclear physics conventions and are highlighted in the code.
Equations in the paper are written with $\hbar = c = k_B = 1$. 

LINX is publicly available at \url{https://github.com/cgiovanetti/LINX} \githubicon under the MIT License. If you use LINX, we suggest you cite this work and the other public codes that provide a foundation for LINX: nudec\_BSM~\cite{Escudero_2019,Abenza_2020}, PRIMAT~\cite{Pitrou_2018}, and PRyMordial~\cite{burns_2023}.

\section{A BBN Primer}\label{sec:BBN}

BBN is the process by which elements are formed through nuclear reactions in the early Universe. 
As the Universe cools from \SI{}{\mega\eV} to \SI{}{\kilo\eV} temperatures, free protons and neutrons  
 combine to form increasingly heavier elements, until the temperature drops below nuclear reaction thresholds, at which point the elemental abundances freeze out. 
By comparing theoretical predictions of these elemental abundances with measurements taken in pristine systems that still have a primordial composition, we can test cosmological models of the early Universe and look for signs of new physics during one of the earliest epochs we have direct access to. 

Here, we provide a brief overview of the main processes occurring during BBN as a precursor to discussing their implementation in LINX\@.  For pedagogical reviews on BBN, see \textit{e.g.} Refs.~\cite[Section~24]{ParticleDataGroup:2022pth},~\cite[Chapter~4]{Kolb:1990vq},~\cite[Chapters~1,4]{Dodelson_2020}, and~\cite{Mukhanov_2004}.

The main epochs of interest during BBN are as follows:
\begin{enumerate}
    \item $T_\gamma \gg \SI{}{\mega\eV}$: \textit{chemical equilibrium between electromagnetic~(EM) and neutrino sectors}. 
    At such high temperatures, weak interactions between neutrinos $\nu$ and the EM sector---composed of neutrons $n$, protons $p$, and electrons $e$ (and positrons)---are sufficiently rapid that all of these particles are in chemical equilibrium with each other.
    This equilibrium state forms the initial conditions for BBN\@. 
    While $n$ and $p$ are in chemical equilibrium, the number density of $n$ is suppressed by the Boltzmann factor $\exp(-Q/T)$, where $Q \equiv m_n - m_p \approx \SI{1.29}{\mega\eV}$ is the mass splitting between $n$ and $p$. 
    \item $T_\gamma \sim \SI{}{\mega\eV}$: \textit{neutrino decoupling}. 
    At this temperature, the interaction rates of the various weak processes maintaining equilibrium between the EM and neutrino sectors drop below the Hubble expansion rate. 
    The neutrino sector thermally decouples from the EM sector.
    Interconversion between $n$ and $p$ through weak processes becomes inefficient; as a result, $n$ and $p$ fall out of chemical equilibrium, and the number density of $n$ is no longer exponentially suppressed.
    \item $T_\gamma\sim \SI{}{\mega\eV} - \SI{80}{\kilo\eV}$: \textit{deuterium bottleneck}. After neutrino decoupling, $n$ and $p$ are constantly interacting with each other to form deuterium $d$, the first step in the formation of heavier nuclei. 
    However, the baryon-to-photon ratio is small; the large number density of photons efficiently splits $d$ back into $n$ and $p$, even when $T_\gamma$ is less than the binding energy of $d$ (approximately \SI{2}{\mega\eV}). 
    Consequently, $d$ cannot form until $T_\gamma \lesssim \SI{80}{\kilo\eV}$. 
    The epoch between neutrino decoupling and $T_\gamma \sim \SI{80}{\kilo\eV}$ has earned the name the ``deuterium bottleneck''; during this period, $n$ and $p$ still exist mainly as free particles and not as bound states, delaying the formation of significant amounts of nuclei.  
    While the deuterium bottleneck, which lasts for about $\sim \SI{e2}{\second}$, is effective, neutrons decay away with a half-life on the order of \SI{e3}{\second}, leading to a small but important depletion of free neutrons.
    
    \item $T_\gamma \lesssim \SI{80}{\kilo\eV}$: \textit{nuclear reactions}. Once the deuterium bottleneck is overcome, nuclear reactions become effective, and heavier elements form rapidly. 
    These reactions continue until the Universe cools below $\sim\SI{10}{\kilo\eV}$,\footnote{Any residual free neutrons decay completely away below this temperature.} leaving only a short window of time for nucleosynthesis to occur. 
    Due to its large binding energy, helium-4 is the most favored nucleus to form; the vast majority of the free neutrons present at the end of the deuterium bottleneck are ultimately locked up in $^4$He. 
    The abundances of all elements freeze out due to nuclear reactions becoming slower than the Hubble expansion rate. 
    
\end{enumerate}
At the end of BBN, we are left with light, stable nuclei with abundances that can be predicted and compared to experimental measurements of the primordial abundance of various light nuclei (such as deuterium and helium-4) in old, pristine astrophysical systems. 
The predicted abundances therefore depend sensitively on the composition of the Universe, the thermodynamic properties of the various components, electroweak physics, and nuclear physics, making BBN a unique probe of early Universe cosmology and new physics.

In this paper, we restrict ourselves to predicting abundances in two scenarios.  The first is standard BBN (henceforth ``SBBN"), or BBN with no new physics.  The second we call SBBN+$\Delta N_\text{eff}$, where we add an inert, relativistic species with an arbitrary---and possibly negative\footnote{While a negative energy density is unphysical, this could be a proxy at late times ($T \ll \SI{}{\mega\eV}$) for models where neutrinos are cooled below their SBBN temperature.}---energy density.  This energy density scales as $\rho\propto a^{-4}$, and increases (decreases) the expansion rate for $\rho>0$ ($\rho<0$).
In these scenarios, there are two cosmological parameters that can be inferred from the primordial abundances of light elements: 
\begin{enumerate} 
    \item \textit{Baryon abundance $\Omega_b h^2$}.\footnote{This quantity is directly proportional to the baryon-to-photon ratio $\eta$.} 
    This quantity strongly affects all nuclear reaction rates, but in particular controls the freezeout of $d$ and its final abundance. 
    State-of-the-art observations of the primordial abundance of deuterium~\cite{Cooke_2018} yield a percent-level determination of $\Omega_b h^2$ using BBN~\cite{schoneberg2024} (see also \citetalias{Giovanetti_2024}); 
    \item $\Delta N_\text{eff}$. 
    Additional relativistic degrees of freedom during neutrino decoupling and the deuterium bottleneck epoch increase the expansion rate of the Universe, strongly impacting the abundance of neutrons at the end of the deuterium bottleneck and, ultimately, the final helium-4 abundance.
    It also affects the freezeout of all the other light elements. 
\end{enumerate}

\begin{figure*}
    \centering
    \begin{tikzpicture}
        % inputs
        \coordinate (inputs) at (-5,0);
        \node[draw, fill=userInputColor!20, rectangle, rounded corners, inner sep=5pt,anchor=west] (inArray) at (inputs) {$\begin{pmatrix}
            \texttt{Delt\_Neff\_init}\\
            \textrm{options} \\
        \end{pmatrix}$};
        \coordinate (dNeffarrowStart) at (inArray.east);

        \node[anchor=north,align=center] (legendUserInputs) at ($(inArray.south)-(-.8,.7)$) {\textcolor{userInputColor}{User Inputs}};

        \node[anchor=north,align=center] (legendUserCalls) at ($(legendUserInputs.south)-(0,.2)$) {\textcolor{userCallColor}{User Calls}};

        \node[anchor=north,align=center] (legendInternalCalls) at ($(legendUserCalls.south)-(0,.2)$) {\textcolor{codeCallColor}{Internal Calls}};
        
        \draw[->,thick,userInputColor] (dNeffarrowStart) -- ++(1,0) coordinate (dNeffArrow);

        \coordinate (abundancesInputs) at ($(inArray.north)+(0,1)$);
        \node[draw, fill=userInputColor!20, rectangle, rounded corners, inner sep=5pt,anchor=center] (abundancesInputBlock) at (abundancesInputs) {$\begin{pmatrix}
            \Omega_b h^2\\
            \tau_n\\
            \textrm{nuclear rates}\\
            \textrm{options} \\
        \end{pmatrix}$};

        % background cycle
        \node[draw, fill=userCallColor!20, rectangle, rounded corners, inner sep=19pt,anchor=west] (background) at (dNeffArrow) {\texttt{background.py}};
        \coordinate (backgroundOutputArrowStart) at ($(background.east)$);

        % \node[anchor=south,align=center] at ($(background.north)+(0,.001)$) {\textcolor{userCallColor}{User}\\ \textcolor{userCallColor}{Calls}};

        \draw[->,thick] (backgroundOutputArrowStart) -- ++(1,0) coordinate (backgroundOutputArrow);

        \node[anchor=west] (backgroundOutputs) at (backgroundOutputArrow) {$\begin{pmatrix}
            t\\
            a\\
            \rho_{\gamma} \\
            \rho_{\nu}\\
            \rho_{\textrm{ext}}
        \end{pmatrix}$};
        \coordinate (backgroundAbundancesInputArrowStart) at ($(backgroundOutputs.east)$);

        \draw[->,thick] (backgroundAbundancesInputArrowStart) -- ++(1.6,0) coordinate (backgroundAbundancesInputArrow);

        % background cycle -- thermo
        \coordinate (thermoOutputs) at ($(backgroundOutputs.south)-(0,1.6)$);
        \node[anchor=center] (thermoOutputColumn) at (thermoOutputs) {$\begin{pmatrix}
            N_{\rm{eff}}\\
            H\\
            T_{\gamma} \\
            T_{\nu}\\
            \vdots
        \end{pmatrix}$};

        \coordinate (GreenThermoArrow) at ($(background.south)+(.5,0)$);

        \coordinate (thermo) at ($(GreenThermoArrow)!(thermoOutputs)!(GreenThermoArrow |- thermoOutputs)$);
        \node[draw, fill=codeCallColor!20, rectangle, rounded corners, inner sep=5pt,anchor=center] (thermoblock) at (thermo) {\texttt{thermo.py}};

        \draw[->,thick,codeCallColor] (GreenThermoArrow) -- (thermoblock.north) ;
        \draw[->,thick] (thermoblock.east) -- (thermoOutputColumn.west) ;

        \coordinate (arrowStart) at ($(thermoOutputColumn.west)-(0,0.7)$) ;
        \coordinate (notArrowTurn) at ($(thermoblock.west)-(.2,0)$);
        \coordinate (arrowTurn) at ($(notArrowTurn)!(arrowStart)!(notArrowTurn |- arrowStart)$);

        \draw[-,thick] (arrowStart) -- (arrowTurn) ;

        \coordinate (arrowHome) at ($(arrowTurn)!(background.south)!(arrowTurn |- background.south)$);
        \draw[->,thick] (arrowTurn) -- (arrowHome) ;

        % abundances cycles
        \node[draw, fill=userCallColor!20, rectangle, rounded corners, inner sep=19pt,anchor=west] (abundances) at (backgroundAbundancesInputArrow) {\texttt{abundances.py}};
        \coordinate (abundancesOutputArrowStart) at ($(abundances.east)$);

        \draw[->,thick] (abundances.east) -- ++(1,0) coordinate (abundanceOutput);

        \node[anchor=west,align=center] (abundanceOutputColumn) at (abundanceOutput) {Primordial\\ Abundances};

        % abundances cycles -- weak rates
        \coordinate (weakRateArrowStart) at ($(abundances.south) - (1.1,0)$);
        \coordinate (weakRateY) at ($(weakRateArrowStart)!(thermoOutputColumn)!(weakRateArrowStart |- thermoOutputColumn)$);
        \node[draw, fill=codeCallColor!20, rectangle, rounded corners, inner sep=5pt,anchor=center] (weakRates) at (weakRateY) {\texttt{weak\_rates.py}};
        \draw[->,thick,codeCallColor] (weakRateArrowStart) -- (weakRates.north);

        \coordinate (weakRateOutputArrowHome) at ($(abundances.south)+(1.1,0)$);

        \node[anchor=center] (weakRateOutput) at ($(weakRateOutputArrowHome)!(weakRates)!(weakRateOutputArrowHome |- weakRates)$) {$\begin{pmatrix}
            \Gamma_{n\rightarrow p}\\
            \Gamma_{p\rightarrow n}\\
        \end{pmatrix}$};
        \draw[->,thick] (weakRates.east) -- (weakRateOutput.west);
        \draw[->,thick] (weakRateOutput.north) -- (weakRateOutputArrowHome);

        % abundances cycles -- nuclear net
        \draw[->,thick,codeCallColor] ($(abundances.north) - (1.1,0)$) -- ++(0,.6) coordinate (nuclearCallArrow);
        \node[draw, fill=codeCallColor!20, rectangle, rounded corners, inner sep=5pt,anchor=south] (nuclear) at (nuclearCallArrow) {\texttt{nuclear.py}};

        \draw[->,thick,codeCallColor] (nuclear.north) -- ++(0,.6) coordinate (reactionsCallArrow);
        \node[draw, fill=codeCallColor!20, rectangle, rounded corners, inner sep=5pt,anchor=south] (reactions) at (reactionsCallArrow) {\texttt{reactions.py}};

        \coordinate (nuclearOutputArrowHome) at ($(abundances.north) + (1.1,0)$);
        \node[anchor=center,align=center] (nuclearOutput) at ($(nuclearOutputArrowHome)!(nuclear)!(nuclearOutputArrowHome |- nuclear)$) {Reaction\\ Network};
        \draw[->,thick] (nuclear.east) -- (nuclearOutput.west);
        \draw[->,thick] (nuclearOutput.south) -- (nuclearOutputArrowHome);

        \node[anchor=center,align=center] (reactionsOutput) at ($(nuclearOutputArrowHome)!(reactions)!(nuclearOutputArrowHome |- reactions)$) {Individual\\ Reactions};
        \draw[->,thick] (reactions.east) -- (reactionsOutput.west);

        \draw[-,thick] (reactionsOutput.south) -- ++(0,-.3) coordinate (reactionArrowOne);

        \coordinate (reactionArrowHome) at ($(nuclear.north) + (.4,0)$);
        
        \draw[-,thick] (reactionArrowOne) -- ($(reactionArrowHome)!(reactionArrowOne)!(reactionArrowHome |- reactionArrowOne)$) coordinate (reactionArrowTwo);
        \draw[->,thick] (reactionArrowTwo) -- (reactionArrowHome);

        % abundances cycles -- thermo cycle
        \coordinate (abundanceThermoArrowPassUpper) at ($(thermoOutputColumn.north)+(0,.25)$);
        \coordinate (pastWeakRatesClose) at ($(weakRates.west) - (.3,0)$);
        \coordinate (pastWeakRatesFar) at ($(weakRates.west) - (.65,0)$);
        \coordinate (thermoCallArrowHome) at ($(thermoblock.north) + (.4,0)$);
        \coordinate (abundanceThermoCallArrowStart) at ($(abundances.west) - (0,.35)$);
        \coordinate (abundanceThermoCallArrowHome) at ($(abundances.west) - (0,.7)$);

        \draw[-,thick,codeCallColor] (abundanceThermoCallArrowStart) -- ($(pastWeakRatesFar)!(abundanceThermoCallArrowStart)!(pastWeakRatesFar |- abundanceThermoCallArrowStart)$) coordinate (callArrowOne);
        \draw[-,thick,codeCallColor] (callArrowOne) -- ($(pastWeakRatesFar)!(abundanceThermoArrowPassUpper)!(pastWeakRatesFar |- abundanceThermoArrowPassUpper)$) coordinate (callArrowTwo);
        \draw[-,thick,codeCallColor] (callArrowTwo) -- ($(thermoCallArrowHome)!(abundanceThermoArrowPassUpper)!(thermoCallArrowHome |- abundanceThermoArrowPassUpper)$) coordinate (callArrowThree);
        \draw[->,thick,codeCallColor] (callArrowThree) -- (thermoCallArrowHome);

        \draw[-,thick] (thermoOutputColumn.east) -- ($(pastWeakRatesClose)!(thermoOutputColumn.east)!(pastWeakRatesClose |- thermoOutputColumn.east)$) coordinate (returnArrowOne);
        \draw[-,thick] (returnArrowOne) -- ($(pastWeakRatesClose)!(abundanceThermoCallArrowHome)!(pastWeakRatesClose |- abundanceThermoCallArrowHome)$) coordinate (returnArrowTwo);
        \draw[->,thick] (returnArrowTwo) -- (abundanceThermoCallArrowHome);

        % arrow from inputs to abundances
        \coordinate (userInputArrowHome) at ($(abundances.west) + (0,.7)$);
        \draw[-,thick,userInputColor] (abundancesInputBlock.east) -- ($(pastWeakRatesClose)!(abundancesInputBlock.east)!(pastWeakRatesClose |- abundancesInputBlock.east)$) coordinate (userInputArrowOne);
        \draw[-,thick,userInputColor] (userInputArrowOne) -- ($(pastWeakRatesClose)!(userInputArrowHome)!(pastWeakRatesClose |- userInputArrowHome)$) coordinate (userInputArrowTwo);
        \draw[->,thick,userInputColor] (userInputArrowTwo) -- (userInputArrowHome);

    \end{tikzpicture}
    \caption{A code block diagram illustrating the functions of different modules of LINX\@.  Blue indicates user inputs, yellow indicates modules the user would typically call, and green indicates modules that the code calls internally.  Blue arrows indicate the flow of user inputs, green arrows indicate internal calls, and black arrows indicate the flow of outputs.  Two additional modules, \texttt{const} and \texttt{special\_funcs}, are not shown; they provide auxiliary constants and special functions to all of the other modules.  Notation is defined  in the main text.  Refer to code documentation for more information about the specific inputs and outputs of each module.}
    \label{fig:codeblockdiagram}
\end{figure*}
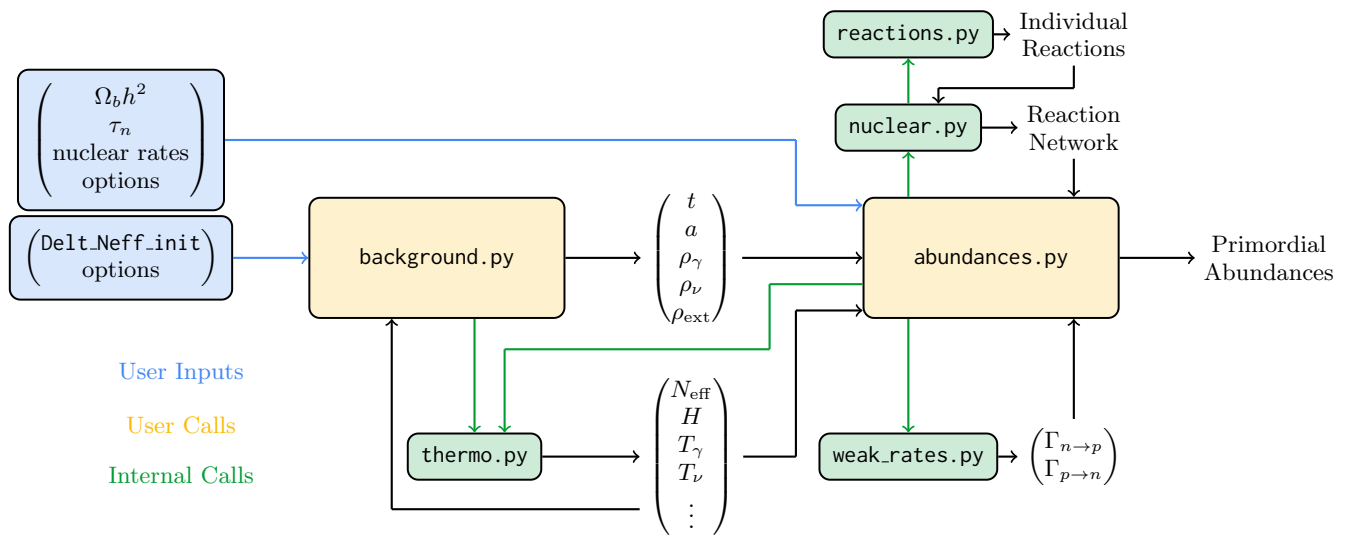

There is a long tradition of BBN codes with numerical implementations of the milestones enumerated above.  The Wagoner code~(1967) was the first to implement the differential equations that needed to be solved to predict element abundances~\cite{1969ApJS...18..247W,Wagoner:1972jh}.  Public options date back to 1992 with the Kawano code NUC123, which updated and modified the Wagoner code~\cite{Kawano_1992}.  The Kawano code inspired both PArthENoPE~(2007)~\cite{Pisanti:2007hk} and AlterBBN~(2011)~\cite{Arbey_2012}, both of which are publicly available and have been updated since their first releases.  Notable private codes include that used in Ref.~\cite{Yeh:2020mgl}, which is also used in the BBN section of the Particle Data Group Review of Particle Physics~\cite{Zyla2020,ParticleDataGroup:2022pth}, and EZ\_BBN~\cite{Coc_2011}, which was the foundation for the public code PRIMAT~(2018)~\cite{Pitrou_2018}.  PRIMAT finally inspired the public code PRyMordial~(2023)~\cite{burns_2023} and, along with PRyMordial, provides a basis for LINX.

Each of the aforementioned public options offers different advantages.  AlterBBN and PRyMordial excel at letting the user explore beyond-Standard-Model scenarios, with PRyMordial providing an accurate treatment of neutrino decoupling and state-of-the-art nuclear rates.  AlterBBN and PArthENoPE stand out for their speed over other options like PRyMordial and PRIMAT\@.  PRIMAT pioneered and includes a precise treatment of proton-neutron interconversion as well as a large reaction network for precise predictions.  PRIMAT and PRyMordial also allow the user to sample the different reaction rates relevant for BBN\@.  We will make frequent comparisons to the PRIMAT and PRyMordial results throughout the text, as these codes are highly accurate and are most similar to LINX\@.  

We will detail a number of LINX's advantages below.  Most relevant here are LINX's speed, which rivals that of AlterBBN and PArthENoPE; its ability to sample the BBN reaction rates quickly, which allows for a more rigorous treatment of nuisance parameters in BBN-only analyses, as well as consistent combination of BBN and CMB analyses; and its differentiability, a feature not present in any other BBN code. The rest of the paper explores these features in more detail.

\section{Structure of LINX}\label{sec:structure} 

LINX harnesses existing results in the literature---especially those of PRIMAT, PRyMordial, and the thermodynamics code nudec\_BSM~\cite{Escudero_2019,Abenza_2020}---to compute the relevant thermodynamics, nuclear reaction rates, and rates of proton/neutron interconversion.  As such, we defer many of the details of the calculation to other sources in the literature and focus instead on describing how the code works and pointing out improvements, as well as other unique aspects of our work, compared to existing codes. While we focus here on the code's default settings, the extended  documentation (available at \url{https://linx.readthedocs.io/en/latest/}) provides more details about optional settings. 

SBBN and SBBN+$\Delta N_\text{eff}$ are the only two scenarios that ship with LINX\@.  However, the modular structure of the code, discussed further below, as well as the use of Python, make it easy to modify LINX to explore other scenarios.  Depending on the scenario, the scope of changes may be limited, possibly only to the built-in thermodynamics module used to set up the computation of background quantities.  Scenarios where certain reaction rates are impacted are also easy to accommodate in the flexible, modular structure of LINX\@.  

To predict primordial elemental abundances, LINX adopts the same philosophy as existing BBN codes by dividing this task into two parts.  Changes in baryonic and elemental abundances do not impact background quantities during the radiation-dominated epoch of BBN, so evaluating background quantities and abundances can be performed independently.  
Thus, the code first determines the relevant background cosmological and thermodynamic quantities as a function of time,  and then uses the background quantities as an input for a separate calculation of the elemental abundances.  Figure~\ref{fig:codeblockdiagram} is a code block diagram detailing the modules used for the different phases of this calculation, which will each be detailed in this Section following a brief discussion of the additional features of the code gained from programming in JAX.

\subsection{Compilation, Vectorization, and Automatic Differentiation with JAX}

LINX is written in Python with JAX~\cite{jax2018github,deepmind2020jax}, a library used to perform highly efficient numerical computations involving arrays.  Just-In-Time (JIT) compilation can be performed on JAX programs using the Accelerated Linear Algebra (XLA) compiler, allowing computations to be rapidly executed on CPUs and GPUs. 
This means users can write code with a high-level syntax, but with a performance that matches that of low-level languages.  
Calling functions and methods in LINX for the first time triggers compilation to this optimized code; subsequent calls using the same input array sizes and data types use the compiled code, greatly improving runtime.

JAX allows for any function to be easily promoted to its vectorized version using the \texttt{vmap} transformation.
Functions can therefore always be written assuming that a given input is a scalar and then later extended to transformed versions that can accept arrays for any of its arguments. 
These vectorized functions can also be compiled, leading to highly efficient array computation without the use of explicit \texttt{for} loops. 
In the examples provided in Section~\ref{sec:examples}, vectorization with the function \texttt{vmap} is used extensively in this manner. 

JAX also makes use of automatic differentiation~(AD), a powerful technique for computing derivatives of functions.
This lies at the heart of the efficient gradient-based inference that is made possible with LINX\@. AD calculates exact derivative values by applying the chain rule to elementary operations as the program executes; this is as opposed to symbolic differentiation, which derives analytical expressions for derivatives, and finite differences, which approximates derivatives through small perturbations. See Ref.~\cite{blondel2024elements} for a recent and comprehensive review of AD\@.

The power of AD lies in its ability to compose program components with ease, making it particularly suited for modular codes like LINX\@. Modern AD-capable libraries, including JAX, typically employ reverse-mode differentiation, which applies the chain rule by propagating derivatives backwards through the computational graph, from outputs to inputs. This approach offers a significant computational advantage when the dimensionality of the output is much smaller than that of the input: the cost of evaluating the gradient of some function $f(\theta)$ can be only a constant factor higher than evaluating $f(\theta)$ itself.

To appreciate the scalability of AD, consider computing $\partial f / \partial \theta_i$ for a function $f: \mathbb{R}^n \to \mathbb{R}$ (such as a likelihood) with $n$ input parameters $\theta_i$. Finite difference methods require at least $n+1$ function evaluations, leading to a computational complexity of $\mathcal{O}(n)$. In contrast, reverse-mode AD computes the full gradient in just two passes through the computational graph,\footnote{The first pass is a forward pass where the function is broken down into elementary operations and evaluated at these intermediate values. The second pass works backwards through these intermediate values, applying the chain rule to compute derivatives with respect to each input parameter in a single pass.} resulting in $\mathcal{O}(1)$ complexity relative to the number of parameters. This efficiency becomes crucial as the number of parameters grows.

All in all, the combination of AD, JIT compilation, and vectorizability makes JAX particularly effective as an ingredient for scalable Bayesian inference over a high-dimensional space, which is precisely what LINX is designed for.

In conjunction with JAX, we also use equinox~\cite{kidger2021equinox} modules as the main objects that are instantiated and called in LINX\@. 
These objects are very similar to Python classes, but can have JAX transformations such as compilation and vectorization applied to them easily. 
LINX also avoids pitfalls related to flag setting for different code options by eliminating global flags and having all options either set during initialization or passed to the equinox module instance during the call.

\subsection{Background Quantities}
\label{sec:background_quantities}

The relevant background quantities in SBBN and SBBN+$\Delta N_\text{eff}$ are the physical properties of the EM and neutrino fluids, the energy density of the inert relativistic species, and the scale factor $a$, all as a function of time. 
The EM sector is in thermal equilibrium and can be described by a single temperature $T_\gamma$.  
The scenarios discussed here are not impacted by physics at temperatures $T_\gamma \gtrsim \SI{10}{\mega\eV}$, and therefore the EM sector has an energy density that is dominated by photons, electrons and positrons. 

A precise treatment of neutrino decoupling is required to compute these background quantities, and all existing public BBN codes have some version of noninstantaneous decoupling implemented. 
PArthENoPE and earlier versions of PRIMAT use a fitting function for the energy transfer between the two fluids, which is suitable only for SBBN~\cite{Pisanti:2007hk,Pitrou_2018}. 
More recently, PRIMAT has improved the accuracy of their treatment, including secondary effects such as nonthermal distortions to the neutrino spectra~\cite{Froustey:2020mcq,Froustey_2022a,Froustey_2022_asymmetry}.
This approach has still, however, only been carefully studied in the context of SBBN\@. 

In order to ensure that our code is both sufficiently accurate for SBBN and easily extended to include new-physics scenarios, we follow the procedure laid out in nudec\_BSM~\cite{Escudero_2019,Abenza_2020}, which was written with beyond-the-Standard-Model scenarios in mind and is also used by PRyMordial for these calculations.
In this approach, all three neutrino species are taken to be a single fluid with a common temperature $T_\nu$. 
In so doing, we neglect nonthermal corrections to neutrinos as they decouple from the other Standard Model~(SM) particles, and we neglect neutrino oscillations. 
These are good approximations within the context of SBBN, at the current levels of precision for D/H, Y$_\mathrm{P}$ and $N_\text{eff}$. 
LINX currently allows electron neutrinos to have a different temperature from muon and tau neutrinos and for any fluid to have a nonzero chemical potential; however, in the context of SBBN and SBBN\texttt{+}$\Delta N_\text{eff}$, these are not important, and all fluids are taken to have zero chemical potential from here on.

There are two equations governing the evolution of $a$, $T_\gamma$, and $T_\nu$.  The first is the relation between the Hubble parameter $H$ and total energy density $\rho_\text{tot}$ (assumed to be dominated by radiation), 
\begin{alignat}{1}
    H \equiv \frac{\dot{a}}{a} = \left( \frac{8 \pi G_\text{N}}{3} \rho_\text{tot}\right)^{1/2} \,,
    \label{eq:Hubble}
\end{alignat}
where dots represent derivatives with respect to proper time and $G_\text{N}$ is Newton's gravitational constant.  The second is the following set of relations: 
\begin{alignat}{1}
    \dot{T}_\gamma &= \left( \frac{d \rho_\text{EM}}{d T_\gamma} \right)^{-1} \dot{\rho}_\text{EM} \,, \quad \dot{T}_\nu = \left( \frac{d \rho_\nu}{d T_\nu} \right)^{-1} \dot{\rho}_\nu \,,
    \label{eq:T_evolution}
\end{alignat}
where the rate of change $\dot{\rho}$ for each sector is determined by energy-momentum conservation, 
\begin{alignat}{1}
    \dot{\rho}_\text{EM} &= -3 H (\rho_\text{EM} + p_\text{EM}) - C_{\rho;\gamma \to \nu} \,, \nonumber \\
    \dot{\rho}_\nu &= -3 H (\rho_\nu + p_\nu) + C_{\rho;\gamma \to \nu} \,.
\end{alignat}
Here, $p$ denotes pressure, while $C_{\rho;\gamma \to \nu}$ is the rate of energy transferred from the EM fluid to the neutrino fluid due to processes coupling the two fluids. 
Finally, we have $\rho_\text{tot} = \rho_\text{EM} + \rho_\nu$ for SBBN and $\rho_\text{tot} = \rho_\text{EM} + \rho_\nu + \rho_\text{ext}$ for SBBN\texttt{+}$\Delta N_\text{eff}$, with $\rho_\text{ext} \propto a^{-4}$ being the inert relativistic species that we add.
$\rho_\text{EM}$ and $p_\text{EM}$ include not only the usual contributions from the thermal distributions of photons and electrons, but also include next-to-leading order QED corrections derived in Ref.~\cite{Bennett_2020} by default. 
The collision terms are likewise computed following Refs.~\cite{Escudero_2019,Abenza_2020}, taking into account all relevant processes that transfer energy between the two sectors. 
Corrections associated with the finite electron mass and using a Fermi-Dirac distribution for neutrinos instead of a Maxwell-Boltzmann distribution for the collision terms are included by default. 

One advantage of using a differentiable paradigm in LINX is that derivatives of any function can be obtained easily and accurately. 
In Eq.~\eqref{eq:T_evolution}, for example, LINX contains a function that computes $\rho_\text{EM}$ as a function of $T_\gamma$; using the built-in JAX function \texttt{jax.grad} yields $d \rho_\text{EM} / d T_\gamma$ with a single line of code. 

All expressions of thermodynamic variables can be found in the LINX \texttt{thermo} module. 
To compute $\rho$, $p$, and number densities $n$ of bosons and fermions, with or without mass, several important numerical innovations are implemented in LINX: 
\begin{enumerate}
    \item For massless particles in thermal equilibrium, the most general form of $\rho$, $p$, and $n$ including nonzero chemical potentials involve polylogarithms, which are not implemented in JAX at this time. 
    We implement a differentiable version of polylogarithms for arguments that are relevant here (\texttt{special\_funcs.Li}), and a version of the gamma function (\texttt{special\_funcs.gamma}) that supports imaginary arguments.  Details of how these functions are implemented, which may be useful in other works,  are given in Appendix~\ref{app:special_functions}.
    \item For massive particles, there is no analytic form for any thermodynamic quantity; they can only be expressed as an integral over the distribution. 
    Since JAX does not implement numerical quadrature, we approximate these integrals as a series involving modified Bessel functions of the second kind $K_n(x)$, a method described in Ref.~\cite{Giovanetti:2021izc} and in Appendix~\ref{app:series}. 
    These functions are also currently not part of JAX; we have implemented a differentiable version of some of these functions (\texttt{special\_funcs.K0}, \texttt{K1} and \texttt{K2}) as part of LINX\@. 
\end{enumerate}

The set of differential equations given by Eqs.~\eqref{eq:Hubble} and~\eqref{eq:T_evolution} are set up and solved in the \texttt{BackgroundModel} equinox module found in the LINX \texttt{background} module, which can be used for both SBBN and SBBN\texttt{+}$\Delta N_\text{eff}$.
At this time, if the user is interested in other new-physics models, they will need to write a variation of this class for their model, using a similar framework.  
By default, integration begins at $T_\gamma = T_\nu \approx \SI{10}{MeV}$, $a = 1$, and $t = 0$,  to ensure the thermodynamic history is computed starting from well before neutrino decoupling, when both sectors are in chemical equilibrium.
Integration is performed using the JAX differential equation solver package Diffrax~\cite{kidger2021} using Tsitouras' 5/4 method~\cite{tsitouras2011runge}, a general-purpose Runge-Kutta method, with solver and adaptive stepsize implementation from Ref.~\cite{hairer_2008}. 
LINX does not terminate the calculation at a predetermined $t$, but instead ends the integration of thermodynamic quantities once $T_{\gamma}\lesssim\SI{5}{keV}$ by default.\footnote{Even though such a termination condition is used, a maximum number of steps to be taken must still be specified, as the size of all arrays must be fixed at the point of compilation.}  
After the integration is complete, $a(t)$ is rescaled by a factor $(T_0 / T_{\gamma,\rm{end}}) a_{\rm{end}}$, where $T_0$ is the CMB temperature today as measured by FIRAS \cite{Mather_1999} and the subscript ``end" denotes values at the end of the calculation.  
This normalizes the scale factor to $a_\text{end}=1$ today, provided there are no further entropy dumps into the EM sector below $T_{\gamma,\text{end}}$ (\textit{i.e.}\ $T_\gamma \propto 1/a$ after $T_{\gamma,\text{end}}$), which is true for both SBBN and SBBN\texttt{+}$\Delta N_\text{eff}$. 

An instance of the \texttt{BackgroundModel} class is initialized by default without any arguments. 
Once initialized, the instance can be called by passing a single argument \texttt{Delt\_Neff\_init}, which is defined through the relation
\begin{alignat}{1}
    \Delta N_{\text{eff},i} \equiv \frac{8}{7} \left( \frac{11}{4} \right)^{4/3} \frac{\rho_{\text{ext},i}}{\rho_{\gamma,i}} \,.
    \label{eq:Neff_definition}
\end{alignat}
Here, $\rho_{\gamma,i}$ and $\rho_{\text{ext},i}$ are the initial energy densities of the photons and the inert relativistic species respectively. 
We stress that $\Delta N_{\text{eff},i}$ is \textit{not} $\Delta N_\text{eff}$, which is defined similarly through Eq.~\eqref{eq:Neff_definition}, but with all quantities on the right-hand side evaluated at $T_\gamma = T_{\gamma,\text{end}}$. 
For SBBN or SBBN\texttt{+}$\Delta N_\text{eff}$, this means after $e^+e^-$ annihilation is complete. 

The output from calling the instance of \texttt{BackgroundModel} is a tuple of one-dimensional arrays, all of the same length. 
These arrays are: 
\begin{enumerate}
    \item \texttt{t\_vec}: an array of times; 
    \item \texttt{a\_vec}: an array of scale factors, correctly normalized so that $a = 1$ when $T_\gamma = T_0$, assuming $T_\gamma \propto 1/a$ after; 
    \item \texttt{rho\_g\_vec}: the energy density of photons; 
    \item \texttt{rho\_nu\_vec}: the energy density of one neutrino species; 
    \item \texttt{rho\_extra\_vec}: the energy density of the inert, relativistic species; 
    \item \texttt{P\_extra\_vec}: the pressure of the inert, relativistic species, assumed to be $p_\text{ext} = \rho_\text{ext} / 3$, and 
    \item \texttt{Neff\_vec}: the value of $N_\text{eff}$ throughout. 
\end{enumerate}
These are the background quantities that form the basis of the abundance calculation. 

A few remarks are in order regarding the method used here, compared to others in the literature. 
First, both LINX and PRyMordial explicitly calculate the energy transfer term between the EM and neutrino fluids, $C_{\rho;\gamma \to \nu}$, in Eq.~\eqref{eq:T_evolution}. 
$C_{\rho;\gamma \to \nu}$ is recomputed for any new-physics model beyond SBBN, \textit{e.g.}\ SBBN\texttt{+}$N_\text{eff}$; the effect of new physics on the incomplete decoupling of neutrinos is therefore always consistently accounted for. 
This is in contrast to the approach adopted in PArthENoPE~\cite{Gariazzo:2021iiu}, which provides a fixed fitting function for $C_{\rho;\gamma \to \nu}$ obtained assuming SBBN, and to PRIMAT, which has a more sophisticated calculation of $C_{\rho;\gamma \to \nu}$~\cite{Froustey:2020mcq,Froustey_2022a,Froustey_2022_asymmetry}, but only for SBBN\@. 

Finally, we note that in our method, there is no need to track the entropy density of the EM plasma, which provides redundant information if the energy densities of all fluids and the energy transfer between them are already known. This is in contrast to PRIMAT and PRyMordial, which uses the entropy density evolution to obtain $a(t)$. 

Table~\ref{tab:background_assumptions_and_Neff} shows the calculated value of $N_\text{eff}$ within SBBN, toggling between various assumptions used in calculating the background quantities.
With all of the important corrections included, our calculated value of $N_\text{eff}$ is 3.044, in good agreement with Ref.~\cite{Abenza_2020}, which forms the basis for our calculation. The size of each correction is also almost identical to those reported in Ref.~\cite{Abenza_2020}. Our value is in agreement with the state-of-the-art, detailed determination of $N_\text{eff}$ in SBBN~\cite{Drewes:2024wbw}.

\renewcommand{\arraystretch}{1.5}
\begin{table}[t]
    \centering
    \begin{tabular}{p{4cm}>{\centering\arraybackslash}p{1.5cm}>{\centering\arraybackslash}p{2.5cm}}

         \textbf{Background Correction} & \textbf{$\boldsymbol{N}_\text{eff}$} & \shortstack{ \textbf{\% change} \\ \textbf{ from ID}} \\
         \hline
         ID & 3.001 & 0.00 \\
         ID\texttt{+}LO-QED & 3.012 & 0.35\\
         ID\texttt{+}NLO-QED & 3.011 & 0.32\\
         MB$\nu$ & 3.042 & 1.36\\
         MB$\nu$\texttt{+}NLO-QED & 3.051 & 1.67\\
         FD$\nu$ & 3.038 & 1.24\\
         FD$\nu$\texttt{+}NLO-QED & 3.048 & 1.55\\
         FD$\nu$\texttt{+}$m_e$ & 3.035 & 1.14\\
         FD$\nu$\texttt{+}$m_e$\texttt{+}LO-QED & 3.045 & 1.48\\
         FD$\nu$\texttt{+}$m_e$\texttt{+}NLO-QED & 3.044 & 1.45\\
         \hline

    \end{tabular}
    \caption{$N_{\rm{eff}}$ prediction including different combinations of options for the computation of background rates. The options are ID: instantaneous decoupling of neutrinos, LO(NLO)-QED: leading (next-to-leading) order QED corrections to the thermodynamic properties of the EM plasma, MB$\nu$: Maxwell-Boltzmann statistics for neutrinos, FD$\nu$: Fermi-Dirac statistics for neutrinos, and $m_e$: finite-mass corrections in collision terms.  LINX uses FD$\nu$\texttt{+}$m_e$\texttt{+}NLO-QED by default.}
    \label{tab:background_assumptions_and_Neff}
\end{table}

\subsection{Nuclear Abundances}\label{sec:nuclear_abundances_in_LINX}

After the relevant background quantities are computed, LINX uses these results to compute primordial abundances. 
There are three components to this calculation: \textit{1)} the evaluation of the $n \leftrightarrow p$ conversion rates, involving weak processes such as $p + e^- \leftrightarrow n + \nu_e$, implemented in the \texttt{weak\_rates} module; \textit{2)} the construction of the nuclear network used to evaluate the abundance of light elements, implemented in the \texttt{reactions} and \texttt{nuclear} modules; and \textit{3)} integration of the differential equation governing the evolution of the abundances, given the background evolution, weak rates, and nuclear rates, implemented in the \texttt{abundances} module.  We detail each of these components in turn.

\subsubsection{Weak Rates}
\label{sec:weak_rates}

At temperatures of $T_\gamma \gtrsim \SI{1}{\mega\eV}$, all of the baryons are, to an excellent approximation, composed of free neutrons and protons. 
In the initial phase of the calculation, it is therefore important to predict the $n \leftrightarrow p$ interconversion rates accurately. 
We adopt the same approach described in detail by PRIMAT~\cite{Pitrou_2018}, which is also used in PRyMordial~\cite{burns_2023}.  We would like to calculate the rates $\Gamma_{n \to p}$ and $\Gamma_{p \to n}$ such that
\begin{alignat}{1}
    \dot{n}_n + 3 H n_n &= - n_n \Gamma_{n \to p} + n_p \Gamma_{p \to n} \,, \nonumber \\
    \dot{n}_p + 3 H n_p &= - n_p \Gamma_{p \to n} + n_n \Gamma_{n \to p} \,.
\end{alignat}
These rates are dependent on experimentally-measured electroweak parameters and therefore have related uncertainties. 
However, one can show that these rates are all proportional to $\tau_n^{-1}$, where $\tau_n$ is the free neutron decay lifetime~\cite{Pitrou_2018}. 
We thus compute these rates normalized to $\tau_n^{-1}$, and treat $\tau_n$ as a single, experimentally-determined nuisance parameter that determines the overall magnitude of these rates.\footnote{Other experimental anchors for the normalization of these rates are possible, but are currently regarded as less precise~\cite{Pitrou_2018}. PRyMordial allows users to switch between various choices~\cite{burns_2023}.}

A full accounting of how to evaluate these rates, starting from the Born approximation and moving on to increasingly sophisticated corrections, is given in Ref.~\cite{Pitrou_2018}. 
These calculations are implemented in the equinox module \texttt{weak\_rates.WeakRates}. 
By default, LINX includes the following corrections derived in Ref.~\cite{Pitrou_2018}, all of which can be toggled in the equinox module: 
\begin{enumerate}
    \item Virtual photon radiative corrections, labeled as `RC0' in Ref.~\cite{Pitrou_2018}. These correspond to including one-loop diagrams with a photon loop, and no emission of real photons; 
    \item Finite temperature radiative corrections and Bremsstrahlung corrections, labeled as `ThRC\texttt{+}BS' in Ref.~\cite{Pitrou_2018} (we do not use separate thermal and Bremsstrahlung corrections). 
    These corrections are included only as a table of values obtained using PRyMordial, assuming SBBN\@. A proper calculation of these corrections for arbitrary background evolution histories requires quadrature in two dimensions, which cannot be straightforwardly performed in JAX\@. However, using PRyMordial, we have verified that the difference between using the SBBN correction for SBBN+$\Delta N_\text{eff}$ and using the true correction is negligible.
    \item Finite nucleon mass corrections, labeled as `FM' in Ref.~\cite{Pitrou_2018}. If used in combination with virtual photon radiative corrections, we do not use separate `RC0' and `FM' corrections; we use instead corrections labeled `RC\texttt{+}FM' in Ref.~\cite{Pitrou_2018}; and
    \item The weak magnetism correction, denoted `WM' in Ref.~\cite{Pitrou_2018}.
\end{enumerate}
Calculating these rates involves integrating over the phase-space distributions of neutrinos and electrons. 
We evaluate these rates using the trapezoidal rule---the only numerical quadrature method implemented in JAX---by calling the JAX built-in function \texttt{jax.numpy.trapz}. This is unlike what is done in both PRyMordial and PRIMAT, which uses more sophisticated numerical quadrature techniques. 
We find, however, no meaningful loss in accuracy by making this substitution, as we show in Section~\ref{sec:comparisons}.

We illustrate the effects of these corrections on the SBBN prediction for Y$_{\rm{P}}$ in Table~\ref{tab:weak_rates}.  We also calculate all corrections on the background quantities discussed earlier in Section~\ref{sec:background_quantities} to compute the values in this table.
These corrections to the Born approximation amount to a shift of approximately 2\% to Y$_\mathrm{P}$; the size of each correction agrees well with those reported in Ref.~\cite{Pitrou_2018}. 
Table~\ref{tab:weak_rates} also shows that the `ThRC\texttt{+}BS' correction is much smaller than 1\%. 

\renewcommand{\arraystretch}{1.5}
\begin{table}[t]
    \centering
    \begin{tabular}{p{4cm}>{\centering\arraybackslash}p{1.5cm}>{\centering\arraybackslash}p{2.5cm}}

         \textbf{Weak Rates Correction} & \textbf{Y}$_{\rm{\textbf{P}}}$ & \shortstack{ \textbf{\% change} \\ \textbf{ from Born}} \\
         \hline
         Born & 0.24279 & 0.00 \\
         Born\texttt{+}FM & 0.24389 & 0.45 \\
         Born\texttt{+}FM\texttt{+}WM & 0.24401 & 0.50 \\
         RC0 & 0.24593 & 1.29 \\
         RC\texttt{+}FM\texttt{+}WM & 0.24715 & 1.79 \\
         RC\texttt{+}FM\texttt{+}WM\texttt{+}ThRC\texttt{+}BS & 0.24714 & 1.79\\
         \hline
         
    \end{tabular}
    \caption{Y$_{\rm{P}}$ prediction in SBBN including different combinations of options for the computation of weak rates.  Notation is the same as that used in Ref.~\cite{Pitrou_2018}, discussed in text---``Born" indicates no corrections are included, ``RC0" includes radiative corrections, ``FM" includes finite mass effects, ``RC+FM'' includes both radiative and finite mass corrections, ``WM" includes the weak magnetism correction, and ``ThRC\texttt{+}BS" includes finite temperature and Bremsstrahlung corrections.  LINX uses the RC\texttt{+}FM\texttt{+}WM\texttt{+}ThRC\texttt{+}BS by default.}
    \label{tab:weak_rates}
\end{table}

Once initialized, a \texttt{WeakRates} class can be called, with the following quantities passed as arguments: 
\begin{enumerate}
    \item \texttt{T\_vec\_ref}: a tuple of arrays, the first entry containing an array of $T_\gamma$ and the second entry containing an array of $T_\nu$.
    This information determines the background quantities for evaluating the weak rates; 
    \item \texttt{T\_start}: the highest value of $T_\gamma$ for evaluating the $n \leftrightarrow p$ rates; 
    \item \texttt{T\_end}: the lowest value of $T_\gamma$ for evaluating the rates; and
    \item \texttt{sampling\_nTOp}: the number of points between \texttt{T\_start} and \texttt{T\_end} inclusive for evaluating the $n \leftrightarrow p$ rates. 
\end{enumerate}
This will then return a tuple containing three elements: the abscissa for $T_\gamma$, followed by $\Gamma_{n \to p}$, and $\Gamma_{p \to n}$ at the temperatures in the abscissa. 

\subsubsection{Nuclear Reactions}

Once $T_\gamma \lesssim \SI{1}{\mega\eV}$, neutrons and protons start the process of combining into the lightest elements, and nuclear reactions start to become important. 
These reactions form a network between the various nuclear species, converting one species to another. 
In any BBN calculation, a choice of which reactions to include in the network is made.  
This ultimately leads to a coupled system of differential equations of the form~\cite{Fowler:1967ty,1969ApJS...18..247W,Pitrou_2018}
\begin{multline}
    \dot{Y}_{i_1} = \sum_{\substack{i_2 \cdots i_p \\ j_1 \cdots j_q}} \bigg( S_{j_1 \cdots j_q \to i_1 \cdots i_p} \Gamma_{j_1 \cdots j_q \to i_1 \cdots i_p}Y_{j_1}^{N_{j_1}} \cdots Y_{j_q}^{N_{j_q}} \\
    - S_{i_1 \cdots i_p \to j_1 \cdots j_q} \Gamma_{i_1 \cdots i_p \to j_1 \cdots j_q} Y_{i_1}^{N_{i_1}} \cdots Y_{i_p}^{N_{i_p}} \bigg) \,,
    \label{eq:abundance_ODE}
\end{multline}
for each species $i_1$. Here, $\Gamma_{i_1 \cdots i_p \to j_1 \cdots j_q}$ is the rate of the reaction $i_1 \cdots i_p \to j_1 \cdots j_q$, and $N_k$ is the stoichiometric coefficient of species $k$ for the reaction. 
$S$ is a symmetry factor associated with each direction of a reaction, given by
\begin{alignat}{1}
    S_{j_1 \cdots j_q \to i_1 \cdots i_p} &= \frac{N_{i_1}}{N_{j_1}! \cdots N_{j_q}!} \,, \nonumber \\
    S_{i_1 \cdots i_p \to j_1 \cdots j_q} &= \frac{N_{i_1}}{N_{i_1}! \cdots N_{i_p}!} \,.
\end{alignat}
$\Gamma$ is given by the decay width for a $1 \to n$ process; if $n_{\rm{b}}$ is the baryon number density, this is $n_\text{b} \langle \sigma v \rangle_{ij \to \cdots}$ for the process $ij \to \cdots$, $n_\text{b}^2 \langle \sigma v^2 \rangle_{ijk \to \cdots}$ for the process $ijk \to \cdots$, and so on.

Detailed balance enforces a relation between $\Gamma_{j_1 \cdots j_q \to i_1 \cdots i_p}$ and $\Gamma_{i_1 \cdots i_p \to j_1 \cdots j_q}$. 
Defining
\begin{alignat}{1}
    \Gamma_{i_1 \cdots i_p \to j_1 \cdots j_q} = \rho_\text{b}^{N_{i_1} + \cdots + N_{i_p} - 1} \lambda_{i_1 \cdots i_p \to j_1 \cdots j_q} \,, 
\end{alignat}
where $\rho_\text{b} = n_\text{b} u$ and $u$ is the atomic mass unit,\footnote{For a discussion on this convention, see Ref.~\cite{Pitrou_2018}} the relationship between forward and backward reaction rates can be parameterized as~\cite{Pitrou_2018}
\begin{alignat}{1}
    \frac{\lambda_{j_1 \cdots j_q \to i_1 \cdots i_p}}{\lambda_{i_1 \cdots i_p \to j_1 \cdots j_q}} = \alpha \left( \frac{T}{\SI{e9}{\kelvin}} \right)^\beta \exp \left( \frac{\gamma}{T / \SI{e9}{\kelvin}} \right) \,,
\end{alignat}
where $\alpha$, $\beta$ and $\gamma$ are constants related only to nuclear masses and spins, quantities that are measured to high precision or known. 
For each reaction, we therefore only need to specify the value of $\lambda$ in one direction.

Following Refs.~\cite{Pitrou_2018,Fields:2019pfx}, we take $\lambda(T)$ to be log-normally distributed, with a mean value $\overline{\lambda}(T)$ and standard deviation $\sigma(T)$, so that 
\begin{alignat}{1}
    \log \lambda(T) = \log \overline{\lambda}(T) + q \sigma(T) \,,
    \label{eq:q_definition}
\end{alignat}
with $q$ being a unit Gaussian random variable. 
Further discussion on the mean reaction rates and uncertainties can be found in Section~\ref{sec:nuclear_network}.

All of the information for each reaction required to evaluate Eq.~\eqref{eq:abundance_ODE} is stored inside an instance of the \texttt{reactions.Reaction} equinox module, which allows for a flexible treatment of how these reactions are included in the network. 
The information for a single reaction---both forward and backward---stored in \texttt{reactions.Reaction} includes:
\begin{enumerate} 
    \item The incoming and outgoing states, stored as tuples of integers, each representing a different species. Our convention is \texttt{0}:$n$, \texttt{1}:$p$, \texttt{2}:$d$, \texttt{3}:$t$, \texttt{4}:$^3$He, \texttt{5}:$\alpha$, \texttt{6}:$^7$Li, \texttt{7}:$^7$Be, \texttt{8}:$^6$He, \texttt{9}:$^6$Li, \texttt{10}:$^8$Li, \texttt{11}:$^8$B, which can be extended to include more species as necessary; 
    \item $\alpha$, $\beta$, and $\gamma$; and 
    \item Either a table of $\overline{\lambda}(T)$ and $\sigma(T)$ in the range \SIrange{e6}{e10}{\kelvin} that we use for linear interpolation, or a function that takes in $T$ and $q$ and returns $\lambda$ that is valid in the same temperature range. 
\end{enumerate}
With these quantities, each term in Eq.~\eqref{eq:abundance_ODE} is fully specified. 

\subsubsection{Nuclear Network}\label{sec:nuclear_network}

A collection of nuclear reactions---including the $n \leftrightarrow p$ weak processes---forms a nuclear network. 
Nuclear networks in LINX are stored as an instance of the equinox module \texttt{nuclear.NuclearRates}, which can be initialized in one of two ways: either with a keyword specifying one of several lists of reactions that form a pre-defined nuclear network already included in LINX, or by specifying a custom list of \texttt{Reaction} instances. 
The \texttt{NuclearRates} class will read the information stored in each \texttt{Reaction} instance and construct Eq.~\eqref{eq:abundance_ODE} for evaluation. 
This avoids the need to hard-code the reaction network and allows the user to easily build custom nuclear networks, an advantage that LINX has over other BBN codes. 

In the literature, there are multiple choices for $\overline{\lambda}(T)$ and $\sigma(T)$ for many of the reactions, some of which are not necessarily in agreement. 
LINX includes several canonical choices of networks, based on results from three different groups. 
In each source, a fit to the $S$-factor for each reaction is first performed; for a $2 \to 2$ reaction $ij \to kl$, this quantity is defined as~\cite{Pisanti:2020efz}
\begin{alignat}{1}
    S_{ij \to kl}(E) \equiv \sigma_{ij \to kl} (E) e^{\sqrt{E_{G,ij}/E}} \,,
\end{alignat}
where $\sigma_{ij \to kl}(E)$ is the experimentally-measured cross section of the reaction as a function of energy $E$. 
$E_{G,ij}  \equiv 2 \pi^2 \mu_{ij} (Z_i Z_j \alpha_\text{EM})^2$ is the Gamow energy for the reaction, where $\mu_{ij}$ is the reduced mass of the $ij$ system, $\alpha_\text{EM}$ is the fine-structure constant, and $Z_i$, $Z_j$ are the atomic numbers of the species $i$ and $j$, respectively. 
Obtaining $\sigma_{ij \to kl}(E)$ typically involves collating many different experimental results across a wide range of energies. 
Once a choice has been made regarding which experimental results to use, a fit to $S_{ij \to kl}(E)$ is performed, accounting for experimental uncertainty. 
The $S$-factor can then be related to the thermally-averaged cross section via~\cite{Pisanti:2020efz} 
\begin{multline}
    \langle \sigma_{ij \to kl} \rangle (T) = \sqrt{\frac{8}{\pi \mu_{ij} }} T^{-3/2}  \\
    \times \int_0^\infty dE \, E S_{ij \to kl}(E) e^{-\sqrt{E_{G,ij}/E}} e^{-E/T} \,.
\end{multline}
For a $2 \to n$ reaction, the reaction rate is then $\lambda = u^{-1} \langle \sigma v \rangle$. 
Based on the fit result for $S_{ij \to kl}$, one can thus obtain $\overline{\lambda}(T)$ and an estimate for $\sigma(T)$. 

The three main approaches that we adopt for $\overline{\lambda}(T)$ and $\sigma(T)$ are as follows:
\begin{enumerate}
    \item \textit{PRIMAT rates}. These rates are used in the PRIMAT code~\cite{Pitrou_2018}. 
    The functional form used to fit each $S_{ij \to kl}(E)$ is motivated by \textit{ab initio} calculations using the $R$-matrix method and frequently employs Bayesian inference methods (see \textit{e.g.}\ Ref.~\cite{Moscoso:2021xog}); 
    \item \textit{PArthENoPE rates}. These are used in the PArthENoPE code~\cite{Gariazzo:2021iiu}. The fit procedure is purely phenomenological, using low-degree polynomials to fit for $S_{ij \to kl}$ with a modified least-squares procedure; and 
    \item \textit{YOF rates}. These are rates used by Ref.~\cite{Yeh:2020mgl} and related papers from some of the authors of this reference. 
    They are based mostly on the NACREII compilation~\cite{Xu:2013fha}, which obtains $S_{ij \to kl}$ by fitting a phenomenological nuclear potential governing the interaction between the two incoming nuclei to cross section data. 
\end{enumerate}
These different sets of rates give quantitatively different BBN predictions.  
Using PRIMAT rates gives D/H predictions that are approximately $2\sigma$ below the experimentally-measured value; because of this, inference of $\Omega_b h^2$ with PRIMAT rates also leads to values that are roughly $2\sigma$ below the value obtained from the CMB power spectrum by Planck (see Ref.~\cite{Pitrou:2020etk} and also~\citetalias{Giovanetti_2024}).
On the other hand, PArthENoPE and YOF rates produce results that are more in line with current measurements~\cite{Pisanti:2020efz,Yeh:2020mgl} and with the CMB results, although YOF rates tend to give significantly larger uncertainties.
We refer to reader to Refs~\cite{Pitrou:2020etk,Pisanti:2020efz,Yeh:2020mgl} for a discussion of why these discrepancies occur.

\renewcommand{\arraystretch}{1.5} 
\begin{table*}[!]
\centering
\begin{tabular}{>{\centering\arraybackslash}m{3.5cm}>{\centering\arraybackslash}m{5cm}>{\centering\arraybackslash}m{3.25cm}>{\centering\arraybackslash}m{3.25cm}}
\textbf{Reaction} & \texttt{key\_PRIMAT\_2023} & \texttt{key\_PRIMAT\_2018} & \texttt{key\_YOF} \\
\hline
$p(n,\gamma)d$ & \cite{Ando:2005cz}& \cite{Ando:2005cz} & \cite{Ando:2005cz}\\
$d(p,\gamma)^3$He & $\leq \SI{4e9}{\kelvin}$:~\cite{Moscoso:2021xog}, $> \SI{4e9}{\kelvin}$:~\cite{Coc:2015bhi} & \cite{Mossa:2020gjc}   & \cite{Mossa:2020gjc}\\
$d(d,p)t$ & \cite{InestaGomez:2017eya}& \cite{Iliadis:2016vkw}  & \cite{Xu:2013fha}\\
$d(d,n)^3$He & \cite{InestaGomez:2017eya} & \cite{Iliadis:2016vkw} & \cite{Xu:2013fha}\\
$^3$He$(n,p)t$ & \cite{Descouvemont:2004cw} & \cite{Descouvemont:2004cw} & \cite{Cyburt:2004cq}\\
$^3$He$(d,p)^4$He & \cite{deSouza:2018gdx}  & \cite{Descouvemont:2004cw} & \cite{Xu:2013fha}\\
$t(d,n)^4$He & \cite{deSouza:2019pmr}& \cite{Descouvemont:2004cw}& \cite{Xu:2013fha}\\
$^3$He$(\alpha,\gamma)^7$Be & \cite{Iliadis:2016vkw} & \cite{Iliadis:2016vkw} & \cite{Xu:2013fha}\\
$t(\alpha,\gamma)^7$Li & \cite{Descouvemont:2004cw} & \cite{Descouvemont:2004cw} & \cite{Xu:2013fha}\\
$^7$Be$(n,p)^7$Li & $< \SI{e9}{\kelvin}$: ~\cite{Descouvemont:2004cw}, $\geq \SI{e9}{\kelvin}$:~\cite{deSouza:2019tpj}& \cite{Descouvemont:2004cw} & \cite{Fields:2019pfx}\\
$^7$Li$(p,\alpha)^4$He & \cite{Descouvemont:2004cw} & \cite{Descouvemont:2004cw} & \cite{Xu:2013fha}\\
$t(p,\gamma)^4$He & \cite{Serpico:2004gx} & \cite{Serpico:2004gx}  & \cite{Serpico:2004gx}\\
\hline
\end{tabular}
\caption{List of references for $\overline{\lambda}(T)$ and $\sigma(T)$ in three of the ``key" networks that we use. Note that the PArthENoPE ``key" network is entirely based on the latest version of the PArthENoPE code~\cite{Gariazzo:2021iiu}. The YOF ``key" network uses rates primarily from Ref.~\cite{Xu:2013fha}, which presents the NACREII collaboration results.}
\label{tab:reaction-rates}
\end{table*}

\begin{figure}
    \begin{tikzpicture}[x=0.75pt,y=0.75pt,yscale=-1,xscale=1]
        %uncomment if require: \path (0,365); %set diagram left start at 0, and has height of 365
        
        %Straight Lines [id:da8539315327933412] 
        \draw    (247,314.25) -- (247,276.5) ;
        \draw [shift={(247,274.5)}, rotate = 90] [color={rgb, 255:red, 0; green, 0; blue, 0 }  ][line width=0.75]    (10.93,-3.29) .. controls (6.95,-1.4) and (3.31,-0.3) .. (0,0) .. controls (3.31,0.3) and (6.95,1.4) .. (10.93,3.29)   ;
        %Straight Lines [id:da7175927259212356] 
        \draw    (266,255.5) -- (297,255.03) ;
        \draw [shift={(299,255)}, rotate = 179.13] [color={rgb, 255:red, 0; green, 0; blue, 0 }  ][line width=0.75]    (10.93,-3.29) .. controls (6.95,-1.4) and (3.31,-0.3) .. (0,0) .. controls (3.31,0.3) and (6.95,1.4) .. (10.93,3.29)   ;
        %Shape: Circle [id:dp4154064811697209] 
        \draw   (372,255.5) .. controls (372,245.01) and (380.51,236.5) .. (391,236.5) .. controls (401.49,236.5) and (410,245.01) .. (410,255.5) .. controls (410,265.99) and (401.49,274.5) .. (391,274.5) .. controls (380.51,274.5) and (372,265.99) .. (372,255.5) -- cycle ;
        %Straight Lines [id:da02108533205456742] 
        \draw [color={rgb, 255:red, 255; green, 0; blue, 0 }  ,draw opacity=1 ]   (337,256) -- (370,255.53) ;
        \draw [shift={(372,255.5)}, rotate = 179.18] [color={rgb, 255:red, 255; green, 0; blue, 0 }  ,draw opacity=1 ][line width=0.75]    (10.93,-3.29) .. controls (6.95,-1.4) and (3.31,-0.3) .. (0,0) .. controls (3.31,0.3) and (6.95,1.4) .. (10.93,3.29)   ;
        %Shape: Circle [id:dp7009888467237113] 
        \draw   (299,255) .. controls (299,244.51) and (307.51,236) .. (318,236) .. controls (328.49,236) and (337,244.51) .. (337,255) .. controls (337,265.49) and (328.49,274) .. (318,274) .. controls (307.51,274) and (299,265.49) .. (299,255) -- cycle ;
        %Shape: Circle [id:dp755976809388597] 
        \draw   (228,255.5) .. controls (228,245.01) and (236.51,236.5) .. (247,236.5) .. controls (257.49,236.5) and (266,245.01) .. (266,255.5) .. controls (266,265.99) and (257.49,274.5) .. (247,274.5) .. controls (236.51,274.5) and (228,265.99) .. (228,255.5) -- cycle ;
        %Shape: Circle [id:dp4078559992430708] 
        \draw   (228,333.25) .. controls (228,322.76) and (236.51,314.25) .. (247,314.25) .. controls (257.49,314.25) and (266,322.76) .. (266,333.25) .. controls (266,343.74) and (257.49,352.25) .. (247,352.25) .. controls (236.51,352.25) and (228,343.74) .. (228,333.25) -- cycle ;
        %Shape: Circle [id:dp3071360088364332] 
        \draw   (371.5,175.5) .. controls (371.5,165.01) and (380.01,156.5) .. (390.5,156.5) .. controls (400.99,156.5) and (409.5,165.01) .. (409.5,175.5) .. controls (409.5,185.99) and (400.99,194.5) .. (390.5,194.5) .. controls (380.01,194.5) and (371.5,185.99) .. (371.5,175.5) -- cycle ;
        %Straight Lines [id:da5427858358156707] 
        \draw    (337,175.5) -- (369.5,175.5) ;
        \draw [shift={(371.5,175.5)}, rotate = 180] [color={rgb, 255:red, 0; green, 0; blue, 0 }  ][line width=0.75]    (10.93,-3.29) .. controls (6.95,-1.4) and (3.31,-0.3) .. (0,0) .. controls (3.31,0.3) and (6.95,1.4) .. (10.93,3.29)   ;
        %Shape: Circle [id:dp14878849029575592] 
        \draw   (299,175.5) .. controls (299,165.01) and (307.51,156.5) .. (318,156.5) .. controls (328.49,156.5) and (337,165.01) .. (337,175.5) .. controls (337,185.99) and (328.49,194.5) .. (318,194.5) .. controls (307.51,194.5) and (299,185.99) .. (299,175.5) -- cycle ;
        %Straight Lines [id:da4361092392520052] 
        \draw [color={rgb, 255:red, 255; green, 0; blue, 0 }  ,draw opacity=1 ]   (311.5,236.5) -- (311.5,194.75) ;
        \draw [shift={(311.5,192.75)}, rotate = 90] [color={rgb, 255:red, 255; green, 0; blue, 0 }  ,draw opacity=1 ][line width=0.75]    (10.93,-3.29) .. controls (6.95,-1.4) and (3.31,-0.3) .. (0,0) .. controls (3.31,0.3) and (6.95,1.4) .. (10.93,3.29)   ;
        %Straight Lines [id:da4686201934678178] 
        \draw    (406.5,184.25) -- (449,184.73) ;
        \draw [shift={(451,184.75)}, rotate = 180.64] [color={rgb, 255:red, 0; green, 0; blue, 0 }  ][line width=0.75]    (10.93,-3.29) .. controls (6.95,-1.4) and (3.31,-0.3) .. (0,0) .. controls (3.31,0.3) and (6.95,1.4) .. (10.93,3.29)   ;
        %Shape: Circle [id:dp6451998492107114] 
        \draw   (447.5,175) .. controls (447.5,164.51) and (456.01,156) .. (466.5,156) .. controls (476.99,156) and (485.5,164.51) .. (485.5,175) .. controls (485.5,185.49) and (476.99,194) .. (466.5,194) .. controls (456.01,194) and (447.5,185.49) .. (447.5,175) -- cycle ;
        %Straight Lines [id:da22465825549024876] 
        \draw    (332,189.75) -- (378.63,239.29) ;
        \draw [shift={(380,240.75)}, rotate = 226.74] [color={rgb, 255:red, 0; green, 0; blue, 0 }  ][line width=0.75]    (10.93,-3.29) .. controls (6.95,-1.4) and (3.31,-0.3) .. (0,0) .. controls (3.31,0.3) and (6.95,1.4) .. (10.93,3.29)   ;
        %Shape: Circle [id:dp6096144996363775] 
        \draw   (448,95) .. controls (448,84.51) and (456.51,76) .. (467,76) .. controls (477.49,76) and (486,84.51) .. (486,95) .. controls (486,105.49) and (477.49,114) .. (467,114) .. controls (456.51,114) and (448,105.49) .. (448,95) -- cycle ;
        %Straight Lines [id:da5228533336125574] 
        \draw    (466.52,154) -- (467,114) ;
        \draw [shift={(466.5,156)}, rotate = 270.68] [color={rgb, 255:red, 0; green, 0; blue, 0 }  ][line width=0.75]    (10.93,-3.29) .. controls (6.95,-1.4) and (3.31,-0.3) .. (0,0) .. controls (3.31,0.3) and (6.95,1.4) .. (10.93,3.29)   ;
        %Straight Lines [id:da8536135137776169] 
        \draw    (448.5,168.25) -- (409.5,167.77) ;
        \draw [shift={(407.5,167.75)}, rotate = 0.7] [color={rgb, 255:red, 0; green, 0; blue, 0 }  ][line width=0.75]    (10.93,-3.29) .. controls (6.95,-1.4) and (3.31,-0.3) .. (0,0) .. controls (3.31,0.3) and (6.95,1.4) .. (10.93,3.29)   ;
        %Straight Lines [id:da024108480310112812] 
        \draw    (451.58,109.41) -- (401,159.75) ;
        \draw [shift={(453,108)}, rotate = 135.14] [color={rgb, 255:red, 0; green, 0; blue, 0 }  ][line width=0.75]    (10.93,-3.29) .. controls (6.95,-1.4) and (3.31,-0.3) .. (0,0) .. controls (3.31,0.3) and (6.95,1.4) .. (10.93,3.29)   ;
        %Straight Lines [id:da2494767276156853] 
        \draw [color={rgb, 255:red, 255; green, 0; blue, 0 }  ,draw opacity=1 ]   (325,236.5) -- (325,194.25) ;
        \draw [shift={(325,192.25)}, rotate = 90] [color={rgb, 255:red, 255; green, 0; blue, 0 }  ,draw opacity=1 ][line width=0.75]    (10.93,-3.29) .. controls (6.95,-1.4) and (3.31,-0.3) .. (0,0) .. controls (3.31,0.3) and (6.95,1.4) .. (10.93,3.29)   ;
        %Straight Lines [id:da03772618674888184] 
        \draw    (385,236.5) -- (385,194.25) ;
        \draw [shift={(385,192.25)}, rotate = 90] [color={rgb, 255:red, 0; green, 0; blue, 0 }  ][line width=0.75]    (10.93,-3.29) .. controls (6.95,-1.4) and (3.31,-0.3) .. (0,0) .. controls (3.31,0.3) and (6.95,1.4) .. (10.93,3.29)   ;
        %Straight Lines [id:da9180273739929294] 
        \draw    (398,237) -- (398,194.75) ;
        \draw [shift={(398,192.75)}, rotate = 90] [color={rgb, 255:red, 0; green, 0; blue, 0 }  ][line width=0.75]    (10.93,-3.29) .. controls (6.95,-1.4) and (3.31,-0.3) .. (0,0) .. controls (3.31,0.3) and (6.95,1.4) .. (10.93,3.29)   ;
        
        % Text Node
        \draw (242,329.4) node [anchor=north west][inner sep=0.75pt]    {$n$};
        % Text Node
        \draw (242.5,249.9) node [anchor=north west][inner sep=0.75pt]    {$p$};
        % Text Node
        \draw (313,247.9) node [anchor=north west][inner sep=0.75pt]    {$d$};
        % Text Node
        \draw (380.5,247.4) node [anchor=north west][inner sep=0.75pt]    {$^{3}\mathrm{H}$};
        % Text Node
        \draw (384,171.9) node [anchor=north west][inner sep=0.75pt]    {$\alpha $};
        % Text Node
        \draw (305,167.4) node [anchor=north west][inner sep=0.75pt]    {$^{3}\mathrm{He}$};
        % Text Node
        \draw (455,167.9) node [anchor=north west][inner sep=0.75pt]    {$^{7}\mathrm{Li}$};
        % Text Node
        \draw (454,86.9) node [anchor=north west][inner sep=0.75pt]    {$^{7}\mathrm{Be}$};
        % Text Node
        \draw (268,258.9) node [anchor=north west][inner sep=0.75pt]  [font=\scriptsize]  {$( n,\gamma )$};
        % Text Node
        \draw (339,259.4) node [anchor=north west][inner sep=0.75pt]  [font=\scriptsize,color={rgb, 255:red, 255; green, 0; blue, 0 }  ,opacity=1 ]  {$( d,p)$};
        % Text Node
        \draw (280.5,223.9) node [anchor=north west][inner sep=0.75pt]  [font=\scriptsize,color={rgb, 255:red, 255; green, 0; blue, 0 }  ,opacity=1 ]  {$( d,n)$};
        % Text Node
        \draw (280.5,205.4) node [anchor=north west][inner sep=0.75pt]  [font=\scriptsize,color={rgb, 255:red, 255; green, 0; blue, 0 }  ,opacity=1 ]  {$( p,\gamma )$};
        % Text Node
        \draw (330,214.9) node [anchor=north west][inner sep=0.75pt]  [font=\scriptsize]  {$( n,p)$};
        % Text Node
        \draw (338,158.4) node [anchor=north west][inner sep=0.75pt]  [font=\scriptsize]  {$( d,p)$};
        % Text Node
        \draw (400.5,205.4) node [anchor=north west][inner sep=0.75pt]  [font=\scriptsize]  {$( p,\gamma )$};
        % Text Node
        \draw (400.5,224.4) node [anchor=north west][inner sep=0.75pt]  [font=\scriptsize]  {$( d,n)$};
        % Text Node
        \draw (404.5,186.4) node [anchor=north west][inner sep=0.75pt]  [font=\scriptsize]  {$(^{3}\mathrm{H} ,\gamma )$};
        % Text Node
        \draw (419,152.4) node [anchor=north west][inner sep=0.75pt]  [font=\scriptsize]  {$( p,\alpha )$};
        % Text Node
        \draw (468.5,127.9) node [anchor=north west][inner sep=0.75pt]  [font=\scriptsize]  {$( n,p)$};
        % Text Node
        \draw (392,113.4) node [anchor=north west][inner sep=0.75pt]  [font=\scriptsize]  {$(^{3}\mathrm{He} ,\gamma )$};
        % Text Node
        \draw (203.33,283.57) node [anchor=north west][inner sep=0.75pt]  [font=\scriptsize]  {$\mathrm{weak}$};
        % Text Node
        \draw (194,295.57) node [anchor=north west][inner sep=0.75pt]  [font=\scriptsize]  {$\mathrm{processes}$};
        
    \end{tikzpicture}

    \caption{A diagram representing the ``key'' network, showing the nuclei that are tracked and the reactions between them. Shown in red are deuterium reactions that dominate the uncertainty in the prediction of D/H~\cite{Pisanti:2020efz,Pitrou:2021vqr}.}
    \label{fig:key_network}

\end{figure}
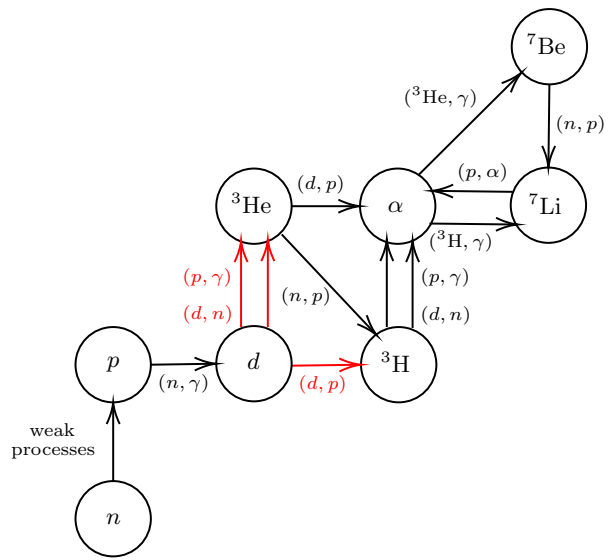

In the meantime, LINX takes an agnostic approach to the different sets of rates discussed above. 
\citetalias{Giovanetti_2024} provides our results for the inference of SBBN and SBBN \texttt{+} $\Delta N_\text{eff}$ using the three sets of rates. 
LINX allows users to choose between nuclear networks constructed from these three different approaches by setting the keyword \texttt{nuclear\_net} in \texttt{nuclear.NuclearRates} appropriately, instead of supplying a list of \texttt{Reaction} instances.
Our choices for these canonical networks are similar to the choices made by PRyMordial~\cite{burns_2023}. 
First, we define a ``key" network of 12 nuclear processes (on top of the weak processes governing proton-neutron interconversion) which have the largest impact on BBN predictions.   
The ``key'' network comprises the same key nuclear reactions adopted by PRyMordial, and is identical to the reduced network used by the PRIMAT code; these are: the $n \leftrightarrow p$ weak processes, $p(n,\gamma)d$, $d(p,\gamma)$$^3$He, $d(d,p)t$, $d(d,n)$$^3$He, $^3$He$(n,p)t$, $^3$He$(d,p)$$^4$He, $t(d,n)$$^4$He, $^3$He$(\alpha,\gamma)$$^7$Be, $t(\alpha,\gamma)$$^7$Li, $^7$Be$(n,p)$$^7$Li, $^7$Li$(p,\alpha)$$^4$He, and $t(p,\gamma)$$^4$He. 
We therefore track eight nuclear species: $n$, $p$, $d$, $t$, $^3$He, $^4$He, $^7$Li and $^7$Be for ``key'' networks. 
All but the last reaction is identified in Ref.~\cite{Iliadis:2020jtc} as the most important processes for an accurate BBN prediction of D/H and Y$_\text{P}$.  This ``key'' network is illustrated in Figure~\ref{fig:key_network}. 
In red, we show the three deuterium burning processes whose rates are the main sources of uncertainty in the prediction of D/H (see our discussion below in Section~\ref{sec:uncert_impact} and Table~\ref{tab:sigmas}); these are $d(p,\gamma)$$^3$He, $d(d,p)t$, and $d(d,n)$$^3$He. Disagreements in the $S$-factor used in the PRIMAT, PArthENoPE and YOF for these reactions are thought to be responsible for the discrepancy in the D/H predictions between these different approaches~\cite{Pitrou:2021vqr}.

Next, we define a ``full" network of 61 reactions, including the 12 reactions listed above, and an additional 49 reactions, involving $^6$He, $^6$Li, $^8$Li, and $^8$B, for a total of 12 nuclear species to track. 
These reactions are identical to those listed in Table 2 of Ref.~\cite{burns_2023},\footnote{There are two reactions that are listed twice in the table: $^7$Be$(^3\text{He}, p) p\alpha\alpha$ and $^7$Li$(t, n)n\alpha\alpha$.}  but with the data taken directly from tables provided by PRIMAT~\cite{Pitrou_2018}.

LINX offers the user the flexibility of choosing between canonical sets of $\lambda(T)$ and $\sigma(T)$ easily. 
The choices that have already been included by default in LINX are:
\begin{enumerate} 
    \item \texttt{`np\_only'}: This choice only includes the $n \leftrightarrow p$ weak processes and no processes including heavier nuclei.
    This is appropriate in the regime where $T_\gamma \gtrsim \SI{1}{\mega\eV}$; 
    \item \texttt{`key\_PRIMAT\_2023'}: This is a ``PRIMAT'' network based on tables of rates obtained from the reduced network in PRIMAT v0.2.2~\cite{Pitrou_2018};  
    
    \item \texttt{`full\_PRIMAT\_2023'}: This uses PRIMAT v0.2.2 rates for the ``full" network; 
    \item \texttt{`key\_PRIMAT\_2018'}: This is based on a set of tables used by PRyMordial, which in turn is based on an earlier version of PRIMAT\@. 
    The $d(p,\gamma)$$^3$He, $^3$He$(d,p)$$^4$He, $t(d,n)$$^4$He, and $^7$Be$(n,p)$$^7$Li rates have all been updated in \texttt{`key\_PRIMAT\_2023'}; 
    \item \texttt{`key\_YOF'}: These are the ``key" network rates that we call ``YOF'' in~\citetalias{Giovanetti_2024}. 
    They are also based on a set of tables used by PRyMordial, which compiled these rates based on Ref.~\cite{Yeh:2020mgl}.   
    Compared to Ref.~\cite{Yeh:2020mgl}, our network leaves out two reactions ($^7$Be$(n,\alpha)$$^4$He and $^7$Be$(d,p)$$^4$He$^4$He) and adds $t(p,\gamma)$$^4$He. 
    The neglected reactions were shown to be inconsequential for D/H and Y$_\text{P}$ predictions; and
    \item \texttt{`key\_PArthENoPE'}: These are the ``key" network rates that we call ``PArthENoPE'' in~\citetalias{Giovanetti_2024}. The rates we use consist of a list of functions extracted from the PArthENoPE code. 
    For each reaction, PArthENoPE provides a function returning their $-1\sigma$, median, and $+1\sigma$ values of $\lambda(T)$. We take $\overline{\lambda}(T)$ to be given by the median and estimate the uncertainty $\sigma(T)$ by taking the $+1\sigma$ values. 
\end{enumerate}
For clarity, Table~\ref{tab:reaction-rates} shows the references for all of the rates used in the \texttt{key\_PRIMAT\_2023}, \texttt{key\_PRIMAT\_2018} and \texttt{key\_YOF} ``key" networks.
Rates for \texttt{key\_PArthENoPE} are all taken from the latest version of the PArthENoPE code~\cite{Gariazzo:2021iiu}. 
Rates in \texttt{full\_PRIMAT\_2023}, other than those already in \texttt{key\_PRIMAT\_2023}, use the same sources as those listed in Ref.~\cite{burns_2023}.

For users that are interested only in accurate predictions of D/H and Y$_\mathrm{P}$, choosing any of the ``key" networks is sufficient, in line with recommendations in Ref.~\cite{Iliadis:2020jtc}.
For $^7$Li/H, however, a ``full'' network run is necessary for both improved accuracy and a complete treatment of nuclear rate uncertainties, which can be as large as $\sim 15\%$, depending on the choice of network (see \textit{e.g.}\ Ref.~\cite{Fields:2022mpw}).
Whether a percent-level accuracy for $^7$Li/H is necessary at this stage remains unclear. 
Measurements of $^7$Li/H from both metal-poor stars (with a reported precision of $\sim 20\%$)~\cite{Sbordone:2010zi} and low-metallicity interstellar gas in the Small Magellanic Cloud (with a reported precision of $\sim 5\%$)~\cite{Molaro:2024wxa} report consistent abundances that are significantly lower than expected from SBBN predictions. While lithium destruction in stellar atmospheres could reconcile this difference in stars~\cite{Fields:2022mpw}, this argument does not apply to interstellar gas~\cite{Molaro:2024wxa}.

QED corrections to nuclear reaction rates have been computed in Ref.~\cite{Pitrou:2019pqh}, which found that the impact of such corrections on primordial abundances are at the 0.2\% level, well below current measurement uncertainties. 
While they can certainly be incorporated into LINX, we neglect these corrections at the present time. 

The rates compiled by LINX are stored in the \texttt{data/nuclear\_rates} directory, with each network stored in its own subdirectory.

Once initialized, a \texttt{NuclearRates} instance can be called by passing the following arguments to the instance:
\begin{enumerate}
    \item \texttt{Y}: an array containing the current abundances of all species; 
    \item \texttt{T\_t}: the current $T_\gamma$; 
    \item \texttt{rhoBBN}: $\rho_\text{b}$, given in \SI{}{\gram\per\centi\meter\cubed} to conform with the nuclear physics units convention; 
    \item \texttt{T\_interval}: an abscissa along which to specify $n \leftrightarrow p$ rates; 
    \item \texttt{nTOp\_frwrd\_vec}: $\Gamma_{n \to p}$, evaluated at points given by \texttt{T\_interval}, computed by \texttt{WeakRates}; 
    \item \texttt{nTOp\_bkwrd\_vec}: $\Gamma_{p \to n}$, evaluated at points given by \texttt{T\_interval}, computed by \texttt{WeakRates}; 
    \item \texttt{eta\_fac}: a parameter that rescales the baryon-to-photon ratio with respect to its default value, stored in \texttt{const.Omegabh2};
    \item \texttt{tau\_n\_fac}: a parameter that rescales the neutron lifetime with respect to its default value, stored in \texttt{const.tau\_n}, and
    \item \texttt{nuclear\_rates\_q}: an array specifying the value of $q_i$ for each reaction in the nuclear network.
\end{enumerate}
This class instance returns an array containing $\dot{Y}_i$ for each of the species being tracked, with the same dimensions as the input parameter \texttt{Y}. 

\subsubsection{Abundance Calculation}

With the \texttt{WeakRates} and \texttt{NuclearRates} classes, we now have all the infrastructure necessary to perform the integration of Eq.~\eqref{eq:abundance_ODE}, given the evolution of background quantities computed in an instance of \texttt{BackgroundModel}.
This is done within the \texttt{abundances.AbundanceModel} equinox module, which can be initialized by supplying an instance of \texttt{NuclearRates} containing the choice of reaction network, and an instance of \texttt{WeakRates}, which calculates $\Gamma_{n \leftrightarrow p}$.\footnote{Specifying \texttt{WeakRates} is optional and if not specified, it is initialized using the default settings.}
Any instance of \texttt{AbundanceModel} can be called by providing the following arguments, which are outputs of \texttt{BackgroundModel}, and all have the same length: 
\begin{enumerate}
    \item \texttt{rho\_g\_vec}: an array containing the energy density of photons; 
    \item \texttt{rho\_nu\_vec}: an array containing the energy density of one species of neutrinos; 
    \item \texttt{rho\_NP\_vec}: an array containing the energy density of the extra relativistic degrees of freedom; and
    \item \texttt{P\_NP\_vec}: an array containing the pressure of the extra relativistic degrees of freedom, which we take to be $p_\text{ext} = \rho_\text{ext} / 3$ in SBBN\texttt{+}$\Delta N_\text{eff}$.
\end{enumerate}
The user can optionally specify \texttt{a\_vec} and \texttt{t\_vec}; these specify the corresponding scale factors and times, respectively, and must be arrays of the same length as the above quantities.
These are outputs of \texttt{BackgroundModel}, but if they are not specified, they are calculated by solving $dt/d\rho_\text{tot}$ and $da/d\rho_\text{tot}$ using energy-momentum conservation, 
\begin{alignat}{1}
    \dot{\rho}_\text{tot} + 3 H (\rho_\text{tot} + p_\text{tot}) = 0 \,,
\end{alignat}
combined with Eq.~\eqref{eq:Hubble}.\footnote{The time $t$ can be written as a function of $\rho_\text{tot}$ as long as $\rho_\text{tot}$ decreases monotonically, which is true unless we are dealing with fluids with $w < -1$, where $p = w \rho$.}

Without passing further options, an instance of \texttt{AbundanceModel} will compute elemental abundances with default values of $100 \Omega_b h^2 = 2.242^{+0.014}_{-0.014}$ (Planck 2018 TTTEEE+lowE+lensing+BAO~\cite{Planck:2018vyg}), $\tau_n = 879.4^{+0.6}_{-0.6}~\SI{}{\second}$ (a mean of lifetime measurements based on trap experiments from Ref.~\cite{Czarnecki:2018okw}), and $q_i = 0$.
The user is free to change these defaults. 
To scan over different values of these parameters, the user should use the following optional keyword arguments when calling an instance of \texttt{AbundanceModel}: 
\begin{enumerate}
    \item \texttt{eta\_fac}: a factor to multiply the default value of $\Omega_b h^2$ by; 
    \item \texttt{tau\_n\_fac}: a factor to multiply the default value of $\tau_n$ by; and
    \item \texttt{nuclear\_rates\_q}: an array with length equal to the number of reactions being used in the instance of \texttt{NuclearRates}, indicating the value of $q_i$ for all of the nuclear rates. 
\end{enumerate}
To perform a scan over these parameters, for example, the user simply needs to call the instance of \texttt{AbundanceModel} with different values for these optional keyword arguments. 

Additional optional keyword arguments control the initial conditions to use when solving Eq.~\eqref{eq:abundance_ODE}: 
\begin{enumerate}
    \item \texttt{Y\_i}: an array containing the initial abundances of all species considered in the nuclear network; 
    \item \texttt{T\_start}: the initial temperature at which to start the integration of Eq.~\eqref{eq:abundance_ODE}; and
    \item \texttt{T\_end}: the final temperature at which to finish the integration of Eq.~\eqref{eq:abundance_ODE}.
\end{enumerate}
If none of these arguments is used, the solver defaults to starting the integral at the temperature stored in \texttt{const.T\_start} and ending the integration at \texttt{const.T\_end}. 
All nuclei are taken to be in nuclear statistical equilibrium, with abundances $Y_{i,0}$, explicitly given by $Y_{n,0} = \Gamma_{p \to n} / (\Gamma_{n \to p} + \Gamma_{p \to n})$, $Y_{p,0} = 1 - Y_{n,0}$, and~\cite{Pitrou_2018}
\begin{alignat}{2}
    Y_{i,0} &=&& (2 s_i + 1) \zeta(3)^{A_i - 1} 2^{\frac{3A_i - 5}{2}} \pi^{\frac{1 - A_i}{2}} \nonumber \\
    & && \times \left( \frac{m_i T^{A_i - 1}}{m_p^{Z_i} m_n^{A_i - Z_i}} \right)^{3/2} \eta^{A_i - 1} Y_{p,0}^{Z_i} Y_{n,0}^{A_i - Z_i} e^{B_i / T} \,,
\end{alignat}
where $s_i$ is the spin of the nucleus, $\zeta$ is the Riemann zeta function, $Z_i$ is the atomic number of the species, $A_i$ is the atomic mass number of the species, $m_i$ is the mass of the species, $\eta$ is the baryon-to-photon ratio, and $B_i$ is the binding energy of the species. 
Note that neutrons and protons start to fall out of chemical equilibrium at the neutrino decoupling temperature ($T \sim \SI{2}{\mega\eV}$), and so initial conditions must be specified carefully at or below this temperature. 

By design, \texttt{AbundanceModel} can integrate Eq.~\eqref{eq:abundance_ODE} starting from any specified initial conditions, allowing the user to switch between nuclear networks during the integration. 
If a small network---such as the ``key" networks consisting of only 12 nuclear reactions plus weak processes---is used for the abundance predictions, the user can simply specify a single \texttt{AbundanceModel} to perform the integration. 
For larger networks, such as the ``full" network consisting of 61 reactions, we recommend following the prescriptions used in both PRIMAT~\cite{Pitrou_2018} and PRyMordial~\cite{burns_2023} of integrating progressively larger networks, for greater computational efficiency. 
This involves solving the abundances using a ``key" network and tracking 8 nuclear species initially, before switching over to the ``full" network and tracking 12 species. 
This can be done by using two \texttt{AbundanceModel} instances with different nuclear networks, and passing the final output from the first as the initial state of the second.
The temperature at which this switch occurs is by default \SI{0.108}{\mega\eV} and is stored in \texttt{const.T\_switch}. 

Other keywords in \texttt{AbundanceModel} allow the user to control properties of the integrator and whether to return just the final abundance or the full abundance evolution as a function of time. 

We once again use Diffrax~\cite{kidger2021} to solve Eq.~\eqref{eq:abundance_ODE}, this time using Kv{\ae}rn{\o}'s 3/2 method by default, which is designed for stiff equation solving~\cite{kvaerno2004singly}. 
The default output of a call to an instance of \texttt{AbundanceModel} is simply an array of abundances for all species considered by the network.

\section{Validation and Performance}\label{sec:validate_limitate}
\subsection{Validation of LINX}\label{sec:comparisons}
To validate LINX, we compare the standard BBN primordial abundances of elements predicted by LINX with those predicted by PRyMordial and PRIMAT\@. 
We perform this validation by comparing predicted $N_\mathrm{eff}$, Y$_\mathrm{P}$, D/H, $^3$He/H (including tritium, which decays to $^3$He with a half-life of 12.32~years~\cite{Kondev:2021lzi}) and $^7$Li/H (including $^7$Be, which decays to $^7$Li with a half-life of 53.3~days~\cite{Kondev:2021lzi}). 
The current experimental uncertainties for $N_\text{eff}$ are at the $\sim 10\%$ level~\cite{Planck:2018vyg} during the CMB epoch, while uncertainties on direct measurements of primordial D/H and Y$_\mathrm{P}$ are at the $\sim 1\%$ level~\cite{Aver_2015,Cooke_2018}. 
Finally, $^7$Li/H has been measured with an uncertainty of $\sim 20\%$~\cite{Sbordone:2010zi} in metal-poor stars and $\sim 5\%$~\cite{Molaro:2024wxa} in low-metallicity gas, although as discussed, whether these measurements truly reflect the primordial abundance of $^7$Li is still somewhat unclear. 
Ideally, we would like the results from LINX, PRyMordial, and PRIMAT---given the same cosmological parameters and nuclear rates---to be in agreement at a level well below the experimental uncertainties discussed above. 

The main features of LINX that we would like to validate by comparing our results to existing code packages are as follows: 
\begin{enumerate}

    \item We develop custom implementations of the gamma function, polylogarithms, and modified Bessel functions of the second kind, and use autodifferentiation to obtain derivatives in our code;

    \item We perform all numerical quadrature either by transforming the integrals into a series (used for solving the background quantities discussed in Section~\ref{sec:background_quantities}), or by switching to the trapezium rule (used for evaluating the weak rates in Section~\ref{sec:weak_rates}). It is important that these new methods do not compromise the accuracy of the calculation; 

    \item As described in Section~\ref{sec:background_quantities}, we solve for the thermodynamics by integrating with respect to time, and not the photon temperature, and do not make use of entropy conservation at all. Instead, LINX performs its integration in time with no preset end point and terminates the integration once the photon temperature drops below some critical value (chosen by default to be \SI{5.2}{\kilo\eV} in LINX). This is a mathematically equivalent approach to integrating with respect to photon temperature, but numerically should be validated;

    \item We use differential equation solvers offered by the package \texttt{Diffrax} that are different from those used by PRIMAT and PRyMordial; and 

    \item Once a network of reactions is specified, we construct the differential equations governing the evolution of the abundance of light elements as a function of time ``on the fly", rather than hardcoding them.
\end{enumerate}

Throughout this Section, we fix $\Omega_b h^2 = 0.02242$ and $\tau_n = \SI{879.4}{\second}$ for all three packages. 
Furthermore, we use only linear interpolation over nuclear reaction rates for \textit{all packages}, since switching between linear and cubic interpolation in PRIMAT can change the prediction of $^7$Li/H by up to 2\%. 
Note that changing the relative and absolute tolerances in the differential equation solvers can also lead to nontrivial shifts in predicted quantities; we use the default settings in the solvers used in all three codes. 
All other settings in LINX are set to their default values. 

\begin{table*}[!t]
    \centering
    \renewcommand{\arraystretch}{1.5} 
    \begin{tabular}{cccccc}
        \multicolumn{1}{p{2.5cm}}{\centering \textbf{Code}} & \multicolumn{1}{p{2.5cm}}{\centering \textbf{$\boldsymbol{N}_\text{eff}$}} & \multicolumn{1}{p{2.5cm}}{\centering \textbf{Y$_{\textrm{P}}$}} & \multicolumn{1}{p{2.5cm}}{\centering \textbf{D/H}} & \multicolumn{1}{p{2.5cm}}{\centering \textbf{$^{\mathbf{3}}$He/H $\mathbf{\times 10^{5}}$}} & \multicolumn{1}{p{2.5cm}}{\centering $^{\mathbf{7}}$\textbf{Li/H} $\mathbf{\times 10^{10}}$} 
         \\

        \hline
        \multicolumn{6}{c}{\centering \textbf{LINX vs. PRyMordial ``key" network}} \\
        PRyMordial & 3.0444 & 0.2471 & 2.4373 & 1.0385 & 5.5570 \\
        LINX & 3.0444 & 0.2471 & 2.4386 & 1.0385 & 5.5505 \\
        \% Difference & -0.00 & \texttt{+}0.00 & \texttt{+}0.05 & -0.00 & -0.12 \\
        
        \hline

        \multicolumn{6}{c}{\centering \textbf{LINX vs. PRyMordial, ``full" network}} \\

        PRyMordial & 3.0444 & 0.2471 & 2.4397 & 1.0387 & 5.4925 \\
        LINX & 3.0444 & 0.2471 & 2.4390 & 1.0383 & 5.4937 \\
        \% Difference & -0.00 & -0.01 & -0.03 & -0.04 & \texttt{+}0.02 \\
       
        \hline

        \multicolumn{6}{c}{\centering \textbf{LINX vs. PRIMAT, ``key" network}} \\

        PRIMAT & 3.0435 & 0.2471 & 2.4379 & 1.0362 & 5.6875 \\
        LINX & 3.0444 & 0.2471 & 2.4346 & 1.0355 & 5.6998 \\
        \% Difference & \texttt{+}0.03 & \texttt{+}0.01 & -0.13 & -0.06 & \texttt{+}0.22 \\

        \hline

        \multicolumn{6}{c}{\centering \textbf{LINX vs. PRIMAT, ``full" network}} \\

        PRIMAT & 3.0435 & 0.2471 & 2.4386 & 1.0360 & 5.6299 \\
        LINX & 3.0444 & 0.2471 & 2.4351 & 1.0353 & 5.6422 \\
        \% Difference & \texttt{+}0.03 & -0.00 & -0.14 & -0.06 & \texttt{+}0.22 \\
        \hline 
        
    \end{tabular}
    \caption{Comparison of primordial abundance predictions between LINX and established BBN codes PRyMordial and PRIMAT\@. The percentage differences are calculated as $(\text{LINX} - \text{reference}) / \text{reference}$, where the reference is either PRyMordial or PRIMAT\@. All calculations use $\Omega_b h^2 = 0.02242$ and $\tau_n = 879.4$ s.}
    \label{tab:full_comparison}
\end{table*}

Table~\ref{tab:full_comparison} compares the values of $N_\text{eff}$, Y$_\mathrm{P}$, D/H, $^3$He/H, and $^7$Li/H predicted by LINX with those predicted by either PRyMordial or PRIMAT, assuming either a ``key" network or a ``full" network. 
We also show the percentage difference between predictions, computed by taking (LINX - reference) / reference, where reference is either PRyMordial or PRIMAT\@. 

For the ``key" network comparison, we compare LINX's \texttt{key\_PRIMAT\_2018} results with the PRyMordial output obtained by setting \texttt{nacreii\_flag = False} and \texttt{smallnet\_flag = True}. 
Meanwhile, for the ``full" network comparison, we use LINX's \texttt{full\_PRIMAT\_2018} results and set \texttt{nacreii\_flag = False} and \texttt{smallnet\_flag = False} in PRyMordial. 
The choice of reaction networks is identical in this case, and the rates are taken ultimately from the same source. 
There is excellent agreement between the packages, with at most a 0.12\% disagreement between predicted abundances. 
For the experimentally-measured quantities $N_\text{eff}$, Y$_\mathrm{P}$, and D/H, the difference in prediction is less than 0.05\%, well below experimental uncertainties. 

We next consider the comparison between LINX and PRIMAT \texttt{v0.2.2}, released in 2023.  
To perform a consistent comparison, we set the following options in PRIMAT: 
\begin{enumerate}
    \item We set \texttt{InterpOrder=1}, so that all rates are interpolated linearly; 
    \item We neglect QED corrections in the rates of nuclear processes with photons in the final state, pointed out in Ref.~\cite{Pitrou:2019pqh} by setting \texttt{\$NuclearRatesQEDCorrections = False}, which only impacts D/H at the 0.2\% level; and
    \item We use an approach used by earlier versions of PRIMAT to account for incomplete decoupling of neutrinos, which uses a fitting function to parameterize the energy exchange between the EM and neutrino fluids. 
    This is accomplished by setting \texttt{\$NEVO = False}, which neglects subleading effects such as neutrino spectral distortionss~\cite{Froustey:2020mcq}, leading to a change in the abundances at the level of less than 0.1\%. 
\end{enumerate}

\begin{figure}[t]
    \centering
    \includegraphics[width=\linewidth]{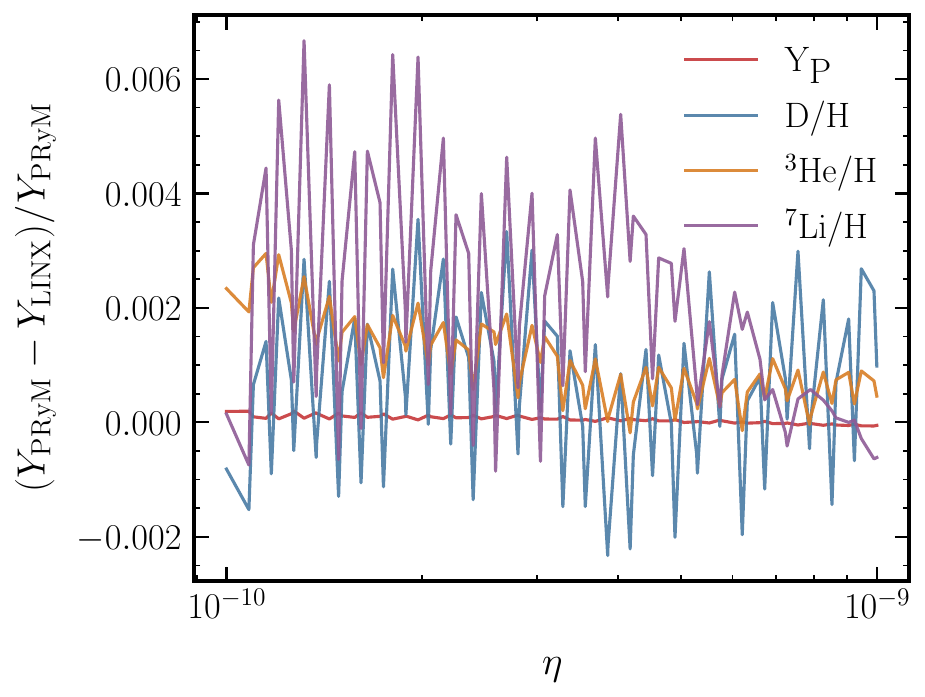}
    \caption{Residuals between PRyMordial and LINX for four primordial element abundances as a function of the baryon-to-photon ratio $\eta$.  For this comparison, we set $\tau_n=\SI{879.4}{s}$ in both codes, use the \texttt{key\_PRIMAT\_2018} network in LINX, and set \texttt{smallnet\_flag=True} in PRyMordial.  The agreement between the two codes is $\ll 1\%$ throughout the range of $\eta$ for D/H and $Y_\mathrm{P}$.  The jagged behavior is likely due to numerical noise associated with the differential equation solvers in either code.}
    \label{fig:residualsPRyM}
\end{figure}

For the ``key" network comparison, we use \texttt{key\_PRIMAT\_2023} in LINX and set \texttt{\$ReducedNetwork = True} in PRIMAT, which leads to the same nuclear network and rates between the two packages. 
Meanwhile, for the ``full" network comparison, LINX uses \texttt{key\_PRIMAT\_2023} up till \texttt{const.T\_switch}, before switching to \texttt{full\_PRIMAT\_2023} at lower temperatures in LINX\@. 
In PRIMAT, we simply set \texttt{\$ReducedNetwork = False}. 
In this case, both reaction networks are identical until \texttt{const.T\_switch}, after which the PRIMAT reaction network becomes much larger, tracking a total of 422 reactions, and nuclear species up to $^{23}$Na. 
Table~\ref{tab:full_comparison} indicates that the agreement for $N_\text{eff}$, Y$_\mathrm{P}$, and D/H is better than 0.15\%, while the agreement for $^7$Li/H is at the level of 0.22\%, once again well below experimental precision. 
This result also demonstrates that our ``full" network of 61 reactions, rather than PRIMAT's 422 reactions, is sufficiently accurate for predicting elemental abundances up to and including $^7$Li.

Finally, Figure~\ref{fig:residualsPRyM} provides a comparison of the output of LINX to that of PRyMordial, as a function of the baryon-to-photon ratio $\eta$ in the ``key" network.  The agreement between these two codes is $\ll 1\%$ for D/H and $Y_\mathrm{P}$ over the entire range of $\eta$ probed.  

\subsection{Performance}

On a single core of an M1 Mac, using the ``key" reaction networks, LINX computes thermodynamics and abundances in about a third of a second, with the abundance calculation taking tens of milliseconds to a few tenths of a second depending on the reaction network used.  Exact runtimes will vary based on hardware.  These are the evaluation times after the code compiles; compiling both the background thermodynamics and abundance modules takes roughly one minute.  When the ``full" networks are used, runtime is typically about half a second for a calculation of both the background thermodynamics and the abundances, following a compilation of a few minutes.  This renders LINX appropriate for MC sampling methods where millions of samples are required to estimate the relevant posteriors.  

For analyses involving gradients, where the gradient is computed over one input to the code, the ``key" networks compile in roughly a minute and run in about half a second, making them suitable for gradient-based sampling methods where thousands of samples may be computed.  Gradients involving the ``full" networks are more expensive, compiling in roughly ten minutes and running in roughly a second after compiling. 
The compilation times, while not significant when performing MC calculations, can still be inconvenient. 
This is due in part to the fact that the XLA compiler is not ideally suited for many control flows inherent to solving differential equations (\textit{e.g.}\ loops and conditionals). 
Since both the background and abundance calculations involve sequential operations of integrating coupled differential equations, this leads to relatively long compilation times~\cite{kidger2024}.

\begin{figure}[t]
    \centering
    \includegraphics[width=\linewidth]{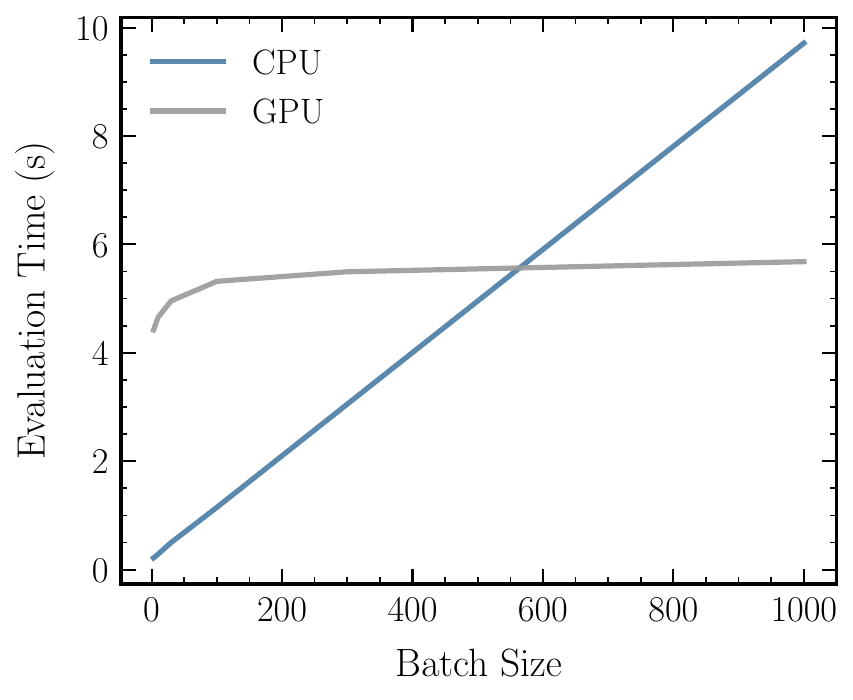}
    \caption{Scaling of evaluation time with batch size when LINX is run on a CPU versus a GPU\@.  For small batch sizes, the CPU is more efficient, but for batch sizes $\gtrsim 550$, the GPU becomes more efficient due to the more favorable scaling.}
    \label{fig:timings}
\end{figure}

Code written in JAX is often well-suited for running on a GPU, and so one might expect to see a speedup from running LINX on a GPU\@.  However, we find this is only true for large batch sizes.  Vectorizing computations (\textit{e.g.}\ evaluating the likelihood for multiple parameter points simultaneously) is significantly more efficient on GPUs, and so any speed advantage only becomes apparent with very large batch sizes. In practice, this means that for typical BBN calculations, CPU computation remains more efficient unless large batches $\gtrsim 550$ are used.  The scaling of the CPU and GPU evaluation times with batch size is depicted in Figure~\ref{fig:timings}.  
The reason for the significant GPU overhead with small batch sizes is that LINX encounters frequent host synchronization operations (\textit{i.e.}\ data transfer between CPU and GPU) when running on GPU\@. As a result, contrary to what might be expected given the use of JAX for the core code, GPU acceleration does not at present offer a performance advantage by default for most common LINX computations.  For example, when running an MC analysis with LINX, samples must be evaluated sequentially, and so a CPU is more desirable for these types of analyses.  However, if one wishes to use LINX to generate an interpolation table of primordial abundances that incorporates effects from nuclear rate uncertainties, memory would likely become a limiting factor well before speed if this computation were run on a GPU\@.

That said, compile times are somewhat faster on a GPU, and a GPU may also be more efficient if the forward model incorporates components that could benefit from parallelizable computations (\textit{e.g.}\ neural networks or Gaussian processes), in which case the aforementioned bottlenecks may no longer dominate the runtime. 

\section{Examples}\label{sec:examples}
In this Section, we provide two pedagogical examples to familiarize the user with LINX\@. 
In addition to demonstrating the use of different LINX modules, these examples also introduce JIT and vectorization with \texttt{vmap}, which are built into JAX\@.  These examples and more are also available as Jupyter notebooks in the \href{https://github.com/cgiovanetti/LINX}{LINX GitHub} \githubicon.

\subsection{The Impact of Nuclear Rate Uncertainties}\label{sec:uncert_impact}
A significant advantage of LINX is the ease with which it allows the user to sample the uncertainties of nuclear rates active during BBN\@.  In this example, we explore how those nuclear rates may be varied and the resulting impact on predicted abundances.  First, we will illustrate how the abundances of D/H and Y$_{\rm{P}}$ can vary as a function of the $d(p,\gamma)^3$He rate and the neutron lifetime, two of the more important nuclear rates determining SBBN abundances.  
Then, to give readers an idea of how important each rate is to BBN predictions, we will explore how separately varying each reaction rate in the ``key" network by one standard  deviation impacts the abundances of D/H, Y$_{\rm{P}}$, $^3$He/H, and $^7$Li/H.

To begin, we need to calculate the background thermodynamic evolution. This is done using the \texttt{background.BackgroundModel} equinox module, which we initialize in the following manner: 
\begin{lstlisting}[language=Python]
from linx.background import BackgroundModel

SBBN_DNeff_bkg = BackgroundModel()
\end{lstlisting}
\texttt{BackgroundModel}, initialized this way, includes all the default options for computing the background quantities, including fully tracking neutrino decoupling, correctly taking the neutrinos to obey a Fermi-Dirac distribution (as opposed to a Maxwell-Boltzmann distribution), finite electron mass corrections, and higher-order QED corrections. 
\texttt{SBBN\_DNeff\_bkg} can now be called with one argument indicating $\Delta N_{\text{eff},i}$, which in the case of SBBN is set to zero. 
This integrates the background evolution from \texttt{const.T\_start} to \texttt{const.T\_end}, which are by default set at \SI{8.6}{\mega\eV} and \SI{5.2}{\kilo\eV}, respectively.  It outputs a tuple of arrays containing: the time abscissa \texttt{t\_vec}; the corresponding scale factor \texttt{a\_vec}; the energy density of photons \texttt{rho\_g}; the energy density of one species of neutrinos \texttt{rho\_nu}; the energy density of the inert, relativistic species \texttt{rho\_NP}; the pressure of the same species \texttt{p\_NP}; and finally $N_\text{eff}$ \texttt{Neff}, all as a function of time \texttt{t\_vec}.  Specifically: 
\begin{lstlisting}[language=Python]
t_vec, a_vec, rho_g, rho_nu, rho_NP, p_NP, Neff = SBBN_DNeff_bkg(jnp.asarray(0.))
\end{lstlisting}
We wrap the input to \texttt{SBBN\_DNeff\_bkg} (and any non-array float input to LINX) in \texttt{jnp.asarray()} to pass them as JAX arrays and to allow the equinox module to compile the function for arrays.

Next, we create an instance of the \texttt{abundances.AbundanceModel} equinox module, initializing with the ``\texttt{key\_PRIMAT\_2023}'' set of rates: 
\begin{lstlisting}[language=Python]
from linx.nuclear import NuclearRates
from linx.abundances import AbundanceModel

nuc_rates = NuclearRates(
    nuclear_net='key_PRIMAT_2023'
)

abundance_model = AbundanceModel(nuc_rates)
\end{lstlisting}
\texttt{abundance\_model} is now able to calculate the primordial abundance of elements given the background evolution determined above, as well as the value of $\Omega_b h^2$, $\tau_n$, and $q_i$ for the nuclear rates, by specifying \texttt{eta\_fac} (a float), \texttt{tau\_n\_fac} (another float), and \texttt{nuclear\_rates\_q} (an array of floats).  

Recall that the total rate for the $i^{\rm th}$ reaction in the network is given by $\log \lambda_i(T) = \log \overline{\lambda}_i(T) + q_i\sigma_i(T)$, so the rates $\log \lambda_i(T)$ are varied by specifying $q_i$ in \texttt{nuclear\_rates\_q} for each reaction in \texttt{nuclear.NuclearRates}.  We define a wrapper function that allows us to vary the $d(p,\gamma)^3$He rate, which is critical for determining the deuterium abundance:
\begin{lstlisting}[language=Python]
import jax.numpy as jnp

def get_abundances_PRIMATDH(q_dpHe3g):
    return abundance_model(
        rho_g, # photon energy density
        rho_nu, # neutrino energy density
        # No contribution from relativistic species
        jnp.zeros_like(rho_g), # extra species 
                               # energy density
        jnp.zeros_like(rho_g), # extra species 
                               # pressure
        t_vec=t_vec, # vector of times at which  
                     # above energy densities 
                     # are given
        a_vec=a_vec, # vector of scale factor at 
                     # corresponding times

        # Construct the vector of rates; all 
        # entries other than the second entry 
        # should be zero
        nuclear_rates_q = jnp.concatenate((
            jnp.array([0]),
            jnp.array([q_dpHe3g]),
            jnp.zeros(len(nuc_rates.reactions)-2)
        ))
    )
\end{lstlisting}
Note that in constructing \texttt{nuclear\_rates\_q}, we had to know that $d(p,\gamma)^3$He is the second rate in the network; this can be determined by interrogating \texttt{abundance\_model.nuclear\_net.reactions\_names}.  

\begin{figure*}[ht]
    \centering
    \includegraphics[width=\linewidth]{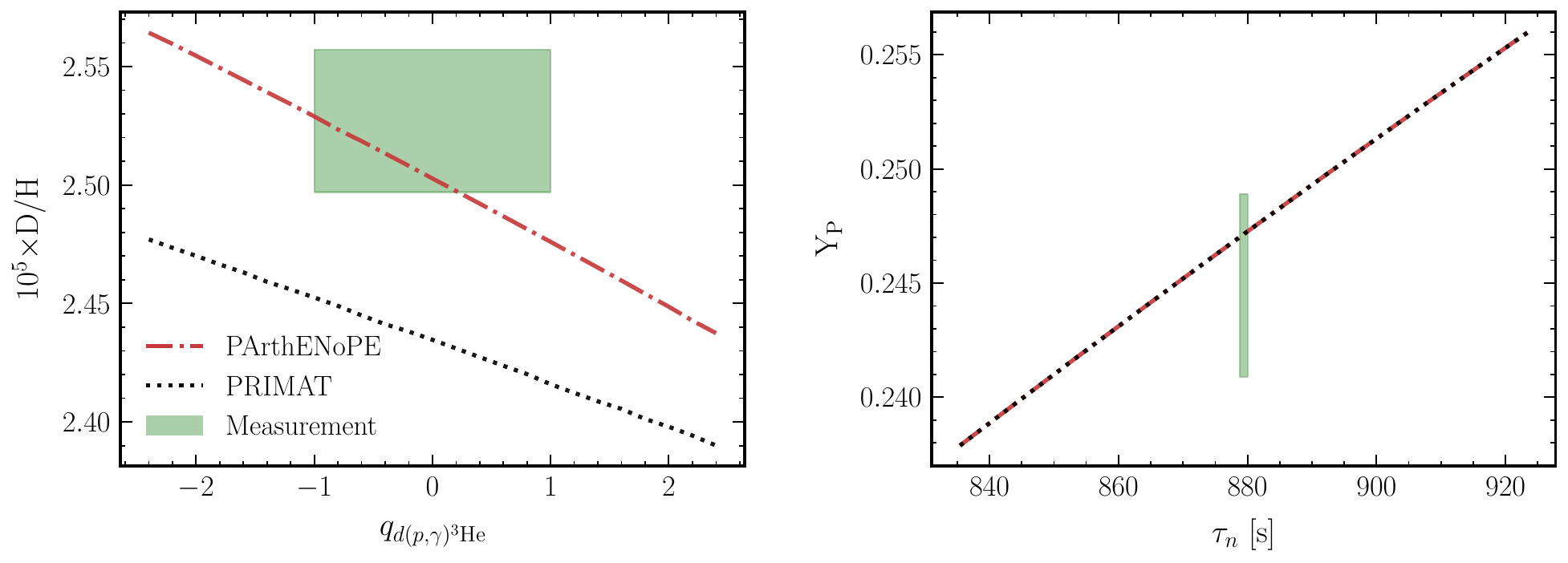}
    \caption{Dependence of the deuterium (\textit{left}) and helium-4 (\textit{right}) predicted abundances on key reaction-rate uncertainties in the network.  The deuterium prediction depends strongly on the rate of $d(p,\gamma)^3$He (as well as $d(d,p)t$ and $d(d,n)^3$He), while the helium-4 prediction depends strongly on the neutron lifetime.  The $x$-axes show the scaling of the $d(p,\gamma)^3$He rate (\textit{left}, see text) or the value of the neutron lifetime (\textit{right}), and the $y$-axes show the corresponding predictions for D/H (\textit{left}) and $Y_\mathrm{P}$ (\textit{right}).  The green regions have a height corresponding to $1\sigma$ the uncertainty of the measurements of D/H~\cite{Cooke_2018} and $Y_\mathrm{P}$~\cite{Aver_2015} and a width corresponding to the $1\sigma$ uncertainty of the quantities on the $x$-axes (defined to be 1 for the left plot or determined by measurement for the right plot \cite{Czarnecki:2018okw}).  The helium-4 prediction does not depend significantly on the choice of reaction network.}
    \label{fig:rates_deuterium_Yp}
\end{figure*}

Next, we vectorize the wrapper function, so that we can pass many values of \texttt{q\_dpHe3g} in at once:
\begin{lstlisting}[language=Python]
from jax import vmap, jit

get_abundances_vmapPRIMATDH = vmap(
    get_abundances_PRIMATDH
)
\end{lstlisting}

We can now set up an array of values of \texttt{q\_dpHe3g} to scan over:
\begin{lstlisting}
q_dpHe3g = jnp.arange(25) * 0.2 - 2.4
resultsPRIMATDH = 
    get_abundances_vmapPRIMATDH(q_dpHe3g)
# transpose results and put them into arrays
Yn, Yp, Yd, Yt, YHe3, Ya, YLi7, YBe7 = resultsPRIMATDH.T
\end{lstlisting}

Executing this code and plotting \texttt{Yd}/\texttt{Yp} as a function of the input value of \texttt{q\_dpHe3g} produces the ``PRIMAT" curve in the left panel of Figure~\ref{fig:rates_deuterium_Yp}.  
The relation between $q$ and the reaction rate is given in Eq.~\eqref{eq:q_definition}.

The other curves in Figure~\ref{fig:rates_deuterium_Yp} can be produced by using the PArthENoPE network instead of the PRIMAT network or by varying the \texttt{abundance\_model} input \texttt{tau\_n\_fac} instead of varying an entry in \texttt{nuclear\_rates\_q} to obtain the Y$_{\rm{P}}$ curves.   The left panel of Figure~\ref{fig:rates_deuterium_Yp} illustrates a known discrepancy between the PRIMAT and PArthENoPE reaction networks~\cite{Pitrou:2021vqr,Moscoso:2021xog,Pisanti:2020efz}. 
Each network produces a different prediction for the primordial deuterium abundance due to different choices made in computing $\overline{\lambda}(T)$ and $\sigma(T)$, especially for the rates $d(p,\gamma)^3\rm{He}$, $d(d,n)^3\rm{He}$ and $d(d,p)t$.
Meanwhile, the neutron lifetime is the dominant source of uncertainty for the Y$_{\rm{P}}$ prediction, as illustrated in the right panel of Figure~\ref{fig:rates_deuterium_Yp}.  This prediction is not sensitive to the choice of reaction network, since it is mostly determined by the neutron abundance at the end of the deuterium bottleneck, which is in turn mostly determined by the $n \leftrightarrow p$ weak conversion processes.

Next, we illustrate the effects of varying each of the rates in the key network by one standard deviation, as well as varying the baryon-to-photon ratio $\eta$ and the neutron lifetime $\tau_n$, on D/H, Y$_{\rm{P}}$, $^3$He/H, and $^7$Li/H\@.  To do so, we need to augment our wrapper function to take \texttt{eta\_fac} and \texttt{tau\_n\_fac}, as well as a vector for \texttt{nuclear\_rates\_q} rather than a single float.  However, we need only vectorize over \texttt{nuclear\_rates\_q}, since we are only interested in one value of \texttt{eta\_fac} and \texttt{tau\_n\_fac}.
\begin{lstlisting}[language=Python]
def get_abundance_eta_tau_q(eta_fac, tau_n_fac, nuclear_rates_q):

    Yn, Yp, Yd, Yt, YHe3, Ya, YLi7, YBe7 = abundance_model(
        rho_g,
        rho_nu,
        jnp.zeros_like(rho_g),
        jnp.zeros_like(rho_g),
        t_vec=t_vec,
        a_vec=a_vec,
        eta_fac=jnp.asarray(eta_fac),
        tau_n_fac=jnp.asarray(tau_n_fac),
        nuclear_rates_q=nuclear_rates_q
    )
    
    return jnp.array([
        Yn, Yp, Yd, Yt, YHe3, Ya, YLi7, YBe7
    ])

get_abundance_v = vmap(
    get_abundance_eta_tau_q, in_axes=(None, None, 0)
)
\end{lstlisting}
Passing \texttt{None} into the \texttt{in\_axes} argument of \texttt{vmap} indicates that input should not be vectorized, while passing an integer indicates the axis of the corresponding input that should be vectorized.  

We can generate fiducial values with 
\begin{lstlisting}[language=Python]
num_reactions = len(nuc_rates.reactions)
fiducial = get_abundance_eta_tau_q(
    1., 1., jnp.zeros(num_reactions)
)
\end{lstlisting}
and compare to the results obtained by varying each input by one sigma, one at a time:
\begin{lstlisting}[language=Python]
eta_vary = get_abundance_eta_tau_q(
    1.006708, 1., jnp.zeros(num_reactions)
)
tau_vary = get_abundance_eta_tau_q(
    1., 1.000682, jnp.zeros(num_reactions)
) 

# vary reaction rates one at a time to one sigma
reac_arrays = jnp.diag(jnp.ones(num_reactions))
reac_vary = get_abundance_v(1., 1., reac_arrays)
\end{lstlisting}
(Scaling the mean values for $\eta$ and $\tau_n$ by the \texttt{eta\_fac} and \texttt{tau\_n\_fac} used above gives the mean value plus one standard deviation reported in Ref.~\cite{Planck:2018vyg} and Ref.~\cite{Czarnecki:2018okw}, respectively.)  The results of this calculation are shown in Table~\ref{tab:sigmas}.

\begin{table}[t]
\begin{tabular}{>{\centering\arraybackslash}p{1.9cm}>{\centering\arraybackslash}p{1.4cm}>{\centering\arraybackslash}p{1.4cm}>{\centering\arraybackslash}p{1.4cm}>{\centering\arraybackslash}p{1.4cm}}

& \multicolumn{4}{c}{\textbf{\% Change}}   \\ \cline{2-5}
\textbf{Parameter} & \textbf{D/H} & \textbf{Y$_{\textrm{P}}$} & \textbf{$^3$He/H} & \textbf{$^7$Li/H} \\
\hline
$\Omega_b h^2$              & -1.1080 & 0.0266 & -0.3858 & -0.7925 \\
$\tau_n$            & 0.0051 & 0.0504 & 0.0081 & 0.1021 \\
\hline
\\

& \multicolumn{4}{c}{\textbf{\% Change}}   \\ \cline{2-5}
\textbf{Reaction} & \textbf{D/H} & \textbf{Y$_{\textrm{P}}$} & \textbf{$^3$He/H} & \textbf{$^7$Li/H} \\

\hline
$p(n,\gamma)d$           & -0.0884 & 0.0005 & 0.0371 & -0.0739 \\
$d(p,\gamma)^3$He        & -0.7700 & -0.0004 & 0.8370 & -0.2358 \\
$d(d,n)^3$He             & -0.5952 & 0.0051 & 0.2222 & -0.0737 \\
$d(d,p)t$                & -0.5000 & 0.0042 & -0.2791 & -0.0517 \\
$t(p,\gamma)^4$He        & -0.0129 & -0.0014 & 0.0708 & -0.4250 \\
$t(d,n)^4$He             & -0.0005 & -0.0016 & 0.0007 & -0.0950 \\
$t(\alpha,\gamma)^7$Li   & -0.0007 & -0.0016 & 0.0006 & 1.9822 \\
$^3$He$(n,p)t$           & 0.0367 & -0.0012 & -0.2088 & -0.0497 \\
$^3$He$(d,p)^4$He        & -0.0389 & -0.0015 & -0.8516 & -0.6109 \\
$^3$He$(\alpha,\gamma)^7$Be & 0.0002 & -0.0016 & 0.0010 & 1.3634 \\
$^7$Be$(n,p)^7$Li        & 0.0003 & -0.0016 & 0.0011 & 0.2689 \\
$^7$Li$(p,\alpha)^4$He   & -0.0158 & -0.0016 & 0.0012 & -3.1029 \\
\hline
\end{tabular}
\caption{Dependence of abundance predictions on uncertainties of various parameters.  The \texttt{key\_PRIMAT\_2023} network is used here for illustration.   \textit{Upper:} Percent change in abundances when the baryon abundance and neutron lifetime are varied within one standard deviation of their mean values (using $\Omega_b h^2$ determined by Ref.~\cite{Planck:2018vyg} and $\tau_n$ determined by Ref.~\cite{Czarnecki:2018okw}), compared to the case where all mean values are used.  \textit{Lower:} Percent change in abundances when each reaction in the ``key" network is varied within one sigma of its mean value.  To obtain these values, we set $q_i$ to 1, $i$ scanning over each reaction in the ``key" network, one at a time.  Then we compare the resulting prediction for the abundances to the result from using all mean values.}
\label{tab:sigmas}
\end{table}

\subsection{The Schramm Plot}\label{sec:schramm}

The Schramm plot (see \textit{e.g.}\ Ref.~\cite{ParticleDataGroup:2022pth}) shows the predicted primordial abundance of various light elements as a function of the baryon abundance $\Omega_b h^2$, in comparison with existing data on Y$_\text{P}$, D/H, and $\Omega_b h^2$ as measured from the CMB anisotropy power spectrum. 
The thickness of each line shown on the Schramm plot indicates the uncertainty of the BBN prediction due to uncertainties in the nuclear rates. 
In this Section, we demonstrate how to use LINX to generate the Schramm plot in SBBN\@.
We calculate the primordial abundances of light elements for 200 discrete values of $\Omega_b h^2$; at each point, we obtain 120 different predictions to determine the uncertainty, each with a slightly different value of $\tau_n$ (drawn from a Gaussian distribution given by $\tau_n = 879.4_{-0.6}^{+0.6}~\SI{}{\second}$) and $q_i$ (with each following a unit Gaussian distribution). 

As in the previous example, we begin by calculating the background thermodynamic evolution:

\begin{lstlisting}[language=Python]
from linx.background import BackgroundModel

SBBN_DNeff_bkg = BackgroundModel()
t_vec, a_vec, rho_g, rho_nu, rho_NP, p_NP, Neff = SBBN_DNeff_bkg(jnp.asarray(0.))
\end{lstlisting}

Once again, we create an instance of \texttt{abundances.AbundanceModel}, using the \texttt{key\_PRIMAT\_2023} network: 
\begin{lstlisting}[language=Python]
from linx.nuclear import NuclearRates
from linx.abundances import AbundanceModel

abundance_model = AbundanceModel(
    NuclearRates(nuclear_net='key_PRIMAT_2023')
)
\end{lstlisting}

Because we want to call this function over 100 different values of \texttt{eta\_fac} and 120 different sets of \texttt{(tau\_n\_fac, nuclear\_rates\_q)}, we define the following wrapper function: 
\begin{lstlisting}[language=Python]
import jax.numpy as jnp

def get_abundance_eta_q(
    eta_fac, tau_n_fac, nuclear_rates_q
):
    Yn, Yp, Yd, Yt, YHe3, Ya, YLi7, YBe7 = abundance_model(
        rho_g_vec, rho_nu_vec, 
        # No contribution from relativistic species
        jnp.zeros_like(rho_g_vec), 
        jnp.zeros_like(rho_g_vec), 
        t_vec=t_vec, a_vec=a_vec,
        eta_fac=jnp.asarray(eta_fac), 
        tau_n_fac=jnp.asarray(tau_n_fac), 
        nuclear_rates_q=nuclear_rates_q
    )
    return jnp.array(
        [Yn, Yp, Yd, Yt, YHe3, Ya, YLi7, YBe7]
    )
\end{lstlisting}
We now use \texttt{jax.vmap} to efficiently produce a vectorized version of the wrapper function: 
\begin{lstlisting}[language=Python]
from jax import vmap, jit

get_abundance_eta = vmap(
    get_abundance_eta_q, in_axes=(None, 0, 0)
)

get_abundance = jit(vmap(
    get_abundance_eta, in_axes=(0, None, None)
))
\end{lstlisting}
The first use of \texttt{vmap} produces a function \texttt{get\_abundance\_eta} that is identical to \texttt{get\_abundance\_eta\_q}, but can now take an array for \texttt{tau\_n\_fac}, and a 2D array for \texttt{nuclear\_rates\_q} (the input for \texttt{eta\_fac} must still be a float).
It outputs an array, which is the result obtained by scanning along the first axis of both \texttt{tau\_n\_fac} and \texttt{nuclear\_rates\_q} simultaneously. 
For example, if we pass a 1D array of length 120 for \texttt{tau\_n\_fac}, and another array of size $120 \times 12$ for \texttt{nuclear\_rates\_q} corresponding to 120 sets of $q_i$ for all 12 reactions in ``\texttt{key\_PRIMAT\_2023}'' (the number of entries in the first axis must be equal), the final output would be a 2D array with dimensions $120 \times 8$, representing the 8 elemental abundances over the 120 different draws of both the nuclear rates and $\tau_n$. 
The second use of \texttt{vmap} produces a function \texttt{get\_abundance} which is separately vectorized for the argument \texttt{eta\_fac}. At this point, we can pass a 1D array of length 200 for \texttt{eta\_fac}, a 1D array of length 120 for \texttt{tau\_n\_fac}, and another array of size $120 \times 12$ for \texttt{nuclear\_rates\_q}, to obtain a 3D array with dimensions $200 \times 120 \times 8$, corresponding to the 8 elemental abundances, indexed by our choice of the 200 values of $\Omega_b h^2$ and 120 different draws of $\tau_n$ and $q_i$. 
Finally, the function \texttt{jax.jit} just-in-time compiles the function, so that subsequent calls to the function with inputs \textit{of the same datatypes and dimensions} will use the compiled version of the function, leading to significant speed-up. 

We are now ready to generate our data.
We begin by defining the abscissa of interest for $\Omega_b h^2$, or rather the ratio of $\Omega_b h^2$ to the default value of $\Omega_b h^2$ stored in \texttt{const.Omegabh2}: 
\begin{lstlisting}[language=Python]
from linx.const import Omegabh2

eta_fac = jnp.logspace(
    jnp.log10(3e-3), jnp.log10(4e-2), num=200
) / Omegabh2
\end{lstlisting}
Next, we generate 120 different draws of $\tau_n$ and $q_i$: 
\begin{lstlisting}[language=Python]
import jax.random as rand

# 0 is the random seed.
key = jax.random.PRNGKey(0)
key, subkey = jax.random.split(key) 

# Should be 12 reactions in the key network
n_reactions = len(
    abundance_model.nuclear_net.reactions
)

sig = 0.6 / 879.4
tau_n_fac = 1. + rand.normal(subkey, (120,)) * sig

nuclear_rates_q = rand.normal(
    subkey, (120, n_reactions)
)
\end{lstlisting}
Finally, we can pass these arguments into \texttt{get\_abundance} to obtain the abundances: 
\begin{lstlisting}[language=Python]
# Dimensions 200 x 120 x 8
res = get_abundance(
    eta_fac, tau_n_fac, nuclear_rates_q
)
\end{lstlisting}
To make the Schramm plot, we obtain the mean value of the abundances over all 120 draws of the nuclear rates and also find the standard deviation: 
\begin{lstlisting}[language=Python]
# Output is Yn, Yp, Yd, Yt, YHe3, Ya, 
# YLi7 and YBe7

Yp = res[:,:,5] * 4
D_H = res[:,:,2] / res[:,:,1]
# T decays to He3 promptly
He3_H = (res[:,:,3]+res[:,:,4])/res[:,:,1]
# Be7 decays to Li7 promptly 
Li7_H = (res[:,:,6]+res[:,:,7])/res[:,:,1]

# Take mean/standard deviation along second 
# axis, i.e. over 120 nuclear rate draws
Yp_mean = jnp.mean(Yp, axis=1)
D_H_mean = jnp.mean(D_H, axis=1)
He3_H_mean = jnp.mean(He3_H, axis=1)
Li7_H_mean = jnp.mean(Li7_H, axis=1)

Yp_sig = jnp.std(Yp, axis=1)
D_H_sig = jnp.std(D_H, axis=1)
He3_H_sig = jnp.std(He3_H, axis=1)
Li7_H_sig = jnp.std(Li7_H, axis=1)
\end{lstlisting}
The exact runtime of this cell will depend on the user's hardware.  Executing all of the code shown above essentially performs 24,000 BBN abundance calculations, but takes approximately 800 seconds on an Apple M3 Macbook Pro, \textit{i.e.}\ about 0.03 seconds per run. 
\begin{figure}[t]
    \centering
    \includegraphics[width=\linewidth]{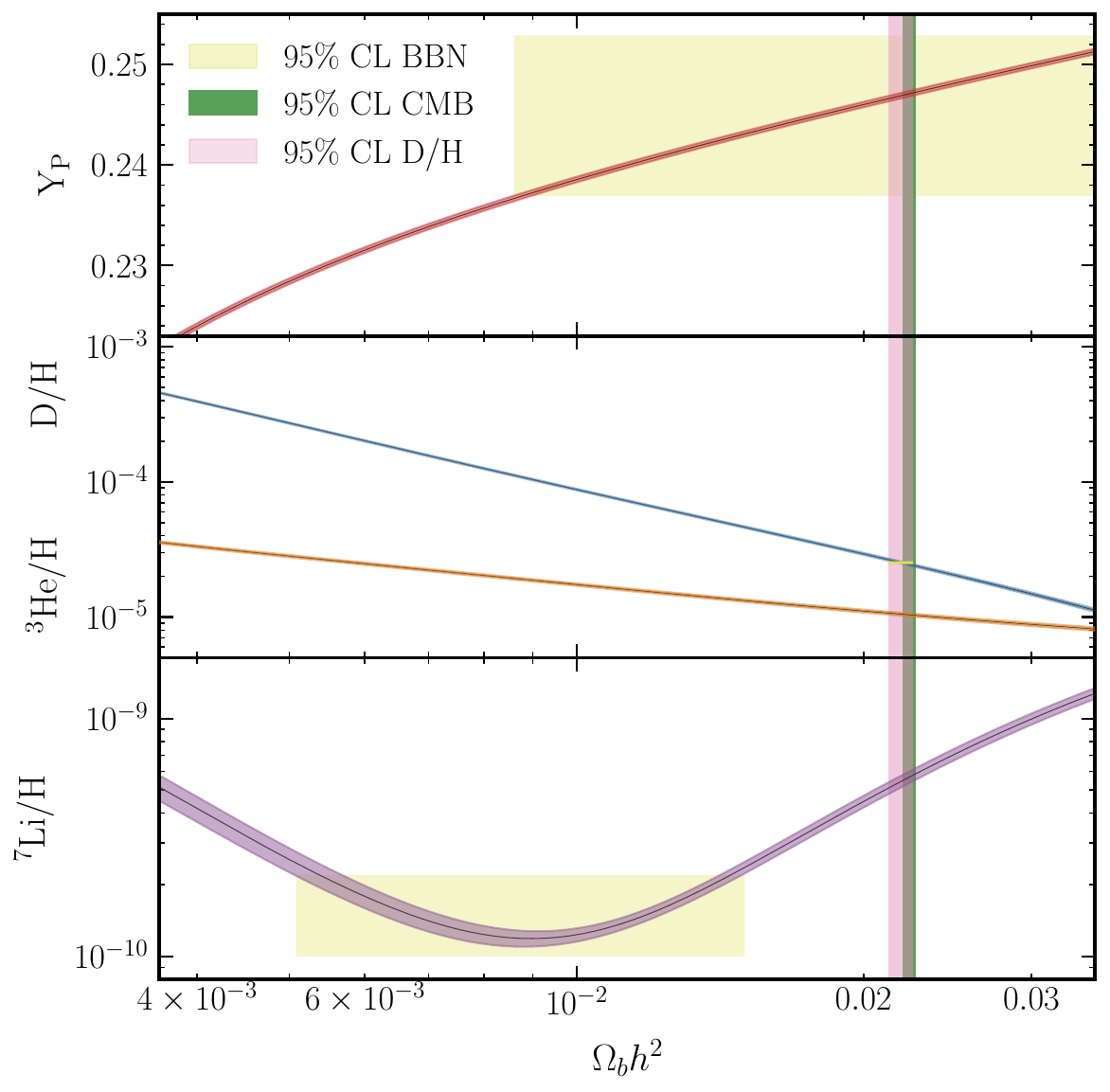}
    \caption{Schramm plot showing the 95\% CL bands for the predicted Y$_\text{P}$ (red), D/H (blue), $^3$He/H (brown, including primordial tritium), and $^7$Li/H (purple, including primordial $^7$Be), with uncertainties set by the nuclear rates, as a function of $\Omega_b h^2$. We use the ``\texttt{key\_PRIMAT\_2023}'' rates for this plot. Yellow boxes indicate the measurement uncertainty at the 95\% CL (denoted by their height) and the corresponding 95\% CL $\Omega_b h^2$ allowed based on the abundance of each element (denoted by their width). The inferred 95\% CL for $\Omega_b h^2$ from Planck TTTEEE+lowE+lensing+BAO is shown in green, while the approximate 95\% CL for $\Omega_b h^2$ from D/H alone is shown in pink (this has the same width as the yellow box for D/H).}
    \label{fig:Schramm}
\end{figure}

Figure~\ref{fig:Schramm} shows the Schramm plot that is obtained from the code detailed above. 
Lines show the predicted primordial abundances Y$_\text{P}$, D/H, $^3$He/H, and $^7$Li/H as a function of $\Omega_b h^2$, with the thickness of the lines showing the uncertainty of the prediction, obtained from the standard deviation. 
To compare these predictions with primordial abundance measurements, we use the following results from Refs.~\cite{Cooke_2018,Aver_2015,ParticleDataGroup:2022pth}:
\begin{alignat}{2}
    \textrm{D/H}^{\rm{obs}} &= 2.527\times10^{-5}\hspace{0.5cm}
    \sigma_{\textrm{D/H}^{\rm{obs}}} &&= 0.030\times10^{-5} \,, \nonumber \\
    \textrm{Y}_{\textrm{P}}^{\rm{obs}} &=0.2449  \hspace{2cm}
    \sigma_{\textrm{Y}_{\textrm{P}}^{\rm{obs}}} &&=0.004 \,, \nonumber \\
    ^7\textrm{Li/H}^{\rm{obs}} &= 1.6 \times 10^{-10}\hspace{0.5cm} 
    \sigma_{^7\textrm{Li/H}^{\rm{obs}}} &&= 0.3 \times 10^{-10} \,,
\end{alignat}
where all uncertainties are taken to be Gaussian and given at a confidence level (CL) of 68\%\@.
The vertical position and height of the yellow boxes indicate the median values and 95\% CL given by the values above (note that the box for D/H is small, but present), while the horizontal span of the yellow boxes approximately show the inferred range of $\Omega_b h^2$ by comparing our prediction to these measurements, including the uncertainty of our predictions. 
D/H is the main source of constraining power for $\Omega_b h^2$, due to the sensitivity of deuterium burning to the baryon abundance.
In green, we show the inferred 95\% CL for $\Omega_b h^2$ from Planck~\cite{Planck:2018vyg} TTTEEE\texttt{+}lowE\texttt{+}lensing\texttt{+}BAO and compare that to the approximate 95\% CL for $\Omega_b h^2$ from D/H alone, shown in pink (this has the same width as the yellow box for D/H). 
The discrepancy between the central value in the green band and the pink band is the $2\sigma$ discrepancy between $\Omega_b h^2$ as inferred from D/H using the ``\texttt{key\_PRIMAT\_2023}'' rates, already pointed out in Ref.~\cite{Pitrou:2020etk}. 

\begin{figure}
    \centering
    \includegraphics[width=\linewidth]{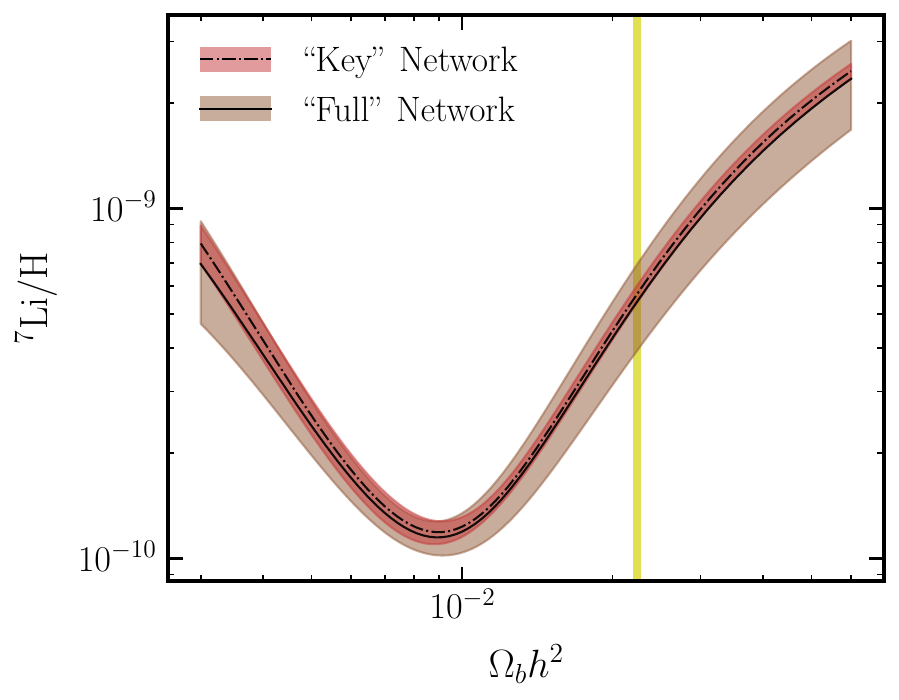}
    \caption{Comparison of the predicted $^7$Li/H abundance (including primordial $^7$Be) using the ``key" network (red) and ``full" network (brown) in LINX, both using the PRIMAT 2023 rates. The shaded regions represent the 95\% confidence intervals due to uncertainties in nuclear reaction rates. The vertical yellow line shows the Planck 2018 best-fit value for $\Omega_b h^2$.}
    \label{fig:smallvfull}
\end{figure}

This example focuses entirely on the ``key" network, though it is also possible to generate the same plot using the ``full" network.  We include a comparison between the LINX ``key" and ``full" networks in Table~\ref{tab:key_vs_full}, noting that there would be a difference in the $^7$Li/H curve in the Schramm plot if the ``full" network were used.  This comparison is highlighted in Figure~\ref{fig:smallvfull}, which shows that using the full network is important for predicting both the $^7$Li/H central value and its uncertainty for all values of $\Omega_b h^2$.

\begin{table*}[t]
    \centering
    \begin{tabular}{ccccc}
         \multicolumn{1}{p{2.5cm}}{\centering \textbf{Network}} & \multicolumn{1}{p{2.5cm}}{\centering \textbf{Y$_{\textrm{P}}$}} & \multicolumn{1}{p{2.5cm}}{\centering \textbf{D/H}} & \multicolumn{1}{p{2.5cm}}{\centering $^{\mathbf{3}}$\textbf{He/H} $\mathbf{\times 10^{5}}$} & \multicolumn{1}{p{2.5cm}}{\centering $^{\mathbf{7}}$\textbf{Li/H} $\mathbf{\times 10^{10}}$} 
         \\
        \hline
          \texttt{key\_PRIMAT\_2023} & 0.2471 & 2.4373 & 1.0385 &  5.5570 \\
         \texttt{full\_PRIMAT\_2023} & 0.2471 & 2.4397 & 1.0387 & 5.4925 \\
         \% Difference & 0 & -0.098 & -0.019 & 1.174\\
         \hline
    \end{tabular}
    \caption{Comparison of primordial abundance predictions from LINX using the ``key" and ``full" networks (the PRIMAT 2023 network is shown for illustration). The percentage differences are calculated using the ``full" network as fiducial.  All calculations use $\Omega_b h^2 = 0.02242$ and $\tau_n = 879.4$ s.}
    \label{tab:key_vs_full}
\end{table*}

\section{Gradient-Assisted Inference}
\label{sec:gradients}

Having introduced an efficient framework for predicting observables downstream of BBN, we can use LINX to constrain physical models through parameter estimation. 
In \citetalias{Giovanetti_2024}, we perform a joint CMB and BBN cosmological parameter inference for the $\Lambda$CDM and $\Lambda$CDM\texttt{+}$N_\text{eff}$ models using nested sampling.  
Here, we discuss gradient-assisted inference and provide an example that leverages some of the capabilities of LINX to perform joint CMB and BBN parameter estimation.

The main advantage of using gradient-assisted methods is efficiency.  While there may be overhead associated with computing the gradient in addition to the likelihood, gradient-assisted inference methods typically require significantly fewer likelihood evaluations than methods that do not use gradients.  Traditional (\textit{e.g.} ``random walk'') MCMC methods are especially susceptible to the curse of dimensionality---sampling quickly becomes inefficient beyond $\mathcal O(20)$ parameters. This is also the case for non-Markov Chain based methods, \textit{e.g.}\ nested sampling~\cite{Skilling_2004,Skilling_2006}. Even with fewer parameters, a computationally-expensive likelihood can benefit substantially from the availability of gradients with respect to the target parameters.

The next subsection describes the gradient-assisted methods used for a fully-differentiable CMB\texttt{+}BBN analysis.  Following that, we detail an analysis that utilizes neural transport reparameterization to perform efficient Hamiltonian Monte Carlo (HMC) sampling of the $\Lambda$CDM\texttt{+}$N_\text{eff}$ parameter posteriors.

\subsection{Methodology}\label{sec:grad_methods}
Here, we provide a heuristic description of each of the components of our gradient-assisted inference.  Appendix~\ref{app:gradients} includes more technical details required to implement these methods.
 
Sampling-based high-dimensional posterior inference can be especially inefficient when parameters are highly correlated; in these cases, a reparameterization of the target parameter space can enhance sampling efficiency.
The technique we use, known as neural transport reparameterization, involves mapping the posterior into a reparameterized space where the geometry is simpler and more amenable to efficient exploration; in particular, we find an invertible transformation between an approximation of the posterior and an uncorrelated multivariate Gaussian.  This will be essential for the Hamiltonian Monte Carlo~(HMC) described later. In what follows, the transformation has a set of parameters $\{\phi\}$ and learned by fitting these parameters such that the transformed distribution provides an initial approximation to the posterior distribution of interest. 

We develop the map between the posterior in the ``target" space (the space in which the parameters live) and this reparameterized space using a normalizing flow~\cite{rezende2015variational} and Stochastic Variational Inference~(SVI) (see \textit{e.g.}\ Ref.~\cite{wainwright2008graphical}).  A normalizing flow is a neural network that transforms distributions through a series of invertible mappings.  SVI is used to learn the optimal weights and biases of the neural network (the parameters $\{\phi\}$) of the normalizing flow, which takes one between the target space and the simple uncorrelated Gaussian distribution in the reparameterized space; more details about this procedure are available in Appendix~\ref{app:gradients}.

Once the mapping to the reparameterized space is learned, we can sample the multivariate Gaussian in this space using Markov Chain-Monte Carlo (MCMC) methods.  After obtaining a new sample in the reparameterized space, one can map it back to the target space using the map learned during neural transport reparameterization.  Finally, we can accept or reject that sample based on its trajectory in the target space, where the likelihood is defined.  This process is repeated until the desired number of samples are accepted.  

\begin{figure*}
    \centering
    \includegraphics[width=\linewidth]{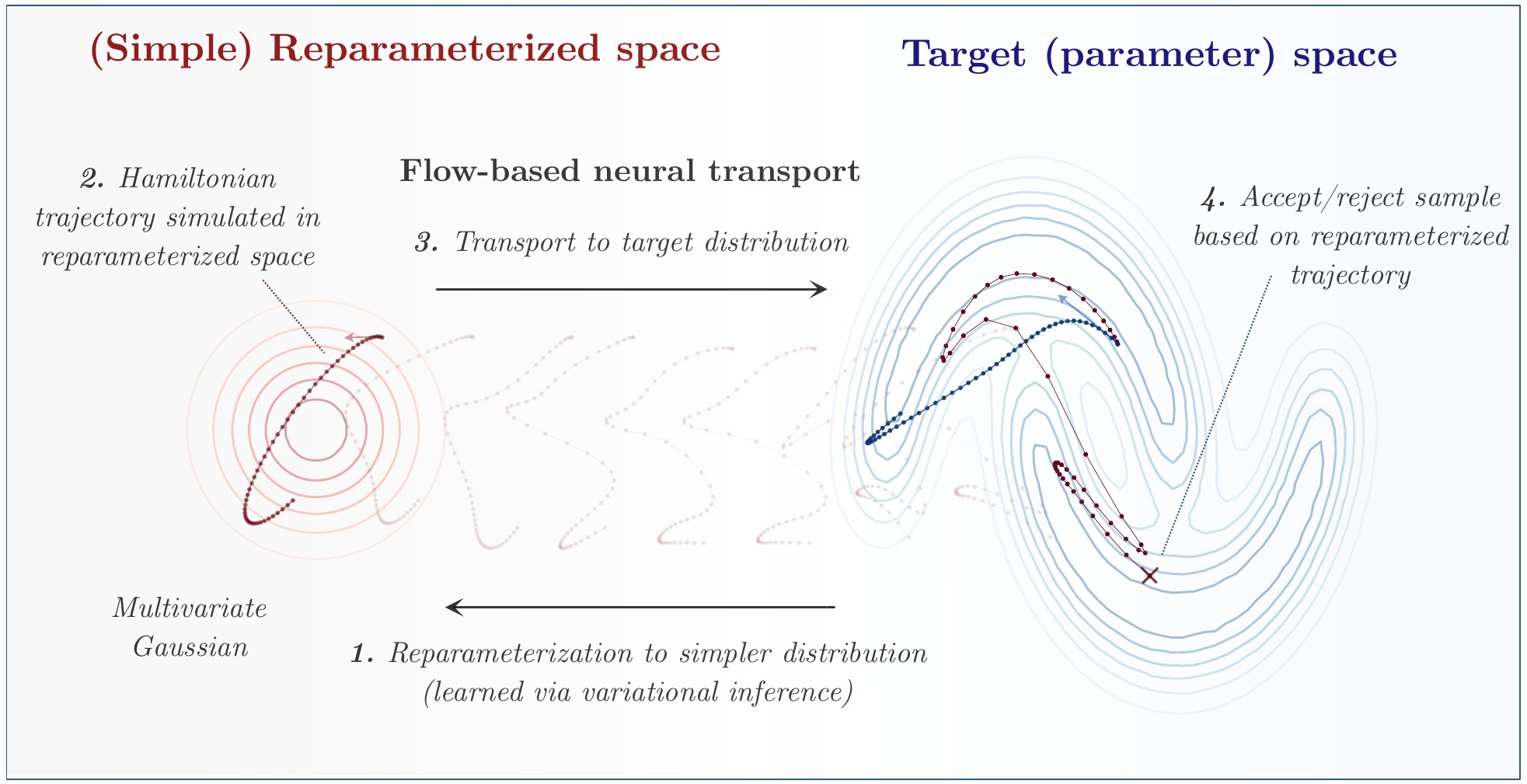}
    \caption{Schematic overview of a Hamiltonian Monte Carlo trajectory with neural transport reparameterization. \emph{(1)} An invertible mapping between a simple multivariate Gaussian distribution and a rough approximation to the posterior is initially learned via variational inference. \emph{(2)} A Hamiltonian trajectory is simulated in the simpler, reparameterized space (left, red points) where the distribution geometry is a simple multivariate Gaussian and \emph{(3)} transported to the target parameter space (right, red points). \emph{(4)} The sample accept/reject is performed using the log-density in the target space. Also shown for comparison is a trajectory natively simulated in the target space (blue points). Reparameterization results in more efficient exploration of the parameter space---the red trajectory on the right traversed the low-density region in the center of the target space, which is difficult for the natively simulated trajectory.}
    \label{fig:neutra}
\end{figure*}

This approach is particularly useful for posteriors with challenging geometries, such as those with multiple modes or strong correlations, where standard MCMC methods might struggle with low acceptance or a suboptimal exploration/exploitation trade-off.  For example, in a joint analysis with the CMB, we expect correlations between cosmological parameters, and we see strong correlations in our results (Figure~\ref{fig:corner_differentiable}, to be discussed below).  Exploring this parameter space with standard MCMC or without first performing neural transport reparametrization could lead to long convergence times, and so the procedure described above, which is depicted in Figure~\ref{fig:neutra}, is advantageous. \\

\noindent
\textbf{Hamiltonian Monte Carlo (HMC):} HMC is a Monte Carlo sampling method that utilizes gradient information to efficiently explore a posterior parameter space. We give a brief overview here as it relates to the analysis in the next subsection; see \textit{e.g.}\ Refs.~\cite{2011hmcm.book..113N,betancourt2015hamiltonian} for further details. 

HMC is based on the Metropolis-Hastings algorithm and uses Hamiltonian dynamics to propose and accept a new sample value. 
It introduces auxiliary ``momentum'' variables $\boldsymbol{q}$ and proposes new states by simulating Hamiltonian dynamics in the extended phase space $\boldsymbol{\theta} \rightarrow (\boldsymbol{\theta}, \boldsymbol{q})$, where $\boldsymbol{\theta}$ are the target parameters of interest (in our case, cosmological parameters). 

The first step of the HMC is to lift the posterior to a joint distribution over the phase space that includes the new ``momentum'' variables.  In other words, we promote the posterior $p(\boldsymbol{\theta}\mid x)$, where $x$ is the data, to $\pi(\boldsymbol{\theta}, \boldsymbol{q}) \equiv \pi(\boldsymbol{q}\mid\boldsymbol{\theta})\, p(\boldsymbol{\theta}\mid x)$. This decomposition into a conditional product ensures that marginalizing out the momentum coordinates (by exploring a large number of momentum configurations) recovers the posterior.  This process is analogous to the equilibration of gas particles in a container; the auxiliary momentum variables play a role similar to the random thermal motion of gas particles, which drives the system towards an equilibrium state that corresponds to the desired posterior distribution.

The resulting Hamiltonian is defined as:
\begin{equation}
H(\boldsymbol{\theta}, \boldsymbol{q}) = \underbrace{-\log p(\boldsymbol{\theta}\mid x)}_{\substack{U(\boldsymbol{\theta}):\\\text{``Potential Energy''}}} + \,\, \underbrace{\left(-\log \pi(\boldsymbol{q}\mid \boldsymbol{\theta})\right)}_{\substack{K(\boldsymbol{q}):\\\text{``Kinetic Energy''}}} \,.
\end{equation}
The ``potential energy" $U(\boldsymbol{\theta})$ is determined by the (un-normalized) target posterior (\textit{i.e.}\ the distribution we want to approximate through samples).  In other words, we treat the posterior as some surface having a potential energy.  Meanwhile the ``kinetic energy" $K(\boldsymbol{q})$ is typically chosen with the implementation and problem in mind.  A common choice, and the one made for this analysis, is $K(\boldsymbol{q}) = \frac{1}{2} \boldsymbol{q}^T M^{-1} \boldsymbol{q}$ for some mass matrix $M$.  $M$ is a hyperparameter that is tuned during a warm-up stage of HMC to achieve the desired acceptance probability of samples drawn from the reparameterized space.

To generate samples, we simulate trajectories in $\boldsymbol{\theta}$-space according to Hamiltonian dynamics.  Simulating these trajectories according to Hamilton's equations requires computing $\nabla_{\boldsymbol{\theta}} U(\boldsymbol{\theta}) = -\nabla_{\boldsymbol{\theta}} \log p\left(\boldsymbol{\theta} \mid x\right)$. This is related to the gradient of the likelihood $p\left(x \mid \boldsymbol{\theta}\right)$ through Bayes' theorem,\footnote{The gradient with respect to the difficult-to-compute evidence $\nabla_{\boldsymbol{\theta}}\log p(x)$ vanishes since it is a constant.} which can be computed directly with LINX\@.  This emphasizes the point that the availability of gradients is crucial for efficient HMC\@.  Each time a new sample is generated, the current sample points are assigned a momentum drawn from a distribution with probability density function $\propto \exp(-K(\boldsymbol{q}))$ and propagated following Hamiltonian dynamics and the gradients of the likelihood provided by LINX\@. 

\begin{figure*}[t]
\begin{lstlisting}[language=Python]
import jax.numpy as jnp

# Probabilistic program primitives
import numpyro.dist as dist
import numpyro.sample as sample
import numpyro.deterministic as deterministic

import linx.const as const

def model_bbn_cmb():
    """ Likelihood model for differentiable joint BBN and CMB analysis."""

    # Prior on ombh2
    eta_fac = sample("eta_fac", dist.Uniform(0.85, 1.2))
    ombh2 = deterministic("ombh2", eta_fac * const.Omegabh2)   

    # Prior on tau_n
    sigma_tau_fac = const.tau_n / const.sigma_tau_n
    tau_fac = sample("tau_fac", dist.Normal(1.0, sigma_tau_fac))

    # Prior on nuclear rates
    nuclear_rates_q = sample("nuclear_rates_q", dist.Normal(0., 1.0).expand([12]))

    # Prior on Neff
    DeltaNeff = sample("DNeff", dist.Uniform(-7, 7))
    params = jnp.array([DeltaNeff, eta_fac, tau_fac, *nuclear_rates_q])

    # BBN likelihood
    log_like, Neff = loglike_bbn(params)
    Neff = deterministic("Neff", Neff)

    if not bbn_only:

        # Additional LCDM priors
        omch2 = sample('omch2', dist.Uniform(0.08, 0.2))
        h = sample('h', dist.Uniform(0.4, 1.0))
        ns = sample('ns', dist.Uniform(0.88, 1.06))
        logA = sample('logA', dist.Uniform(2.5, 3.5))
        tau = sample('tau', dist.TruncatedDistribution(dist.Normal(0.0506, 0.0086), low=1e-4, high=0.15))

        # Add CMB likelihood
        log_like = log_like + loglike_cmb(jnp.array([ombh2, omch2, h, tau, ns, logA, Neff,]))

    # Total likelihood
    numpyro.factor('log_like', log_like)
\end{lstlisting}
\caption{NumPyro model specification for the joint BBN and CMB analysis. This model, specified as a probabilistic program, defines priors on cosmological parameters, computes the BBN likelihood from LINX (\texttt{loglike\_bbn}), and combines it with a differentiable CMB likelihood using the CosmoPower power spectrum emulator (\texttt{loglike\_cmb}). The model is targeted using gradient-based inference methods.}
\label{lst:model}
\end{figure*}

The accept/reject step of HMC differs from that of random walk MCMC, where the acceptance probability depends directly on the ratio of posterior log densities $p(\boldsymbol{\theta}\mid x)$ at the initial and proposed points.  Instead, 
each sample point is accepted or rejected based on the ratio of $\exp(-H(\boldsymbol{\theta}, \boldsymbol{q}))$ of the start and end of its trajectory.
Intuitively, this drives the Markov chain generated by HMC toward an equilibrium, Boltzmann-like distribution, which is precisely the desired posterior (after marginalizing over $\boldsymbol{q}$). 
Since Hamiltonian dynamics conserves the total energy $H$, simulating Hamiltonian trajectories connects points of constant density, up to numerical modeling, allowing for efficient exploration of the parameter space and producing trajectories that connect points that may be far apart in $\boldsymbol{\theta}$-space while maintaining high sample acceptance rates.  An acceptance probability of $\sim0.8$, as used here, is common and can be achieved by tuning hyperparameters (like the mass matrix $M$ described above, step size, and number of steps in the Hamiltonian trajectories) in a warm-up stage of HMC\@.  Acceptance probabilities that are too high could lead to substandard exploration of the full posterior parameter space.

HMC's performance can degrade in the presence of difficult posterior geometry, such as highly correlated parameters or multimodal posteriors, as illustrated by the blue trajectory in Figure~\ref{fig:neutra}. Taking steps to minimize the impact of these hurdles can make order-of-magnitude differences in convergence times, especially if the likelihood computation is slow enough to be considered expensive.  A common strategy to optimize posterior geometry is to reparameterize the target parameters into a space where the correlations are less severe; this is what we achieved with the neural transport reparametrization described above.  Further, the use of adaptive HMC variants, such as the No-U-Turn Sampler (NUTS) \cite{hoffman2014no}, which automatically tunes the step size and number of steps taken during each iteration, can help with these issues. NUTS employs a criterion to stop trajectories when they begin to turn back on themselves, ensuring efficient exploration without wasting computational resources on overly long trajectories. We use NUTS/HMC to approximate the posterior parameter space in the joint BBN and CMB analysis below. 

\subsection{Combined CMB and BBN Analysis}
\label{sec:joint_analysis}
LINX is differentiable and can be used for gradient-based parameter inference in a standalone manner.  Here, we show how it can be combined with differentiable CMB likelihoods to perform joint analyses. 

\citetalias{Giovanetti_2024} provides the inferred $\Lambda$CDM\texttt{+}$N_\text{eff}$ best-fit parameters obtained by combining LINX with the Boltzmann code CLASS~\cite{lesgourgues_2011a}, using a comprehensive Planck CMB power spectrum likelihood with all associated nuisance parameters. This analysis, performed via nested sampling \cite{Skilling_2004,Skilling_2006}, took roughly 8,500 CPU-hours parallelized over 192 processors. A directly analogous, differentiable version of this analysis is not possible at present, due to the unavailability of public differentiable CMB Boltzmann solvers (although see Ref.~\cite{Hahn:2023nvb} for recent progress in this direction). As a proxy, we use the CMB power spectrum emulator CosmoPower~\cite{SpurioMancini2022,Piras23,Piras:2024dml}, which trains differentiable surrogates---simple dense neural networks---on the outputs of the Boltzmann code CAMB~\cite{Lewis:2002ah}, to perform the same analysis. We use a set of high-resolution surrogate models originally developed for the South Pole Telescope Collaboration, which vary over $N_\text{eff}$ in addition to the six base $\Lambda$CDM parameters.\footnote{Models available at \href{https://github.com/alessiospuriomancini/cosmopower/tree/857c605eff3decb54593bf1259ce3df63a6fba76/cosmopower/trained_models/SPT_high_accuracy}{this GitHub link}.} Then, we implement a differentiable version of the Planck ``Lite'' likelihood~\cite{Prince:2019hse}, which includes a Gaussian approximation for the large-scale $TT$ component of the likelihood and a data-informed prior on large-scale polarization implicitly included through a prior on the optical depth to reionization.\footnote{\url{https://github.com/smsharma/cmb-differentiable-emu}.} Differentiable versions of more comprehensive Planck likelihoods including nuisance parameters (instrumental and otherwise) are not at present available.\footnote{Comprehensive differentiable likelihoods are however planned for some current and upcoming CMB analyses---see \textit{e.g.}\ Ref.~\cite{Balkenhol:2024sbv}.}

Meanwhile, the BBN likelihood is given by 
\begin{multline}
    -2 \log\mathcal{L}_{\rm{BBN}}= \left(\frac{\textrm{Y}_{\textrm{P}}^{\rm{pred}}(\Omega_b h^2,N_{\rm{eff}},\boldsymbol{\nu}_{\rm{BBN}})-\textrm{Y}_{\textrm{P}}^{\rm{obs}}}{\sigma_{\textrm{Y}_{\textrm{P}}^{\rm{obs}}}}\right)^2 \\+ 
    \left(\frac{\textrm{D/H}^{\rm{pred}}(\Omega_b h^2,N_{\rm{eff}},\boldsymbol{\nu}_{\rm{BBN}})-\textrm{D/H}^{\rm{obs}}}{\sigma_{\textrm{D/H}^{\rm{obs}}}}\right)^2 \,.
\end{multline}
The observed parameters are given by Ref.~\cite{Cooke_2018} (D/H) and Ref.~\cite{Aver_2015} (Y$_{\rm{P}}$):
\begin{alignat*}{2}
    \textrm{D/H}^{\rm{obs}} &= 2.527\times10^{-5}\hspace{0.5cm}
    \sigma_{\textrm{D/H}^{\rm{obs}}} &&= 0.030\times10^{-5}\\
    \textrm{Y}_{\textrm{P}}^{\rm{obs}} &=0.2449  \hspace{2cm}
    \sigma_{\textrm{Y}_{\textrm{P}}^{\rm{obs}}} &&=0.004 \,.
\end{alignat*}
For a given sample, $\Omega_b h^2$ and $N_{\rm{eff}}$ are the same across the BBN likelihood and the CMB likelihood.  $\boldsymbol \nu_{\rm{BBN}}$ are the BBN nuisance parameters (neutron decay lifetime and nuclear reaction rates), described in Sections~\ref{sec:nuclear_abundances_in_LINX} and~\ref{sec:examples}.  $\textrm{Y}_{\textrm{P}}^{\rm{pred}}(\Omega_b h^2,N_{\rm{eff}},\boldsymbol{\nu}_{\rm{BBN}})$ and $\textrm{D/H}^{\rm{pred}}(\Omega_b h^2,N_{\rm{eff}},\boldsymbol{\nu}_{\rm{BBN}})$ are computed by LINX\@.

\begin{figure*}[!htbp]
    \centering
    \includegraphics[width=\linewidth]{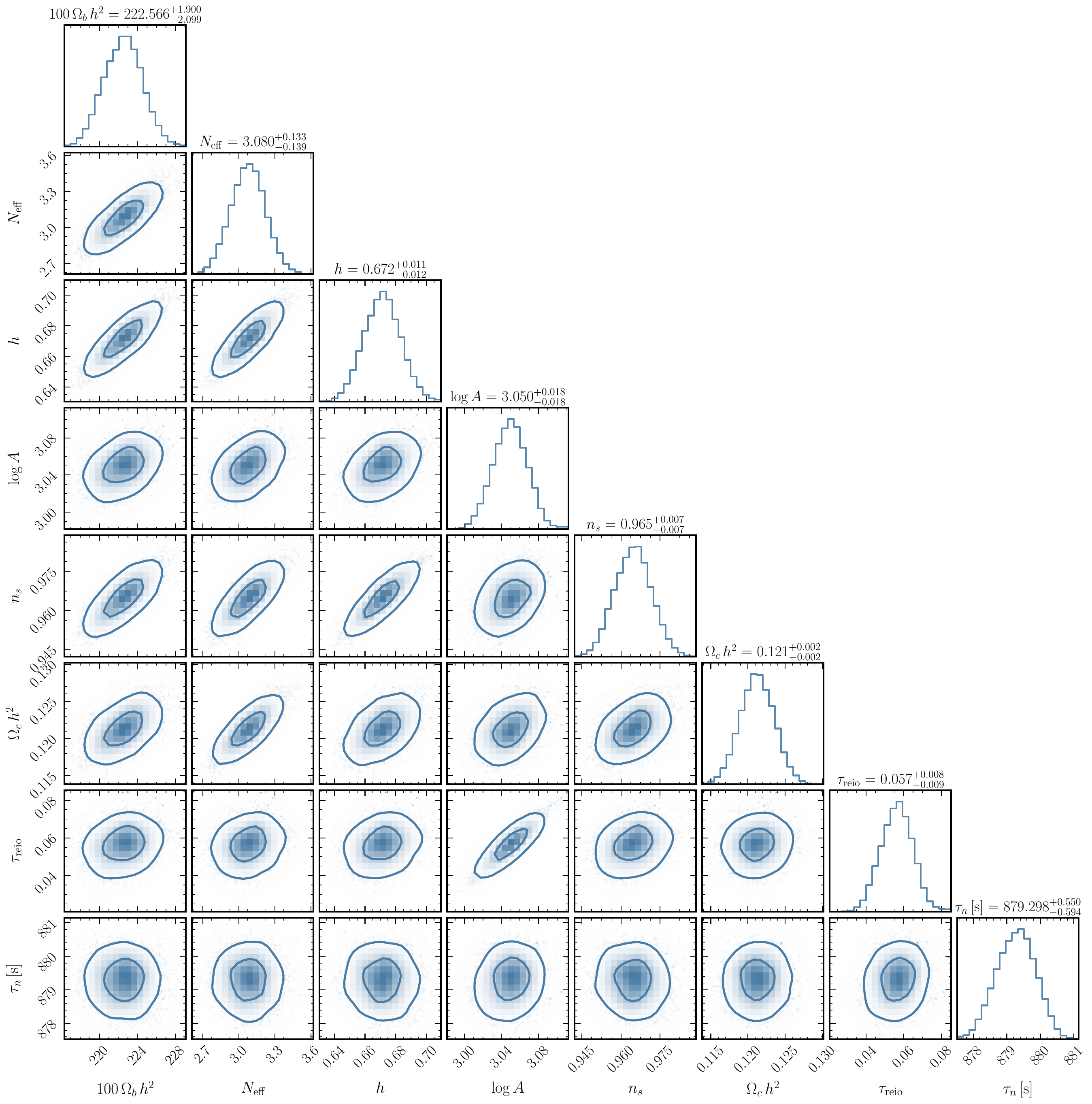}
    \caption{Joint and marginal parameter posterior distributions from the combined BBN and CMB gradient-based analysis. The median and middle-68\% containment values are listed above the corresponding marginals and the contours on the joint distributions correspond to 68/95\% containment regions.}
    \label{fig:corner_differentiable}
\end{figure*}

We implement the joint-likelihood CMB and BBN model using the probabilistic programming framework NumPyro~\cite{phan2019composable,bingham2019pyro}. The model, which contains 21 parameters of interest in total, is shown in Figure~\ref{lst:model} in the format specified using NumPyro. As in \citetalias{Giovanetti_2024}, each of the CMB and BBN likelihoods is computed at the same values of the input parameters $\Omega_b h^2$ and $N_{\rm{eff}}$. However, unlike in \citetalias{Giovanetti_2024}, the Y$_\mathrm{P}^\mathrm{pred}$ output of the BBN solver cannot be used as input to the power spectrum emulator, as it is not varied during its training. A single likelihood and gradient evaluation takes $\sim 2$ seconds on an M1 MacBook Pro.  We then use the neural transport reparameterization described above to learn a map from the target space to a reparameterized space, where we obtain the samples.  More details about the analysis are provided in Appendix~\ref{app:gradients}.

The resulting posterior from the gradient-based analysis is shown in Figure~\ref{fig:corner_differentiable}. We show the posteriors obtained for the six $\Lambda$CDM parameters (the baryon density $\Omega_b h^2$, the dimensionless Hubble parameter $h$, the amplitude of primordial scalar perturbations $A$, the scalar tilt $n_s$, the cold dark matter density $\Omega_c h^2$, and the optical depth to reionization $\tau_{\rm{reio}}$), $N_{\rm{eff}}$, and the neutron lifetime $\tau_n$. All other BBN nuisance parameters are included in the analysis, but not shown for clarity.  Results comparable to those from the full analysis of \citetalias{Giovanetti_2024} are obtained: 
\begin{align}
100 \Omega_b h^2 &= 
\begin{cases}
    &2.223_{-0.018}^{+0.018} \hspace{10pt} \textrm{\citetalias{Giovanetti_2024}}\\
    &2.226^{+0.019}_{-0.021} \hspace{10pt} \textrm{This work}
\end{cases}\\
N_\text{eff} &=
\begin{cases}
    & 3.08_{-0.13}^{+0.13} \hspace{10pt} \textrm{\citetalias{Giovanetti_2024}}\\
    & 3.08^{+0.13}_{-0.14} \hspace{10pt} \textrm{This work}
\end{cases}
\end{align}
We emphasize that this agreement is despite the fact that Y$_{\rm{P}}^{\rm{pred}}$ is not used as an input to the power spectrum emulator and  the CMB nuisance parameters are not sampled in this work.  This differentiable analysis is performed with just $\sim24$ hours of CPU-time on two laptop cores. This is a roughly three orders-of-magnitude gain in resource utilization, enabling efficient BBN analyses even on local compute hardware.

\section{Conclusions}\label{sec:discussion}

LINX adopts an accurate framework for BBN elemental abundance predictions while leveraging the capabilities of JAX to enable sampling of a large number of parameters and gradient-based inference. The key features and findings of this work can be summarized as follows:

\begin{enumerate}
    \item LINX provides accurate predictions of primordial element abundances, showing excellent agreement with established codes;

    \item LINX leverages JIT compilation through JAX, resulting in significantly-faster execution times compared to traditional BBN codes. This speed advantage enables rapid exploration of parameter spaces and efficient integration with other cosmological analysis tools.  Because of the ability to use nested sampling, Bayesian model comparison is also possible with LINX\@.  In~\citetalias{Giovanetti_2024}, we leverage this unique capability to conduct both BBN-only and CMB\texttt{+}BBN inference on cosmological parameters with nested sampling;

    \item LINX is designed with extensibility in mind, allowing users to modify or add nuclear reaction networks or beyond-Standard Model scenarios without hard-coding. This flexibility facilitates the exploration of various new-physics BBN scenarios;

    \item The code is fully differentiable, allowing for the use of gradient-based inference methods such as Hamiltonian Monte Carlo and Stochastic Variational Inference, enabling efficient parameter estimation in BBN studies; and 
    
    \item We have demonstrated the application of LINX by performing a joint CMB and BBN analysis using gradient-based methods. By combining LINX with a differentiable CMB emulator, we obtained results comparable to more computationally-intensive nested sampling approaches in a fraction of the time on standard computing hardware.

\end{enumerate}

A promising direction for future work is extending LINX to incorporate a wider range of new-physics scenarios, leveraging its modular design to explore beyond-Standard Model effects on BBN\@.  As discussed briefly in the conclusions of \citetalias{Giovanetti_2024}, these kinds of extensions can enable both BBN-only and joint CMB\texttt{+}BBN analyses to test models of new physics, including electromagnetically coupled WIMPs~\cite{Sabti:2019mhn,Sabti:2021reh,Giovanetti:2021izc,An:2022sva}, neutrino-coupled WIMPs~\cite{Berlin:2017ftj,Giovanetti:2024orj}, \SI{}{\mega\eV}-scale axion-like particles~\cite{Depta:2020zbh}, and hidden sector particles such as millicharged particles with dark radiation~\cite{Adshead:2022ovo}.

\section*{Acknowledgements}

We thank the PRyMordial team (Anne-Katherine Burns, Tim Tait, and Mauro Valli) for early access to their code and assistance with setup and usage throughout the beta-testing phase.  We thank the JAX team and the individuals who assisted us with JAX and JAX packages, including Dan Foreman-Mackey, Peter Hawkins, and Patrick Kidger.  
We thank Alessio Spurio Mancini for help with CosmoPower.  
We thank Calvin Chen, Chris Dessert, Vera Gluscevic, Marius Kongsore, Annika Peter, Katelin Schutz, and Ken Van Tilburg for testing and debugging LINX installation instructions.  
ML is supported by the Department of Energy~(DOE) under Award Number DE-SC0007968 as well as the Simons Investigator in Physics Award. 
HL was supported by the Kavli Institute for Cosmological Physics and the University of Chicago through an endowment from the Kavli Foundation and its founder Fred Kavli, and Fermilab operated by the Fermi Research Alliance, LLC under contract DE-AC02-07CH11359 with the U.S. Department of Energy, Office of Science, Office of High-Energy Physics.
SM is partly supported by the U.S. Department of Energy, Office of Science, Office of High Energy Physics of U.S. Department of Energy under grant Contract Number  DE-SC0012567. 
JTR is supported by NSF grant PHY-2210498.
This material is based upon work supported by the NSF Graduate Research Fellowship under Grant No.~DGE1839302. 
This work is supported by the National Science Foundation under Cooperative Agreement PHY-2019786 (The NSF AI Institute for Artificial Intelligence and Fundamental Interactions, \href{http://iaifi.org/}{http://iaifi.org/}).
This research was supported in part by Perimeter Institute for Theoretical Physics. Research at Perimeter Institute is supported by the Government of Canada through the Department of Innovation, Science and Economic Development and by the Province of Ontario through the Ministry of Research, Innovation and Science.  
This work was performed in part at the Aspen Center for Physics, which is supported by NSF grants PHY-1607611 and PHY-2210452.
This research was supported in part by grant NSF PHY-2309135 to the Kavli Institute for Theoretical Physics~(KITP).  
The work presented in this paper was performed on computational resources managed and supported by Princeton Research Computing.  This work was supported in part through the NYU IT High Performance Computing resources, services, and staff expertise. 
The authors made use of the Claude AI assistant (Anthropic, PBC) for writing assistance and editing.
This work makes use of CosmoPower~\cite{SpurioMancini2022,Piras23,Piras:2024dml} and the corner~\cite{corner}, diffrax~\cite{kidger2021on}, equinox~\cite{kidger2021equinox}, JAX~\cite{jax2018github,deepmind2020jax}, matplotlib~\cite{Hunter:2007}, numpy~\cite{harris2020array}, NumPyro~\cite{phan2019composable,bingham2019pyro}, and scipy~\cite{2020SciPy-NMeth} Python packages.  LINX documentation was generated using Sphinx~\cite{sphinx}.

\appendix
\section{Special Functions}
\label{app:special_functions}

As part of the development of LINX, we implement several special functions that are important to our calculations in JAX in a manner that is fully differentiable. 
Readers may find these special functions useful for their own projects in JAX\@. 

\subsection{Gamma Function}

The $\Gamma$ function is defined as 
\begin{alignat}{1}
    \Gamma(z) = \int_0^\infty dt\,  t^{z-1} e^{-t} 
\end{alignat}
for $\text{Re}(z) > 0$; this function can then be analytically continued over the entire complex plane.
Although JAX comes with its own version of the $\Gamma$ function under the \texttt{jax.scipy.special} module, their function does not support the use of complex numbers. 
We implement our own version of $\Gamma(z)$ that supports complex numbers, \texttt{special\_funcs.gamma}, using the Lanczos approximation~\cite{Lanczos:1964zz}, 
\begin{alignat}{1}
    \Gamma(z + 1) &\simeq \sqrt{2\pi} \left( z + g + \frac{1}{2} \right)^{z+1/2} e^{-(z + g + 1/2)} A_g(z)
\end{alignat}
for $\text{Re}(z + g + 1/2) > 0$, where $g$ is a real constant that we can choose arbitrarily and
\begin{multline}
    A_g(z) = \frac{1}{2} p_0(g) + p_1(g) \frac{z}{z+1} \\
    + p_2(g) \frac{z(z-1)}{(z+1)(z+2)} + \cdots \,,
\end{multline}
where $p_i(g)$ are a set of coefficients that depends on $g$. 
Methods for calculating the coefficients $p_i(g)$ are detailed in \textit{e.g.}\ Ref.~\cite{numericanaPaulGodfrey}, but we simply use the coefficients for $g = 7$ as provided by the GNU Scientific Library~\cite{Galassi:2019czg} (we refer the reader to our code for the exact values).
Finally, to extend the approximation across the entire complex plane, we use the identity 
\begin{alignat}{1}
    \Gamma(1 - z) \Gamma(z) = \frac{\pi}{\sin (\pi z)} \,.
\end{alignat}

\subsection{Polylogarithms}
\label{app:polylog}

The polylogarithm function is defined through the following series representation:
\begin{alignat}{1}
    \text{Li}_n(z) = \sum_{k=1}^\infty \frac{z^k}{k^n} \,,
    \label{eq:polylog_definition}
\end{alignat}
which is valid for any complex $n$ and complex $z$ with $|z| < 1$; it can then be analytically continued over the rest of the complex plane of $z$. 
Polylogarithms are not implemented in JAX\@.
We implement a differentiable version of polylogarithms for the case where $n > 1$ is an integer and $z$ is a real number with $z \leq 1$, the only regime of interest in thermodynamics. 
Following Ref.~\cite{crandall2006note}, we use three different formulas to perform the calculation in three different regimes of $z$. 
For $|z| < 1/2$, we perform a direct evaluation of the first 60 terms of the series Eq.~\eqref{eq:polylog_definition}. For $|z| > 2$, we use the relation 
    \begin{alignat}{1}
        \text{Li}_n(z) + (-1)^n \text{Li}_n(1/z) = - \frac{(2 \pi i)^n}{n!} B_n \left( \frac{\log z}{2 \pi i} \right) \,,
    \end{alignat}
where $B_n$ is the Bernoulli polynomial, and evaluate $\text{Li}_n(1/z)$ using Eq.~\eqref{eq:polylog_definition}. 
Finally, for the intermediate regime of $1/2 \leq |z| \leq 2$, we have the following expression for the polylogarithm, which is valid for $n > 1$: 
\begin{multline}
    \text{Li}_n (z) = \sum_{m=0}^\infty \! ^\prime \frac{\zeta(n - m)}{m!} \log^m z \\
    + \frac{\log^{n-1}z}{(n-1)!}[H_{n-1} - \log(- \log z)] \,,
\end{multline}
where $H_n$ is the $n$th harmonic number, $\zeta$ is the Riemann zeta function, and the $^\prime$ indicates that the sum excludes $m = n-1$. 

These three regimes are glued together using \texttt{jax.numpy.where} to ensure that the function is differentiable for any $z < 1$. 

\subsection{Modified Bessel Functions of the Second Kind}
\label{app:bessel_funcs}

The modified Bessel function of the second kind is defined as 
\begin{alignat}{1}
    K_\alpha(x) = \lim_{\nu \to \alpha} \frac{\pi}{2} \frac{I_{-\nu}(x) - I_\nu(x)}{\sin \nu \pi} \,,
\end{alignat}
where
\begin{alignat}{1}
    I_\nu(x) = \sum_{m = 0}^\infty \frac{1}{m! \, \Gamma(m + \nu + 1)} \left( \frac{x}{2} \right)^{2m + \nu} \,.
\end{alignat}
Note that as $\nu \to \alpha$ when $\alpha$ is an integer, both the numerator and the denominator tend to zero, making it important to specify the limit in the definition of the function. 
We first begin by describing how we calculate $K_0(x)$, following methods laid out in Ref.~\cite{zhang1996computation}. 
For $z < 9$, we use
\begin{multline}
    K_0(z) = - \left( \log (z/2) + \gamma \right) I_0(z) + \frac{(z/2)^2}{(1!)^2}  \\
    + \left(1 + \frac{1}{2} \right) \frac{(z/2)^4}{(2!)^2} + \left(1 + \frac{1}{2} + \frac{1}{3} \right) \frac{(z/2)^6}{(3!)^2} + \cdots \,,
\end{multline}
where $\gamma \simeq 0.5772$ is the Euler-Mascheroni constant, and $I_0(z)$ is the modified Bessel function of the first kind, implemented in JAX as \texttt{jax.scipy.special.i0}. 
On the other hand, for $z \geq 9$, we use the relation 
\begin{alignat}{1}
    I_0(z) K_0(z) = \frac{1}{2z} \left[1 - \frac{1}{2} \frac{(-1)}{(2z)^2} + \frac{1}{2} \cdot \frac{3}{4} \frac{(-1)(-9)}{(2z)^4} + \cdots \right] \,,
\end{alignat}
with the two regimes joined together by an appropriate use of \texttt{jax.numpy.where} to ensure differentiability. 

Next, to compute $K_1(z)$, we use the Wronskian relation~\cite{watson1995treatise} 
\begin{alignat}{1}
    I_0(z) K_1(z) + I_1(z) K_0(z) = \frac{1}{z} \,,
\end{alignat}
with $I_1(z)$ implemented as \texttt{jax.scipy.special.i1}. 

The last modified Bessel function we require is $K_2(z)$, which can now be easily obtained using recurrence relations:
\begin{alignat}{1}
    K_0(z) - K_2(z) = - \frac{2}{z} K_1(z) \,.
\end{alignat}
$K_0(z)$, $K_1(z)$ and $K_2(z)$ are implemented as \texttt{special\_funcs.K0}, \texttt{K1}, and \texttt{K2}, respectively. 

\section{Thermodynamic Variables with the Series Method}\label{app:series}

In this Appendix, we review how to obtain thermodynamic variables using the series method, first detailed in Ref.~\cite{Giovanetti:2021izc}. 
Evaluating these quantities this way is more expedient with no loss in accuracy and avoids the use of numerical quadrature, which is not implemented in JAX\@. 
For a particle of mass $m$ obeying either a Bose-Einstein or a Fermi-Dirac distribution, we can write its number density $n$, energy density $\rho$, and pressure $p$ as
\begin{alignat}{1}
    n_\pm &= \frac{g}{2 \pi^2 \beta^3} \int_{\beta m}^\infty dx \frac{x \sqrt{x^2 - (\beta m)^2}}{e^{x-\beta \mu} \pm 1} \,, \\
    \rho_\pm &= \frac{g}{2 \pi^2 \beta^4} \int_{\beta m}^\infty dx \frac{x^2 \sqrt{x^2 - (\beta m)^2}}{e^{x - \beta \mu} \pm 1} \,, \\
    p_\pm &= \frac{g}{6 \pi^2 \beta^4} \int_{\beta m}^\infty dx \frac{[x^2 - (\beta m)^2]^{3/2}}{e^{x - \beta \mu} \pm 1} \,,
\end{alignat}
where $g$ is the spin degrees of freedom, $\beta \equiv T^{-1}$, $T$ is the temperature, and $\mu$ is the (dimensionful) chemical potential associated with the equilibrium distribution. The `$+$' (`$-$') is appropriate for a Fermi-Dirac (Bose-Einstein) distribution. 

First, we define the function 
\begin{alignat}{1}
    A_n^\pm (\beta, \mu) = \int_{\beta m}^\infty dx \frac{x^n \sqrt{x^2 - (\beta m)^2}}{e^{x - \beta \mu} \pm 1} \,,
\end{alignat}
so that
\begin{alignat}{1}
    n_\pm &= \frac{g}{2 \pi^2 \beta^3} A_1 (\beta, \mu) \,, \\
    \rho_\pm &= \frac{g}{2 \pi^2 \beta^4} A_2 (\beta, \mu) \,, \\
    p_\pm &= \frac{g}{6 \pi^2 \beta^4} \left[A_2(\beta, \mu) - (\beta m)^2 A_0(\beta, \mu) \right] \,.
\end{alignat}
Noting that $(e^{x-\beta\mu} \pm 1)^{-1}$ can be rewritten as a geometric series, we find
\begin{multline}
    A_n^\pm (\beta, \mu) = \sum_{k=0}^\infty e^{(k+1)\beta \mu} \int_0^\infty dy \, \left[ y^2 + x^2 \right]^{(n-1)/2} \\
    \times y^2  (\mp 1)^k e^{-(k+1) \sqrt{y^2 + x^2}} \,,
\end{multline}
where $x \equiv m/T$. The new integral inside the summation can be related to various combinations of $K_0$, $K_1$, and $K_2$. 
Explicitly, we find
\begin{alignat}{2}
    A_0^\pm(\beta, \mu) &=&& \sum_{k=0}^\infty (\mp 1)^k e^{(k+1)\beta \mu} \frac{x}{k+1} K_1[(k+1) x] \,, \nonumber \\
    A_1^\pm(\beta, \mu) &=&& \sum_{k=0}^\infty (\mp 1)^k e^{(k+1)\beta\mu} \frac{x^2}{k+1} K_2[(k+1) x] \,, \nonumber \\
    A_2^\pm(\beta, \mu) &=&& \sum_{k=0}^\infty (\mp 1)^k e^{(k+1)\beta \mu} \bigg( \frac{3x^2}{(k+1)^2} K_2[(k+1) x] \nonumber \\
    & && \quad + \frac{x^3}{k+1} K_1[(k+1) x] \bigg) \,.
\end{alignat}
For $\mu = 0$, summing the first twenty terms is sufficient for precision of at least $0.1\%$ compared to numerical quadrature, although we warn that higher values of $\mu$ may require more terms to reach convergence.
We use \texttt{special\_funcs.K0}, \texttt{K1}, and \texttt{K2}, described in Appendix~\ref{app:bessel_funcs}, to evaluate these terms. 
For $x < 1/30$, we switch to the relativistic expression for these quantities, which are
\begin{alignat}{1}
    n_\pm &= \mp \frac{g}{\pi^2 \beta^3} \text{Li}_3 \left(\mp e^{\mu/T} \right) \,, \\
    \rho_\pm &= \mp \frac{3g}{\pi^2 \beta^4} \text{Li}_4 \left(\mp e^{\mu/T} \right) \,, \\
    p &= \rho/3 \,.
\end{alignat}
These relativistic expressions can be evaluated using \texttt{special\_funcs.Li}, detailed in App.~\ref{app:polylog}.

All thermodynamic quantities are calculated using functions stored in the \texttt{thermo} module and are fully differentiable. 

\section{Details of Gradient-Assisted Inference Methods}\label{app:gradients}

\setcounter{figure}{0} 
\renewcommand{\thefigure}{C\arabic{figure}}

Section~\ref{sec:grad_methods} outlined the gradient-assisted methods used in the differentiable CMB+BBN analysis.  Here, we first provide pedagogical details about neural transport reparameterization and Stochastic Variational Inference~(SVI), as it serves the neural transport reparameterization performed for the analysis in Section~\ref{sec:joint_analysis}.  We then provide details about additional options used  for that gradient-assisted analysis.
\\

\noindent

\subsection{Stochastic Variational Inference (SVI)} 
Variational inference (see \textit{e.g.}\ Ref.~\cite{wainwright2008graphical}) recasts posterior inference as an optimization, rather than sampling, problem. It introduces a variational family of distributions $q_\phi(\boldsymbol{\theta})$, $\boldsymbol{\theta}$ being the parameters of the model, parameterized by $\{\phi\}$ (hereinafter $\phi$), aiming to find the member of this family closest to the true posterior. The variational family could be a multivariate Gaussian, in which case the parameters $\phi$ would describe mean vector and (symmetric) covariance matrix, or it could be a more flexible distribution like a normalizing flow~\cite{rezende2015variational}. In the latter case, which we use here, the parameters $\phi$ define the neural network associated with the normalizing flow.

The goal of variational inference is to optimize $\phi$ by minimizing the (reverse) Kullbeck-Leibler (KL) divergence between the variational and true posterior distributions $p(\boldsymbol{\theta} \mid x)$,
\begin{equation}
\phi^* = \arg\min_\phi \left[\mathrm{KL}\left(q_\phi(\boldsymbol{\theta}) ||\, p(\boldsymbol{\theta} \mid x)\right)\right].
\end{equation}
The true posterior is however the inference target and \textit{a priori} unknown. A related computable quantity, the Evidence Lower BOund~(ELBO), is defined as
\begin{equation}
\mathrm{ELBO}(\phi) = \mathbb{E}_{q_\phi(\boldsymbol{\theta})}[\log p(x, \boldsymbol{\theta}) - \log q_\phi(\boldsymbol{\theta})] \,,
\end{equation}
where the joint distribution $p(x, \boldsymbol{\theta}) = p(x\mid\boldsymbol{\theta})\,p(\boldsymbol{\theta})$ can be computed via the prior and likelihood product and the expectation $\mathbb{E}_{q_\phi(\boldsymbol{\theta})}[f(\boldsymbol{\theta})] \equiv \int q_\phi(\boldsymbol{\theta}) f(\boldsymbol{\theta}) \, \mathrm{d}\boldsymbol{\theta}$ is estimated via Monte Carlo sampling. 

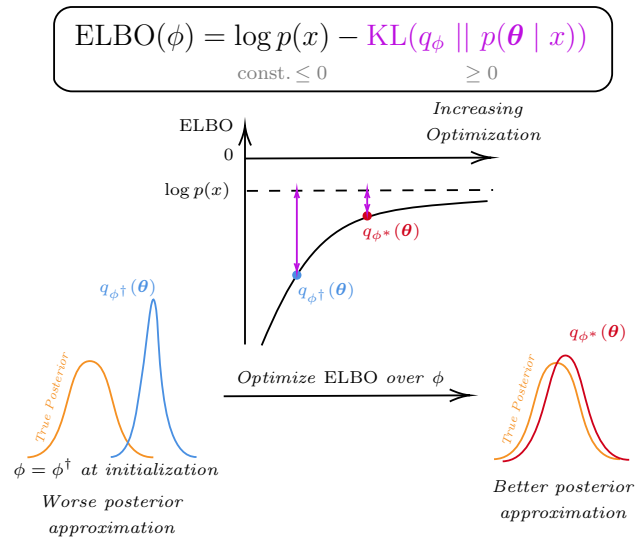
\begin{figure}[t!]
\centering
\resizebox{\columnwidth}{!}{
\begin{tikzpicture}[x=0.75pt,y=0.75pt,yscale=-1,xscale=1]
    
    %Straight Lines [id:da866196607988885] 
    \draw    (229.5,194) -- (229.5,77.5) ;
    \draw [shift={(229.5,75.5)}, rotate = 90] [color={rgb, 255:red, 0; green, 0; blue, 0 }  ][line width=0.75]    (10.93,-3.29) .. controls (6.95,-1.4) and (3.31,-0.3) .. (0,0) .. controls (3.31,0.3) and (6.95,1.4) .. (10.93,3.29)   ;
    %Straight Lines [id:da2380738529941473] 
    \draw    (229.5,93.5) -- (360.5,93.01) ;
    \draw [shift={(362.5,93)}, rotate = 179.78] [color={rgb, 255:red, 0; green, 0; blue, 0 }  ][line width=0.75]    (10.93,-3.29) .. controls (6.95,-1.4) and (3.31,-0.3) .. (0,0) .. controls (3.31,0.3) and (6.95,1.4) .. (10.93,3.29)   ;
    %Straight Lines [id:da1145401243540125] 
    \draw  [dash pattern={on 4.5pt off 4.5pt}]  (230.5,111) -- (363.5,110.5) ;
    %Curve Lines [id:da35673234380740815] 
    \draw    (238.5,194.5) .. controls (268.83,121.67) and (288.83,121.67) .. (360.5,116.5) ;
    %Curve Lines [id:da39807673820791445] 
    \draw [color={rgb, 255:red, 245; green, 145; blue, 35 }  ,draw opacity=1 ]   (112,255) .. controls (136.5,254.25) and (131,202.25) .. (146,203) .. controls (161,203.75) and (155,255.25) .. (180,255) ;
    %Curve Lines [id:da5891180896713824] 
    \draw [color={rgb, 255:red, 74; green, 144; blue, 226 }  ,draw opacity=1 ]   (157,254.83) .. controls (168.48,254.48) and (170.42,232.96) .. (173.67,211.83) .. controls (176.91,190.71) and (178.67,163.83) .. (180.67,170.83) .. controls (182.67,177.83) and (182.09,190.85) .. (184.33,212.17) .. controls (186.57,233.48) and (190.8,254.96) .. (203.33,254.83) ;
    %Straight Lines [id:da1674962211860942] 
    \draw    (218,222) -- (349,221.51) ;
    \draw [shift={(351,221.5)}, rotate = 179.78] [color={rgb, 255:red, 0; green, 0; blue, 0 }  ][line width=0.75]    (10.93,-3.29) .. controls (6.95,-1.4) and (3.31,-0.3) .. (0,0) .. controls (3.31,0.3) and (6.95,1.4) .. (10.93,3.29)   ;
    %Shape: Circle [id:dp025277523854106088] 
    \draw  [color={rgb, 255:red, 74; green, 144; blue, 226 }  ,draw opacity=1 ][fill={rgb, 255:red, 74; green, 144; blue, 226 }  ,fill opacity=1 ] (255.5,156.5) .. controls (255.5,155.4) and (256.4,154.5) .. (257.5,154.5) .. controls (258.6,154.5) and (259.5,155.4) .. (259.5,156.5) .. controls (259.5,157.6) and (258.6,158.5) .. (257.5,158.5) .. controls (256.4,158.5) and (255.5,157.6) .. (255.5,156.5) -- cycle ;
    %Shape: Circle [id:dp9633509342299156] 
    \draw  [color={rgb, 255:red, 208; green, 2; blue, 27 }  ,draw opacity=1 ][fill={rgb, 255:red, 208; green, 2; blue, 27 }  ,fill opacity=1 ] (293.5,124.5) .. controls (293.5,123.4) and (294.4,122.5) .. (295.5,122.5) .. controls (296.6,122.5) and (297.5,123.4) .. (297.5,124.5) .. controls (297.5,125.6) and (296.6,126.5) .. (295.5,126.5) .. controls (294.4,126.5) and (293.5,125.6) .. (293.5,124.5) -- cycle ;
    %Curve Lines [id:da559797716691192] 
    \draw [color={rgb, 255:red, 245; green, 145; blue, 35 }  ,draw opacity=1 ]   (364,256) .. controls (388.5,255.25) and (383,203.25) .. (398,204) .. controls (413,204.75) and (407,256.25) .. (432,256) ;
    %Curve Lines [id:da7149737930704392] 
\draw [color={rgb, 255:red, 208; green, 2; blue, 27 }  ,draw opacity=1 ]   (368.67,257.17) .. controls (393.17,256.42) and (391.67,198.33) .. (403,200) .. controls (414.33,201.67) and (412,256.58) .. (437,256.33) ;
    %Straight Lines [id:da3749291256281291] 
    \draw [color={rgb, 255:red, 189; green, 16; blue, 224 }  ,draw opacity=1 ]   (257.5,112.5) -- (257.5,152.5) ;
    \draw [shift={(257.5,154.5)}, rotate = 270] [color={rgb, 255:red, 189; green, 16; blue, 224 }  ,draw opacity=1 ][line width=0.75]    (4.37,-1.32) .. controls (2.78,-0.56) and (1.32,-0.12) .. (0,0) .. controls (1.32,0.12) and (2.78,0.56) .. (4.37,1.32)   ;
    \draw [shift={(257.5,110.5)}, rotate = 90] [color={rgb, 255:red, 189; green, 16; blue, 224 }  ,draw opacity=1 ][line width=0.75]    (4.37,-1.32) .. controls (2.78,-0.56) and (1.32,-0.12) .. (0,0) .. controls (1.32,0.12) and (2.78,0.56) .. (4.37,1.32)   ;
    %Straight Lines [id:da6555570260571157] 
    \draw [color={rgb, 255:red, 189; green, 16; blue, 224 }  ,draw opacity=1 ]   (295.5,112.5) -- (295.5,120.5) ;
    \draw [shift={(295.5,122.5)}, rotate = 270] [color={rgb, 255:red, 189; green, 16; blue, 224 }  ,draw opacity=1 ][line width=0.75]    (4.37,-1.32) .. controls (2.78,-0.56) and (1.32,-0.12) .. (0,0) .. controls (1.32,0.12) and (2.78,0.56) .. (4.37,1.32)   ;
    \draw [shift={(295.5,110.5)}, rotate = 90] [color={rgb, 255:red, 189; green, 16; blue, 224 }  ,draw opacity=1 ][line width=0.75]    (4.37,-1.32) .. controls (2.78,-0.56) and (1.32,-0.12) .. (0,0) .. controls (1.32,0.12) and (2.78,0.56) .. (4.37,1.32)   ;
    %Rounded Rect [id:dp13945528241657845] 
    \draw   (126.4,21.32) .. controls (126.4,15.95) and (130.75,11.6) .. (136.12,11.6) -- (418.08,11.6) .. controls (423.45,11.6) and (427.8,15.95) .. (427.8,21.32) -- (427.8,47.48) .. controls (427.8,52.85) and (423.45,57.2) .. (418.08,57.2) -- (136.12,57.2) .. controls (130.75,57.2) and (126.4,52.85) .. (126.4,50.48) -- cycle ;
    
    % Text Node
    \draw (216.5,86.9) node [anchor=north west][inner sep=0.75pt]  [font=\scriptsize]  {$0$};
    % Text Node
    \draw (184.5,104.9) node [anchor=north west][inner sep=0.75pt]  [font=\scriptsize]  {$\log p( x)$};
    % Text Node
    \draw (192.5,70.9) node [anchor=north west][inner sep=0.75pt]  [font=\scriptsize]  {$\mathrm{ELBO}$};
    % Text Node
    \draw (114.08,247.95) node [anchor=north west][inner sep=0.75pt]  [font=\tiny,color={rgb, 255:red, 245; green, 166; blue, 35 }  ,opacity=1 ,rotate=-285.97] [align=left] { \textit{\textcolor[rgb]{0.96,0.57,0.14}{True Posterior}}};
    % Text Node
    \draw (325,61.5) node [anchor=north west][inner sep=0.75pt]   [align=left] {\begin{minipage}[lt]{43.19pt}\setlength\topsep{0pt}
    \begin{center}
    {\scriptsize  \textit{Increasing }}\\{\scriptsize  \textit{Optimization}}
    \end{center}
    
    \end{minipage}};
    % Text Node
    \draw (113,274) node [anchor=north west][inner sep=0.75pt]   [align=left] {\begin{minipage}[lt]{64.61pt}\setlength\topsep{0pt}
    \begin{center}
    {\scriptsize \textit{Worse posterior }}\\{\scriptsize  \textit{approximation}}
    \end{center}
    
    \end{minipage}};
    % Text Node
    \draw (149.5,156.9) node [anchor=north west][inner sep=0.75pt]  [font=\scriptsize,color={rgb, 255:red, 74; green, 144; blue, 226 }  ,opacity=1 ]  {$q_{\phi ^{\dagger }}(\boldsymbol{\theta })$};
    % Text Node
    \draw (136,18.4) node [anchor=north west][inner sep=0.75pt] [font=\large]  {$\mathrm{ELBO}( \phi ) =\log p( x) -\textcolor[rgb]{0.74,0.06,0.88}{\mathrm{KL}( q_{\phi } \ ||\ p(\boldsymbol{\theta } \mid x))}$};
    % Text Node
    \draw (220,207) node [anchor=north west][inner sep=0.75pt]   [align=left] {\begin{minipage}[lt]{88.76pt}\setlength\topsep{0pt}
    \begin{center}
    {\scriptsize \textit{Optimize }$\displaystyle \mathrm{ELBO}$ \textit{over }$\displaystyle \phi $}
    \end{center}
    
    \end{minipage}};
    % Text Node
    \draw (256.5,158.9) node [anchor=north west][inner sep=0.75pt]  [font=\scriptsize,color={rgb, 255:red, 74; green, 144; blue, 226 }  ,opacity=1 ]  {$q_{\phi ^{\dagger }}(\boldsymbol{\theta })$};
    % Text Node
    \draw (291.5,126.9) node [anchor=north west][inner sep=0.75pt]  [font=\scriptsize,color={rgb, 255:red, 208; green, 2; blue, 27 }  ,opacity=1 ]  {$q_{\phi ^{*}}(\boldsymbol{\theta })$};
    % Text Node
    \draw (366.08,248.94) node [anchor=north west][inner sep=0.75pt]  [font=\tiny,color={rgb, 255:red, 245; green, 166; blue, 35 }  ,opacity=1 ,rotate=-285.97] [align=left] {\textit{\textcolor[rgb]{0.96,0.57,0.14}{True Posterior}}};
    % Text Node
    \draw (357,266) node [anchor=north west][inner sep=0.75pt]   [align=left] {\begin{minipage}[lt]{64.52pt}\setlength\topsep{0pt}
    \begin{center}
    {\scriptsize \textit{Better posterior }}\\{\scriptsize \textit{approximation}}
    \end{center}
    
    \end{minipage}};
    % Text Node
    \draw (403,181.4) node [anchor=north west][inner sep=0.75pt]  [font=\scriptsize,color={rgb, 255:red, 208; green, 2; blue, 27 }  ,opacity=1 ]  {$q_{\phi ^{*}}(\boldsymbol{\theta })$};
    % Text Node
    \draw (224,42) node [anchor=north west][inner sep=0.75pt]  [font=\footnotesize,color={rgb, 255:red, 128; green, 128; blue, 128 }  ,opacity=1 ] [align=left] { const.$\displaystyle \, \leq 0$};
    % Text Node
    \draw (347,42) node [anchor=north west][inner sep=0.75pt]  [font=\footnotesize,color={rgb, 255:red, 128; green, 128; blue, 128 }  ,opacity=1 ] [align=left] {$\displaystyle \geq 0$};
    % Text Node
    \draw (103,254) node [anchor=north west][inner sep=0.75pt]   [align=left] {{\scriptsize $\displaystyle \phi =\phi ^{\dagger }$ \textit{at initialization}}};

\end{tikzpicture}   
}
\caption{An illustration of the decomposition of the Evidence Lower BOund (ELBO) via the log-evidence $\log p(x)$ and KL divergence between the approximate posterior $q_\phi$ (parameterized by $\phi$) and true posterior $p(\boldsymbol{\theta}\mid x)$. ELBO maximization via gradient ascent over the variational parameters $\phi$ reduces the gap between the ELBO and $p(x)$, leading to a small KL divergence between the approximate and true posteriors: $q_{\phi^*} \approx p(\boldsymbol{\theta}\mid x)$. An un-optimized ELBO on the other hand corresponds to $q_{\phi^{\dagger}} \not\approx p(\boldsymbol{\theta}\mid x)$}
\label{fig:svi}
\end{figure}

The ELBO satisfies the identity (see \textit{e.g.}\ Ref.~\cite{ghojogh2021factor})
\begin{equation}
\log p(x) = \mathrm{ELBO}(\phi) + \mathrm{KL}\left(q_\phi(\boldsymbol{\theta}) ||\, p(\boldsymbol{\theta} \mid x)\right).
\label{eq:elbo}
\end{equation}
Since $p(x)$ is constant for given data, and the KL divergence is strictly positive, maximizing the ELBO is thus equivalent to minimizing the KL divergence between $q_\phi$ and the true posterior.
In practice, $\phi$ is optimized by gradient ascent on the ELBO, and expectations in the ELBO are approximated by Monte Carlo samples from $q_\phi$ (hence ``stochastic''). With the optimal $\phi^*$, $q_{\phi^*}(\boldsymbol{\theta}) \approx p\left(\boldsymbol{\theta} \mid x\right)$ provides an approximation to the posterior. The key advantage is that this turns posterior inference into a more tractable optimization problem, and $q_{\phi^*}$ can then be sampled from to get representative posterior samples. Again, computing the gradient estimates of $\nabla_\phi \mathrm{ELBO}(\phi)$ is necessary for efficient gradient ascent with arbitrary models \cite{kucukelbir2017automatic}. A schematic illustration of ELBO and its decomposition as used for posterior approximation is shown in Figure~\ref{fig:svi}. \\

\subsection{Neural Transport Reparameterization} 

Reparameterizing the target posterior is a common way to improve MCMC sampling efficiency, especially for correlated posterior distributions. A specific method to automatically learn a good reparameterization is \emph{neural transport} reparameterization \cite{hoffman2019neutra}, the key idea behind which is to learn a transformation from a simple base distribution (typically a standard uncorrelated Gaussian) to an approximation of the target posterior using a flexible neural network. This transformation is learned via variational inference (as described above), minimizing the KL divergence between the target distribution and the true posterior via ELBO maximization.

Using a normalizing flow~\cite{rezende2015variational,kingma2016improved} as the variational distribution, we initially perform 2000 gradient ascent steps using the Adam~\cite{kingma2014adam} optimizer with learning rate $6\times 10^{-4}$ using the ELBO in Eq.~\eqref{eq:elbo} as the objective function. The normalizing flow learns a mapping between a simple ``base'' distribution---an uncorrelated, multivariate Gaussian $p(z)$---and an approximation of the target posterior.

Using the simple base distribution $p(z)$ as the reparameterized posterior, we run NUTS/HMC to draw 4000 samples across 2 chains following 500 tuning steps (which adjusts the step size, number of steps, and the dense mass matrix targeting a proposal acceptance rate of 0.8). The resulting MCMC chains show good convergence properties, with split Gelman-Rubin statistics (ratio of intra-chain to inter-chain variance)  $< 1.04$ for all parameters and effective sample sizes $\gtrsim 100$. The initial neural reparameterization was found to be crucial to ensure convergence in the number of steps used; the model shows poor convergence properties without it.

\bibliography{references_formalism}

\appendix

\end{document}